\def\HA{{H$\alpha$}}

\def\peryr{\:\rm{\,yr^{-1}}}

\def\LUM{\:{\rm ergs\:s^{-1}}}
\def\FLUX{\:{\rm ergs\:cm^{-2}\:s^{-1}}}
\def\PHOTFLUX{\:{\rm phot\:cm^{-2}\:s^{-1}}}

\def\sii{[\ion{S}{2}]}

\def\oiii{[\ion{O}{3}]}

\def\hii{\ion{H}{2}}

\documentclass[12pt,preprint]{aastex}

\begin{document}


\newcommand{\MSOL}{\mbox{$\:M_{\sun}$}}  

\newcommand{\EXPN}[2]{\mbox{$#1\times 10^{#2}$}}
\newcommand{\EXPU}[3]{\mbox{\rm $#1 \times 10^{#2} \rm\:#3$}}  
\newcommand{\POW}[2]{\mbox{$\rm10^{#1}\rm\:#2$}}
\newcommand{\SING}[2]{#1$\thinspace \lambda $#2}
\newcommand{\MULT}[2]{#1$\thinspace \lambda \lambda $#2}
\newcommand{\CHINU}{\mbox{$\chi_{\nu}^2$}}
\newcommand{\vsini}{\mbox{$v\:\sin{(i)}$}}

\newcommand{\fuse}{{\em FUSE}}
\newcommand{\hst}{{\em HST}}
\newcommand{\iue}{{\em IUE}}
\newcommand{\euve}{{\em EUVE}}
\newcommand{\einstein}{{\em Einstein}}
\newcommand{\rosat}{{\em ROSAT}}
\newcommand{\chandra}{{\em Chandra}}
\newcommand{\xmm}{{\em XMM-Newton}}
\newcommand{\swift}{{\em Swift}}
\newcommand{\asca}{{\em ASCA}}
\newcommand{\galex}{{\em GALEX}}
\newcommand{\cxo}{CXO\,1337}


\shorttitle{SNRs in M33}
\shortauthors{Long \etal}

\title{
{A Deep \chandra\  ACIS Survey of M83\footnote{
Based on observations made with NASA's \chandra\ X-ray Observatory. 
NASA's \chandra\ Observatory is operated by Smithsonian Astrophysical Observatory 
under contract \# NAS83060 and the data were obtained through program GO1-12115.
}}
}

\author{
Knox S. Long\altaffilmark{2},
Kip D. Kuntz\altaffilmark{3},
William P. Blair\altaffilmark{3},
Leith Godfrey\altaffilmark{4,5},
Paul P. Plucinsky\altaffilmark{6}, 
Roberto Soria\altaffilmark{4}, 
Christopher Stockdale\altaffilmark{7},
and
P. Frank Winkler\altaffilmark{8}
}

\altaffiltext{2}{Space Telescope Science Institute, 3700 San Martin Drive, 
Baltimore, MD, 21218;  long@stsci.edu}
\altaffiltext{3}{The Henry A. Rowland Department of Physics and Astronomy, 
Johns Hopkins University, 3400 N. Charles Street, Baltimore, MD, 21218; 
kuntz@pha.jhu.edu, wpb@pha.jhu.edu}
\altaffiltext{4}{Curtin Institute of Radio Astronomy, Curtin University, 
1 Turner Avenue, Bentley WA6102, Australia; leith.godfrey@icrar.org, roberto.soria@icrar.org}
\altaffiltext{5}{Netherlands Institute for Radio Astronomy (ASTRON), Postbus 2, 7990 AA Dwingeloo, the Netherlands'; godfrey@astron.nl}
\altaffiltext{6}{Harvard-Smithsonian Center for Astrophysics, 60 Garden Street, 
Cambridge, MA 02138; plucinsky@cfa.harvard.edu}
\altaffiltext{7}{Department of Physics, Marquette University, PO Box 1881, Milwaukee, WI 53201-1881, USA; christopher.stockdale@marquette.edu}
\altaffiltext{8}{Department of Physics, Middlebury College, Middlebury, VT, 05753; 
winkler@middlebury.edu}

\begin{abstract}

We have obtained a series of deep X-ray images of the nearby galaxy M83 using \chandra, with a total exposure of 729 ks. Combining the new data with earlier archival observations totaling 61 ks, we find 378 point sources within the D$_{25}$ contour of the galaxy.  We find  80 more sources,  mostly background AGN, outside of the D$_{25}$ contour. Of the X-ray sources, 47 have been detected in a new radio survey of M83 obtained using the Australia Telescope Compact Array.  Of the X-ray sources, at least 87 seem likely to be supernova remnants (SNRs), based on a combination of their  properties in X-rays and at other wavelengths.   We attempt to classify the point source population of M83 through a combination of spectral and temporal analysis. As part of this effort, we carry out an initial spectral analysis of the 29 brightest X-ray sources. The soft X-ray sources in the disk, many of which are SNRs, are associated with the spiral arms, while the harder X-ray sources, mostly X-ray binaries (XRBs), do not appear to be. After eliminating AGN, foreground stars and identified SNRs from the sample, we construct the cumulative luminosity function (CLF) of XRBs brighter  than  \EXPU{8}{35}{\LUM}.  Despite M83's relatively high star formation rate, the CLF indicates that most of the XRBs in the disk are  low mass XRBs.  


Subject Headings: galaxies: individual (M83) -- galaxies: ISM -- supernova remnants -- X-rays: general -- X-rays: binaries -- X-rays: individual (M83) -- SN: individual (SN1923A)

\end{abstract}

\section{Introduction \label{sec_intro}}

M83 (NGC~5236) is a large,  grand-design spiral galaxy  at a distance of 4.61 Mpc \citep{saha06} that has been the site of six historical supernovae (SNe; see \citet{stockdale06} and references therein).     It has very active star formation, estimated by \cite{boissier05} at between 3 and 4$\MSOL\peryr$.  Due to its nearly face-on inclination \cite[i=24$^{\circ}$, ][]{talbot79}, very well defined spiral arms, and location along a line of sight with low Galactic absorption \cite[N$_H$=\EXPU{4}{20}{cm^{-2};}][]{kalberla05}, M83 provides an outstanding laboratory for understanding X-ray source populations in star-forming galaxies.

Given the nature of M83, it is not surprising that it has been  observed many times at X-ray wavelengths.  
\cite{trinchieri85} were the first to obtain spatially resolved images of M83 using the $Einstein~Observatory$ and to discuss its X-ray properties.  \cite{immler99} used the improved sensitivity of the ROSAT HRI to resolve about half of the total luminosity of the galaxy 
 (\EXPU{1.1}{40}{\LUM}, corrected to our assumed distance of 4.61 Mpc) into 21 X-ray point sources, and to show that some of the diffuse emission was associated with the  spiral arms of M83.  However,  detailed study of the galaxy
has  become possible only with the increased sensitivity of \chandra\  (and {\em XMM/Newton}).  Soria \& Wu  (2002, 2003) analyzed  a 51 Ks observation of M83 with \chandra's ACIS-S instrument, identifying 127 sources to a limiting sensitivity of $\sim \EXPU{4.6}{36}{\LUM}$ (also adjusted to a distance of 4.61 Mpc), and showed that many of the discrete sources as well as much of the emission from hot gas and unresolved sources is associated with the spiral arms.  Most of the sources have been tentatively  identified as X-ray binaries (XRBs), based on their spectral characteristics, but a number of supernova remnants (SNRs) and possible supersoft sources (SSS) have been identified as well.

Here we provide an overview of a very deep set of \chandra\ observations of M83 and an analysis of the X-ray point source populations.    We also describe new radio observations of M83 made with the Australia Telescope Compact Array (ATCA), which were obtained to support the \chandra\ observations.  We have previously used these data to report the discovery of a new ultra-luminous X-ray source (ULX) in M83 \citep{soria12}, to report the recovery of the remnant of the historical SN1957D in X-rays in M83 \citep{long12} and more recently to describe the properties of a new micro-quasar in M83 \citep{soria14}.

\section{Observations\label{sec_obs}}

\subsection{X-ray Observations\label{sec_xobs}}

\begin{figure}
\plotone{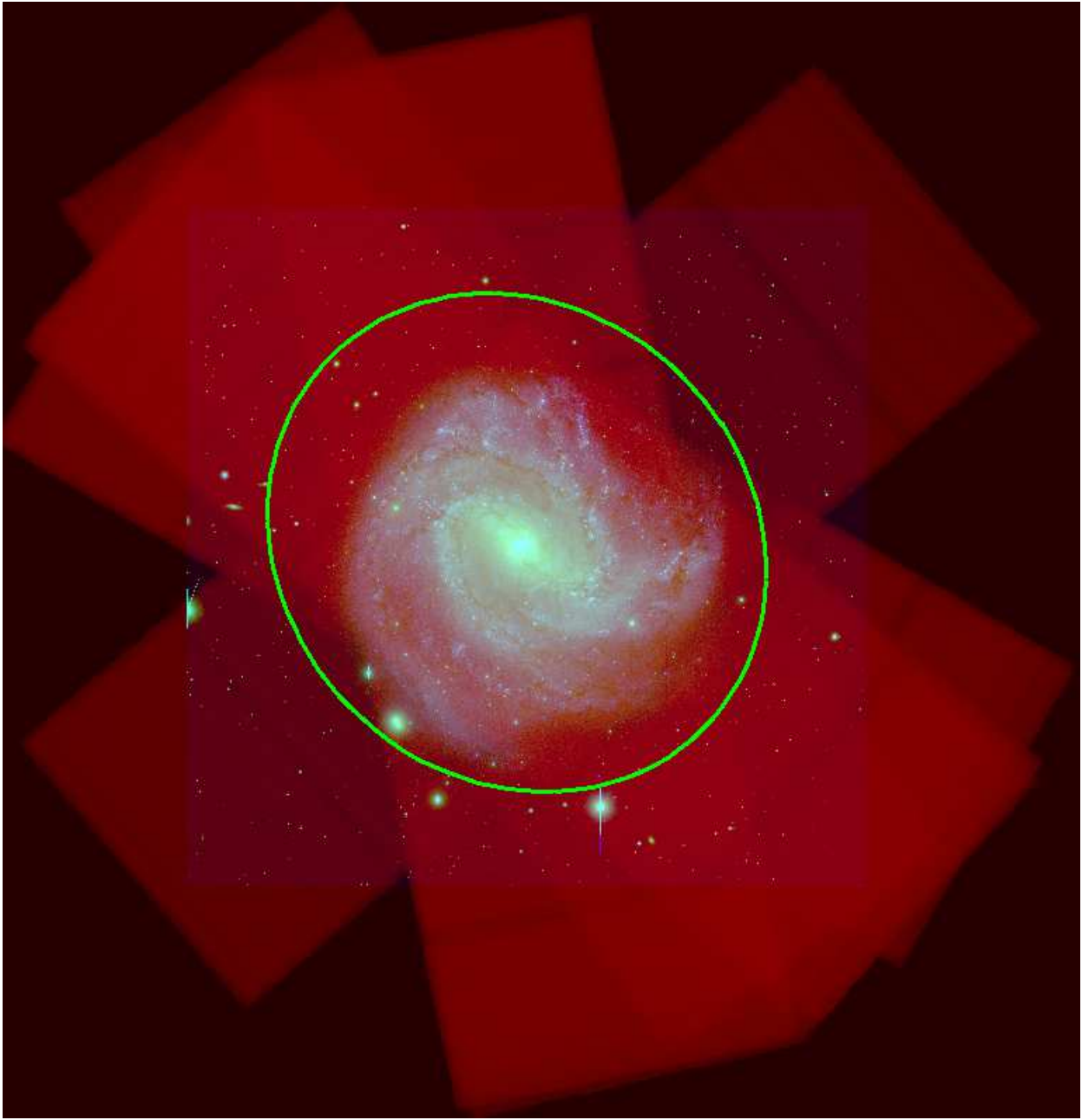}
\figcaption{An exposure map created for the S2, S3, and S4 chips for all of the \chandra\ observations of M83, superposed on an optical image of M83.  The D$_{25}$ size of M83 is shown in green.  Though most of the inner portions of M83 are well covered by our observations, regions to the NW and SE have somewhat lower coverage than elsewhere. \label{fig_exp}}
\end{figure}

The X-ray observations of M83 in our survey were all carried out with ACIS-S in order to maximize the sensitivity to soft X-ray sources, such as SNRs, and to diffuse emission.  The nucleus of M83 was centered in  the field of the back-illuminated S3 chip to provide reasonably uniform coverage of M83. In addition to the S3 chip, data were also obtained from chips S1, S2, S4, I2 and I3.  All of the observations were made in the ``very faint" mode to optimize background subtraction.   Observations were spaced over a period of one year from 2010 December to 2011 December,  as  indicated in Table \ref{table_obs}.  The only difference among observations was the roll orientation of the spacecraft and the differing exposure times.  All of the observations were nominal, and yielded a total of 729 ks of useful data.  In order to maximize our sensitivity and more importantly to improve our ability to identify time variable sources, we included in our analysis earlier  \chandra\ observations of M83 in 2000 and 2001 totaling 61 ks obtained by G. Rieke (Prop ID. 1600489)  and by A. Prestwich (Prop ID.  267005758).  These data were obtained in a very similar manner to that of our survey, and increased the total exposure to 790 ks.   As indicated in Figure\ \ref{fig_exp}, which shows the summed exposure time map superposed on an optical image of the galaxy, most of the optically bright inner disk of the galaxy was included in the observations. 

For the data analysis which follows, we reprocessed the data using tools available in {\small {CIAO}} Version 4.4 \citep{fruscione06}. We used CalDB version 4.5.3 for the analysis, which includes a revised model for the buildup of contamination on the optical blocking filter of ACIS.  After reprocessing the data, we aligned all the images to one another using the positions of sources identified with the source finding routine {\em wavdetect} and the script {\em reproject\_aspect}, which matches the offsets, scale and rotation of each image to a reference image--in our case ObsID 12994, the observation with the longest exposure time.  To reduce the error in this alignment we used a subset of the sources identified with {\em wavdetect}, specifically 49 sources with more than 300 broadband counts in the summed exposures, and located between 1 and 6 kpc from the galactic center, to avoid the confused nuclear region of the galaxy.  The {\em reproject\_aspect} procedure reduced the alignment errors between ObsID 12994 and the other images from 0.5 -- 1 pixels 
to just 0.1 -- 0.2 pixels, or 0\farcs05 -- 0\farcs1. 

This procedure assured that the X-ray images were coaligned, but does not establish absolute astrometry for the X-ray source catalog.  Typically, the absolute astrometry of \chandra\ images is accurate to 0\farcs 6.\footnote{90\% of ACIS-S test sources within 3\arcmin\  of on-axis have positions accurate to within 0\farcs 6; see http://cxc.harvard.edu/cal/ASPECT/celmon/}  An initial comparison of X-ray source positions in the \chandra\ catalog to those in our ATCA survey indicated an average offset of 0$\farcs$4 between the two, mostly in right ascension. A similar offset was seen in comparisons between SNR candidates identified in our Magellan images \citep{blair12}, which are discussed in more detail in Section \ref{sec_snrids}.  Sources in the Magellan data and the ATCA data are well aligned with one another.  The astrometry of the ATCA data is tied to the ICRF2 quasar catalog \citep{ma09}, which is accurate to 0$\farcs$1, while the Magellan astrometry  is tied to high-precision stars from the UCAC3  and USNO-B1.0 catalogs, which produced rms errors $< 80$ mas.  As discussed in Section \ref{sec_radio} below, there are a number of sources that appear both in our X-ray source catalog and in the ATCA data.  
Comparison of the object positions in the two catalogs indicated a small but systematic offset between the two catalogs.   Therefore, in the catalog produced below (Table \ref{table_xray_cat}), we have shifted the X-ray positions by $\Delta$R.A. = -0$\farcs$35 and $\Delta$Decl. = 0$\farcs$08 to compensate for this systematic error and produce the best absolute positions. 

\begin{figure}
\plotone{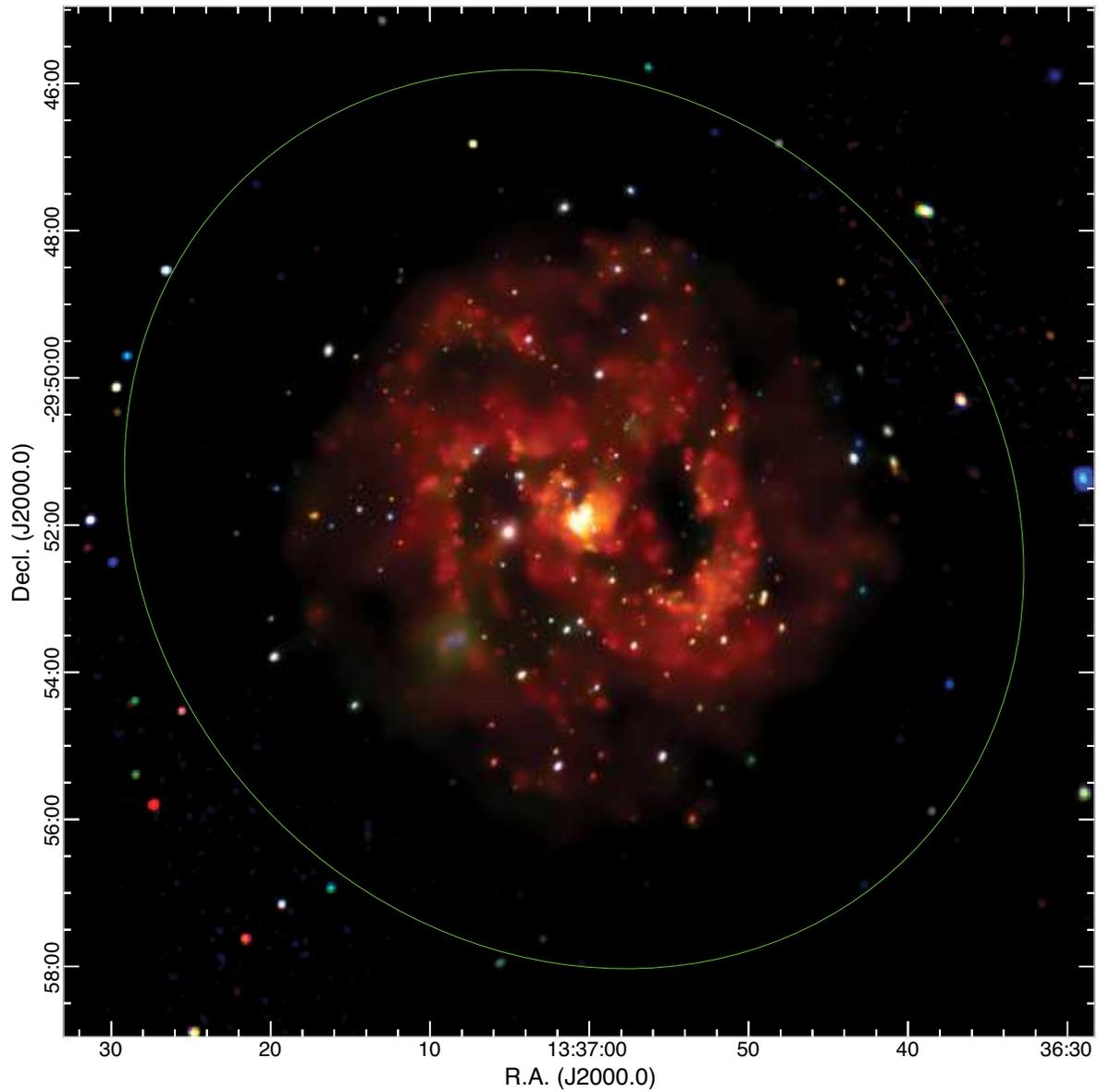}
\figcaption{The inner portion of combined X-ray image obtained from the 2010-2011 \chandra\ observations of M83. Photons from 0.35-1.1 keV, 1.1-2.6 keV, and 2.6-8 keV are shown in red, green, and blue, respectively.    The D$_{25}$ ellipse of M83 is shown in green.  Soft diffuse X-ray emission traces the spiral arms of the galaxy.  The new ultraluminous X-ray source described by \cite{soria12} is in the inter-arm region about 1\arcmin\ east of the nucleus.
\label{fig_xray}}
\end{figure}

A slightly-smoothed three-color image of the combined X-ray data is shown in Figure \ref{fig_xray}. In addition to a large number of point sources, the image contains soft diffuse X-ray emission that traces the spiral arms of the galaxy. There is a patch of extended emission with a hard X- ray spectrum near R.A. 13:37:08, Declination -29:53:40 visible to the SE of the nucleus.  We have not noticed similar features in the X-ray images of other nearby galaxies that we have studied. We suspect that it is a background cluster of galaxies whose X-ray emission is visible through the disk of M83. There is no obvious optical cluster  at this location, but this would could well be due to obscuration as a dust lane runs through this region. Finally, the nuclear region comprises both diffuse emission and about 40 bright point sources.

\subsection{Radio Observations\label{sec_radio}}

To support and extend our X-ray study of M83, we have been carrying out a number of other studies of M83, including optical broad- and narrow-band imaging with the IMACS camera on Magellan \citep{blair12}, optical imaging with the Wide Field Camera 3 (WFC3) on the {\em Hubble Space Telescope} ({\em HST}; W. P. Blair PI, Prop. ID. 12513, \cite{blair14}), and radio imaging with the Jansky Very Large Array (JVLA, C. Stockdale PI, Prog. ID. 12A-335).  The \hst\ and the JVLA results will be described elsewhere. Here we describe new 6 and 3 cm radio imaging we have obtained from ATCA.

M83 was observed with the ATCA in the 6A configuration for 12 hours on each of 2011 April 28, 29 and 30,  for a total observing time of 36 hours, simultaneously recording 2 GHz bandwidths centered on 5.5 GHz and 9 GHz in all four polarization products. Regular scans on the nearby phase calibrator PKS 1313--333 were scheduled throughout the observations, as well as scans on the ATCA primary flux calibrator PKS 1934--638. Standard calibration and editing procedures were carried out using the MIRIAD data analysis package \citep{sault95}. Following calibration, the data were exported for imaging and self-calibration using the Common Astronomy Software Applications (CASA) package \citep{mcmullin07}. Due to the large size of M83 ($\sim 12\arcmin$) relative to the ATCA primary beam width (8\farcm 5 and 5\farcm 2  at 5.5 and 9 GHz respectively), primary beam correction was required to account for the radial dependence of gain in the image. 

The primary beam correction results in position-dependent noise, with the rms noise increasing with distance from the pointing center. This renders most source-finding algorithms unsuitable. We therefore applied the following procedure to produce the source catalog.  First, we identified candidate sources by running the MIRIAD image-finding algorithm {\em imsad}, prior to primary beam correction. The resulting source list was edited based on visual inspection of the candidate sources. After applying the primary beam correction to each image, flux densities were obtained from the CASA task {\em imstat} using an elliptical aperture centered on each candidate source, with the aperture major and minor axes equal to twice the restoring beam FWHM and rotated to the same position angle. For each candidate source, the uncertainty in flux density was calculated using the nearest source-free region of the radio map. In cases where the fit converged, coordinates were obtained by fitting an elliptical Gaussian to the pixels with values greater than half the peak---using the CASA task {\em imfit}; otherwise, the centroid was determined and used. 

\begin{figure}
\plotone{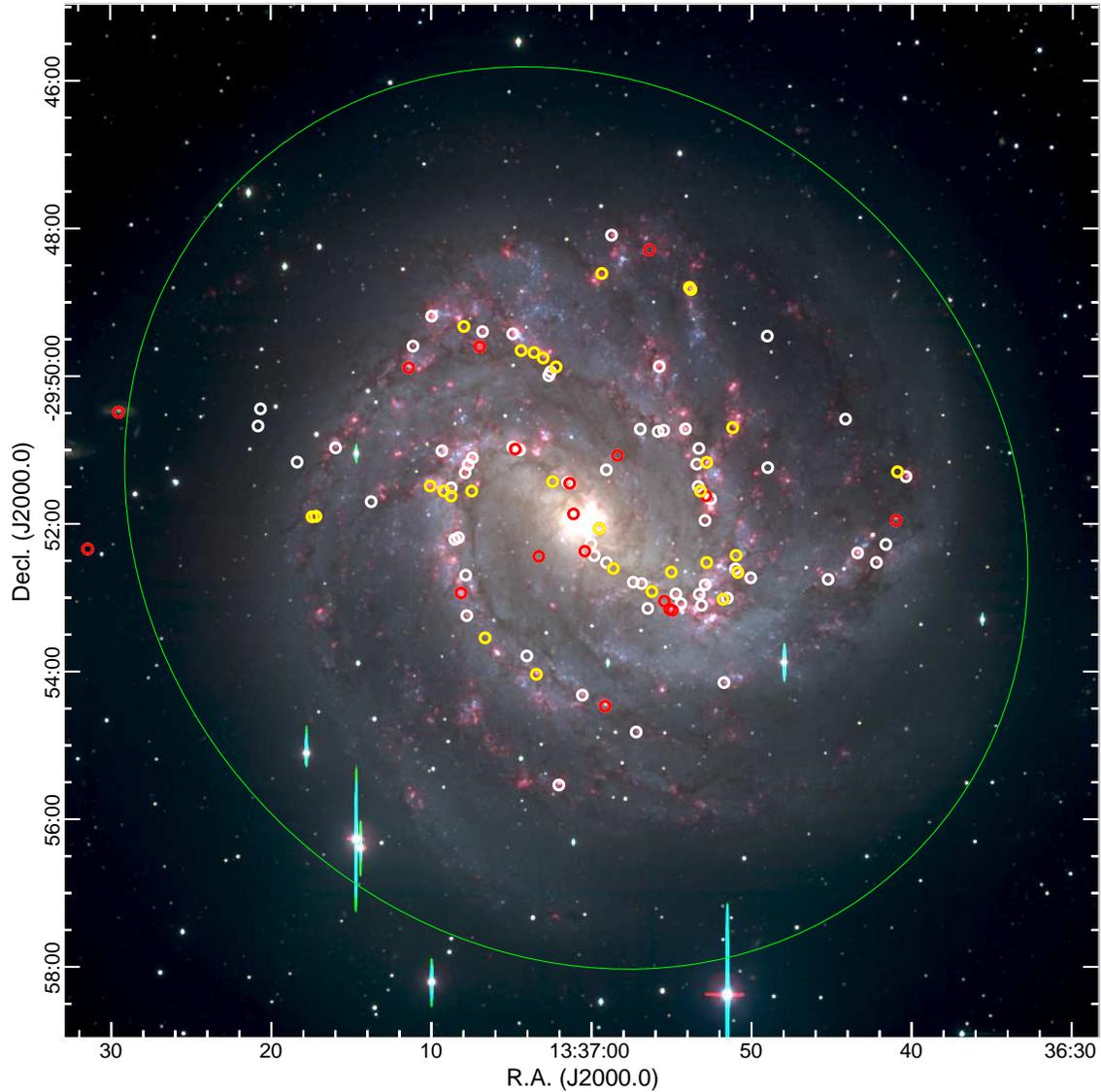}
\figcaption{Radio sources detected in our ATCA survey of M83 are plotted on a 3 color image of M83.  Here \HA\ emission is shown in red, V band in green, and the B band in blue, all from our Magellan survey \citep{blair12}.
The field is identical with that shown in Figures 2 and 4, and the D$_{25}$ ellipse is again shown in green.  Radio sources coincident with \chandra\ X-ray sources are shown in red, ones coincident with SNRs from \citet{blair12} are shown in yellow, and the remaining sources in white.  Most of the sources of all types lie along the spiral arms of the galaxy. The two radio sources outside the D$_{25}$ contour of the galaxy are AGN. 
\label{fig_radio}}
\end{figure}

The results are shown in Table \ref{table_radio} and in Figure \ \ref{fig_radio}.  We find a total of 109 radio sources, mostly associated with the spiral arms of the galaxy.  Of these, 49 of the ATCA sources lie within 2$\arcsec$ of sources in the earlier VLA study of M83 conducted by \cite{maddox06}.\footnote{There is a clear offset in the positions in the two surveys of about  0\farcs6, which we corrected before associating sources with one another.} Only 8 of the 55 VLA sources have no counterparts in our ATCA survey. Two ATCA sources, A063 and A064, are associated with M06-34; two, A077 and A079, are associated with M06-40.  A063 is closer (0$\farcs$5) to M06-34 than is A064 (1$\farcs$5) and is most likely the correct association.  On the other hand, the A71 and A79 are about the same distance (1$\farcs$7 and 1$\farcs$9) from M06-40. 

Two historical SNe, SN1957D and SN1950B, were clearly detected, and both had  been previously identified by \citet[][see also references therein]{maddox06}. The situation with SN1923A is more problematic, however, because the position of the optical SN was not well determined.  \cite{pennington82} went to considerable effort to measure the position on original plate material from 1923, but the SN image itself creates an extended patch of emission and the centroid may be affected by other local emissions near the site.  Precessing their position to J2000, the \cite{pennington82} position is 4$\farcs$2 away (3$\farcs$8 south in declination) from the closest ATCA, A096, which seems too far to be a real association, if the errors on the position given by \cite{pennington82} can be believed.  However, we note that the position determined by \cite{pennington82} for SN1957D is 2$\farcs$4 south from what is now the well-known position of SN1957D (ATCA and \hst), in the same direction as the discrepancy for SN1923A.  This is also outside the stated errors in \cite{pennington82}, and so we suspect the error budget they estimated was too low. 

We have independently used the figure based on the original plate showing SN1923A, provided in \citet[][their Figure 1b]{pennington82}  to reconstruct the SN position on our Magellan images.  Again with the uncertainty that other sources may have been blended with the SN on the original plate, we determine a reconstructed J2000 position of R.A.=13:37:09.19, Dec=$-$29:51:01.39, which is 3\farcs2 almost due north of the Pennington position, and just over 2\arcsec\ from A096 (differing primarily in R.A.). If we consider the difference between the Pennington position and our reconstruction to be more indicative of the real uncertainties,  then an identification with A096 is consistent.

\cite{eck98} have also claimed a radio source near the optical position of SN1923A, although their radio source position had an uncertainty of 5\arcsec. Given that the ATCA radio sources in this region are unconfused, the source reported by \cite{eck98}  is likely to be the same source as A096.  If true, then the changes in flux of this source with time are also consistent with it being due to the young remnant of the SN.  If continued monitoring of A096 confirms the decay in flux with time, this would be a smoking gun for the ID, and the ATCA radio position would provide the best positional information for SN1923A going forward.

Comments and suggested identifications for many additional sources in the ATCA images are provided in Table \ref{table_radio}, but we defer discussion of these until Section \ref{sec_ids}.

\section{The X-ray Point Source Catalog \label{sec_ae}}

As indicated in Section \ref{sec_xobs}, the X-ray emission from  M83 is complex. Not only are there contributions from point sources and extended emission, but the density of point sources varies greatly.  Furthermore,  the \chandra\ point spread function (PSF) also varies with position away from the field center.  As a result, the generation of a point source catalog is not trivial.  We have used a combination of {\small {CIAO}} tools and the package {\em ACIS Extract} \cite[{\em AE},][]{broos10} to carry out this task.  The procedure we used, in brief, was to use {\em wavdetect} to create source lists from each of the individual observations as well as from the combined X-ray images.  We created source lists for a variety of energy bands and combined all of these, eliminating duplicate sources, to yield a total of 847 candidate sources.   We  pared this list down to 458 ``good'' sources using various statistical tests and comparisons among the data subsets.  Finally, we visually inspected the resulting ``good'' source list on the \chandra\ images as a sanity check that our list of candidate sources was not missing any obvious sources.   The following paragraphs provide the details of this procedure.

\subsection{Preliminary source list and positions}

We used the {\small {CIAO}} task {\it wavdetect} to generate initial source lists.  In running {\it wavdetect}, we set the {\it sigthresh} parameter  to $10^{-6}$, which is expected to produce about one false detection per chip.  Because the \chandra\ PSF degrades as one moves off-axis, we set 
the {\it scales} parameter  to all the powers of two up to 32 pixels.
We used five overlapping energy bands,
0.35-1.0 keV, 0.35-8.0 keV, 0.5-2.0 keV, 1.0-2.0 keV, and 2.0-8.0 keV,
which produced 180 individual source lists,
one for each energy band for each ObsID.
Although we ran  {\it wavdetect} individually on the S1, S2, S3, and S4 chips, we first merged
all of the lists for different chips of each ObsID/energy
combination,
after checking for duplicate sources near the chip
gap.  In merging the source lists, we used 
the local PSF size to determine whether nearby
sources
detected in different lists were indeed the same source;
if the centers of a source in one list fell within
the 85\% encircled energy ellipse of a source in another list,
we assumed it to  be the same source.
We then merged the lists for the same ObsID but different energies,
to create a list of unique sources for each ObsID.
Finally, we merged the lists from all of the ObsIDs,
again using the PSF size criterion for matching sources.

In parallel, we ran {\it wavdetect} on the combined data set, 
using the same parameters and energy bands as for the individual observations.  
As before, we merged the lists produced in the different energy bands,
using the above criterion for duplicate sources.
Finally, we combined the lists from the combined and individual observations.
By running the source detection on both the total data set and the individual observations,  
we managed to produce the deepest possible catalog 
while not missing transients that might fall below the significance threshold in the total data set.
 
A visual check of the source positions projected onto a display of the actual data was used as a sanity check.  The only area where we made modifications to the list produced through the {\it wavdetect} process described above was in the nuclear region, where diffuse emission is quite prominent.  In this complicated region it appeared that some of the identified sources were not well centered on visible X-ray peaks, and in a few cases, there were multiple {\it wavdetect} positions identified with a single peak in the X-ray image.  We manually removed duplicate targets and adjusted positions for a handful of sources to produce the final catalog.

\subsection{Final Source List and Count Rates}

With the positions established, we then used {\em AE} to derive net count rates from the sources in various energy bands:  0.35-8 keV (total or T), 0.35-1.1 keV (soft or S), 1.1-2.6 keV (medium or M), 2.6-8 keV (hard or H), 0.5-2 keV ($\mathcal S$), and 2-8 keV ($\mathcal H$).  Our choice of these bands was based on a variety of overlapping goals.  The broad 0.35-8 keV samples the full energy range accessible to \chandra\ observations. The three bands, 0.35-1.1 keV, 1.1-2.6 keV and 2.6-8 keV provide bands intended to classify sources on the basis of their hardness ratios.  The boundary at 1.1 keV, in particular, is just above the region containing strong features due to Ne and Fe seen in the spectra of most SNRs.  The 0.5-2 keV and 2-8 keV bands are needed because number counts of AGN and of X-ray binary populations are normally carried out in these bands, and because the 0.5-2.0 keV band, encompassing the peak of the response curve, provides better statistics for some purposes than S+M. 

The {\em AE}  count rates were used to establish which of the sources in the candidate list were statistically valid. 
We retained any source that had a  probability-of-no-source $<$ \EXPN{5}{-6} in any one of these bands in the total data set.  For our final run of {\em AE}, our  list of source candidates had 847 potential sources.  Of those, we find a total of 458 valid point sources, whose   properties are listed in Table \ref{table_xray_cat}. Most of these (409)  were detected as sources in the 0.35-8 keV energy band.   Totals of  315, 305, and 179  were detected as sources in the soft, medium and hard bands,  respectively.   There were 383 sources detected in the 0.5-2 keV band and 210 sources in the 2-8 keV band. Without the 0.5-2 keV and 2-8 keV sources there would have been 443 sources; adding these two bands resulted in 15 more sources.  Of the 458 point sources,  378 are located within the area defined by the D$_{25}$ ellipse of the galaxy (which we take to have a major axis diameter of 12.9 arcmin), and the remaining 80 are outside this region.  
There were 43 sources in the nuclear region (defined to be any source within a projected radius of 0.5 kpc from the optical nucleus).  

Over the course of analyzing these data, we repeated the process of identifying source candidates and processing them through {\em AE} multiple times, with differing versions of CIAO and the other tools that make up {\em AE}. While the vast majority of sources appear in all of the lists we have developed, it is clear that the faintest 5-10\% of the sources we detect by this procedure vary somewhat.  This is not surprising given the combination of diffuse emission and field crowding that exists in places.  Once again, a visual check was used to inspect the ``good'' sources from {\em AE}.  Most of the sources are readily apparent in the raw images or in a smoothed version of them, but a small number of the  fainter objects, mostly located in regions with bright diffuse emission, are hard to pick out in these images.  The user is cautioned to be wary of the faintest sources, but we have not eliminated these objects from the catalog since they passed the statistical test we applied.  

The complete source list in right ascension order is presented in  Table \ref{table_xray_cat}.   For ease of reference we have labeled the sources  X001 through X458.  Along with this name (col. 1), the Table provides for each source the position (col. 3 and 4), the position error as determined by AE (col. 5), the total exposure time (col. 6), the count rate in the 0.35-8 keV band (col. 7), and two hardness ratios (col. 8 and 9), which are described in the next section.  Count rates for detected sources range over almost four orders of magnitude, from  \EXPU{2}{-5} to  0.11 counts s$^{-1}$.

\section{Derived Source Characteristics}

\subsection{Source Fluxes, Luminosities, and Hardness Ratios \label{sec_fluxes}}

The conversion of source counts to flux of an X-ray source depends on the assumed X-ray spectrum and, for soft sources, the absorbing column.  Although {\em AE} provides several estimates of the fluxes of the sources it reports, it does not permit one to specify the shape of the spectrum.  {\em AE}  does, however, produce nominal X-ray spectra of all of the sources if requested.  We have used these spectra in conjunction with the task  {\em xspec} to estimate band-limited fluxes, assuming a power-law source with a photon index of $-$1.9 and foreground absorption of \EXPU{4}{20}{cm^{-2}}.  This choice of spectrum is appropriate for compact binaries and background AGN, and is typical of that chosen for other studies of the X-ray properties of normal galaxies \cite[see, e. g.][]{tuellmann11}.   Fluxes in units of photons cm$^{-2}$ s$^{-1}$ for the various bands, T, S, M, H, $\mathcal S$ and $\mathcal H$, are presented in Table  \ref{table_fluxes}.  
In addition, the average  0.35-8 keV X-ray luminosities of the sources, calculated from the energy flux in the 0.35-8 keV band, and assuming that all sources are located at the distance of M83, are shown in Table \ref{table_fluxes}.  The luminosities, calculated in this simple way,  range from \EXPU{5}{35}{\LUM} to \EXPU{2.2}{39}{\LUM}. The total luminosity of all the point sources is \EXPU{1.2}{40}{\LUM}, so the brightest source, which is the ULX we described in \cite{soria12}, contributes 18\% of the total point source luminosity of the galaxy.  Broad band (0.35-8 keV) luminosities estimated using flux conversions based on our fiducial power law are fairly insensitive to differences in line of sight absorption less than \EXPU{1}{22}{cm^{-2}} for hard sources like compact binaries.  However, our simple calculation overestimates luminosities for soft, relatively unabsorbed sources, such as many SNRs (by about a factor of 2 for a source with a characteristic plasma temperature of 0.6 keV and low absorption). 

Hardness ratio plots, based on the photon fluxes of the sources, are shown in Figure \ref{fig_hardness}, with different panels of the plot showing different subsets of the sample.    
Of the 458 sources in the sample, 57\% have (M--S)/T ratios greater than -0.5, and 27\% have (H--M)/T ratios greater than 0.   Many of the sources form a central clump in the region with (M--S)/T ratio near $-$0.1 and (H--M)/T ratio near +0.2, and extending toward the S=0 line.  This is the region of the diagram where compact binaries and AGN are both expected.  The upper right panel shows sources located outside the D$_{25}$ radius (green symbols), which indeed are systematically harder and thus consistent with a population dominated by AGN and other background sources. 
The panel at upper left shows sources within the D$_{25}$ contour, which no doubt includes a combination of background AGN and sources intrinsic to M83.  There is a also group of softer X-ray sources, as shown by the red symbols in the lower left panel, along the line with very little emission in the H band.  Many of these sources, as will be discussed in Section \ref{sec_snrids}, are coincident with optical emission nebulae with line ratios that suggest they are SNRs. A number of the additional objects in this region be SNRs as well.  Some of the softest sources, those with very few counts in the M or H band, may be super-soft sources (SSSs -- see below).  The sources in the nuclear region (shown in black, lower right panel) have a wide variety of spectral shapes, consistent with a mixture of XRBs, SNRs, and other thermal emission within the starburst nucleus.   More detailed discussions of specific source identifications are provided in Section \ref{sec_ids}.

\subsection{Variable and Transient Sources \label{sec_variability}}

Variability is an important clue to the nature of the X-ray sources in any X-ray sample, but various authors assess variability is different ways.  Rather than attempt a full analysis of variability in this overview paper, we produce a preliminary assessment of the number of sources that were variable using a single, simple approach; we compare the time averaged fluxes of each of the sources in our source catalog in each of the separate observations listed in Table \ref{table_obs}.  We have not carried out a systematic assessment of whether sources vary within a single observation, although we have identified a few example objects for which this is true (see Appendix \ref{appendix_bright}).

For our assessment, we have used  a standard measure of variability
\begin{equation}
R_{ij}\equiv\frac{F_{i}-F_{j}}{\sqrt{\sigma^2_{F_{i}}+\sigma^2_{F_{j}}}},
\end{equation}
where $F$ is the measured photon flux in an observation, and $\sigma_F$ is its Poisson-based uncertainty \cite[see, e.g.][]{primini93, fridriksson08}. 
Typically, sources are considered to be variable if $R_{ij}$ is greater than some threshold value $R_0$.  To assess whether a source was variable in our set of observations, we first calculated the maximum observed value of $R$, hereafter $R_{max}$, from the multiple observations using equation 1.  To estimate the value of $R_{max}$ expected for a non-variable source, we  created $10^6$ simulations of each observation of the source, allowing for Poisson statistics, variations in the effective area of \chandra\ with time, and the regions used by AE to generate the background. We calculated  $R_{max}$ for each of these simulations, and used this to calculate for a constant source  the probability $P(R_{max}>R_0)$ that  $R_{max}$ would be greater 
than $R_0$.   Not surprisingly, these distribution functions resemble the cumulative distribution function and have no sharp features.  There is no obvious threshold value of $R_0$ in the cumulative distribution function $P(R_{max}>R_0)$ that isolates constant sources from variable sources upon visually inspecting plots of the fluxes in the various observations of individual objects. Thus we have somewhat arbitrarily labeled as variable those sources for which $P(R_{max}>R_0)<0.0027$, roughly the $3\sigma$ level. 

We searched for variability in all of the bands.
The sources that have been determined to be variable 
have been marked in Table~\ref{table_xray_cat}.  Of the 458 sources in our catalog, 96 were found to be variable in one or more bands,
and of these, 83 were found in 0.35-8 keV (T) band.   This is not surprising since the broad band contains all of the counts.  There were 47, 52, 29, 68 and 38 detected in the S, M, H, $\mathcal S$, and $\mathcal H$ bands, respectively.   Of the 13 sources not detected as variable in the 0.38-8 keV band, 8 were found to be variable in the 0.3-1.1 keV (S) band; there were no ``new'' variables detected in the 1.1-2.6 keV (M) band and small numbers in each of the other bands.\footnote{The 13 sources that were  not detected as variable in the 0.35-8 keV (T) band, but were detected as variable in one of the other variables are indicated ``V*'' in column   6 of Table~\ref{table_xray_cat}.}
Our sensitivity to variation is clearly a function of the source brightness.
We can detect variations comparable to the mean photon flux
for sources brighter than $\sim10^{-6} \PHOTFLUX$,
and variations of one third the mean flux for sources
with photon fluxes $\sim1.5\times10^{-5} \PHOTFLUX$.

For sources with 0.35-8.0 keV band fluxes $>10^{-15}\FLUX$ within the D25 contour,  the variable sources have average hardness ratios of (M--S)/T = -0.313$\pm$0.009 and (H--M)/T = -0.0419$\pm$0.0040.  This compares to ratios of  (M--S)/T =  of  -0.539$\pm$0.012 and (H--M)/T = -0.0358$\pm$0.0059 for the non-variable sources.  One would expect the variable sources to be dominated by XRBs and AGN, both of which typically have hard X-ray spectra, while SNRs, which have softer spectra, are non-variable.  For the (M--S)/T ratio, the variable sources are are harder on average, while for the (H--M)/T ratio, the two sources populations are indistinguishable.  An inspection of Table~\ref{table_xray_cat} shows that many of variable sources are identified with binaries or AGN based on other considerations.

Some of the sources in our catalog are transients. 
The definition of a transient source varies between authors
and has been shaped by the depth and cadence of the relevant data 
as well as details about the types of sources being sought.  
The criterion common to all definitions of transients is 
that the range between the maximum and minimum luminosities
be ``large,'' where large ranges from factors of 5 to 10.
Some authors set the problematic criterion that the source be undetectable at its faintest, 
which can in principle
select different types of objects depending on the sensitivity of the data set being considered.
Temporal criteria are often applied to distinguish between 
transients and other types of strongly variable sources,
usually to select sources with very short duty cycles.

We have set two criteria to identify transient objects.  We first require 
that $R_{max}>4$ and $P(R)<0.0027$, which is effectively a statement that transients must have a large difference between the maximum and minimum fluxes.  We then insist that transients be objects for which the bulk of the measurements have fluxes less than one quarter of the maximum flux, which constrains the duty cycle of the bright phase, eliminating  objects that are variable on roughly month time-scales 
(i.e. the approximate cadence of our observations).  
However, these criteria still allow a wide variety of light-curve behaviors.
To better characterize the behaviors, 
we have flagged our transient candidates with further descriptors as follows:
``C'' or ``classical'' transient sources have a single high point and a quiescent level consistent with zero (three sources);
``B'' sources have more than one high point (five sources); and ``N''  vary by a large amount, but never completely disappear (eight sources).  The latter might be seen as classical transients given a less sensitive detection limit.

Also, since all but two of the observations were made over the course of about one year,
and the remaining two(ObsIDs 793 and/or 2064)  were taken almost a decade earlier, 
we must consider how this may have affected the variability assessments.
We flagged with ``A'' those sources that were on (or off) in the archival data, 
and off (or on) in the current data. 
Five sources were very significantly detected in the archival data but consistent with zero in the current data,
while six sources (including the ULX discussed by \citet{soria12}) would have been detected in the archival data
at levels seen in the new observations.  If we exclude the archival data, 21 sources are no longer considered variable.
Interestingly, two non-variable sources (based on the entire data set), 
X030 and  X255, were found to be variable when 
the archival data were excluded, though their variability was just barely significant.   
Of the sources found to be variable only with the addition of the archival data,
half show clear ``events'' in the archival data,
meaning that the increase of baseline allowed detection of more transients.
The other half do not show noticeable transient behavior,
suggesting that the detection of their variability with the addition of the archival data
may be due primarily to the increased count rate statistics.

\section{Source Identifications in the M83 Catalog \label{sec_ids}}

The sources in our new M83 X-ray point source catalog no doubt arise from a variety of types of objects, including XRBs and SNRs, as well as objects not associated with M83, foreground stars and background AGN.  We can, of course, follow two paths toward identifying sources:  a) compare to previously cataloged objects of various kinds, and b) use the observed multiwavelength characteristics of the X-ray sources to make our best determination directly.  In Table \ref{table_ids}, we summarize the results of the more extended discussion below by showing our identifications and cross references as appropriate.  Specifically, column 1 of this Table lists the X-ray source number while columns 2, 3 and 4 show positional coincidences with sources cataloged in recent X-ray, radio, and optical SNR studies,  respectively,  Columns 5 and 6 provide information about the variability and spectral properties of the sources, which as discussed in Sections \ref{sec_variability} and \ref{sec_bright} respectively, are useful for determining the nature of the sources.  Finally, columns 7 and 8 provide our classification of the various source types along with notes that support or qualify the source classifications.  The rationale for the various identifications and classifications is discussed below.

\subsection{Sources from earlier Chandra and XMM studies of M83}

The most detailed analysis of the archival \chandra\ observations of M83 was carried out by \cite{soria03} who identified 127 sources in data obtained with the S3 chip in the 51 ks exposure obtained in 2000. Of these, 124 are located near (within 2\arcsec) of sources identified in our new deeper survey.  This is hardly surprising, since we have included the archival data in our catalog as well.  Since \cite{soria03} analyzed only the S3 data, there are number of sources that we detect in the 2000 observations that they did not report.

There are three sources from \cite{soria03} missing from our catalog.  One of these, the soft source SW03-098, would have been included in our catalog had we chosen to report any source detected in any of the various observations instead of only sources detected in the combined data set.  Located in a spiral arm, this soft, and possibly variable,  source fell below our detection limit in the combined data, but was seen in our analysis of Obs.\  ID 793, the observation discussed by \cite{soria03}. The source SW03-004 is located 2$\farcs$2 from a source X026 in our catalog, and it is likely that this is  in fact the same source as X026 despite being outside the limit of 2$\farcs$0 we set in calculating spatial coincidences. The third source missing from our catalog, SW03-114, was among the faintest sources reported by \cite{soria03}, and appears to be spurious, at least by our detection criteria. 

Recently, \cite{ducci13} have reported the results of a study of M83 from {\em XMM-Newton}.\footnote{Based on a comparison of the source positions in the two catalogs, there is a small offset of about 1$\farcs$0 between the two catalogs. We had to shift the {\em XMM} positions by +0$\farcs$94 in right ascension and -0$\farcs$45 in declination to align the two catalogs.}   Of the 189 {\em XMM} sources identified by \cite{ducci13},  46  are within the D$_{25}$ contours of the galaxy, and 40 of these are aligned (within 3\arcsec) with sources in our \chandra\ catalog.    An additional 37 {\em XMM} sources outside the D$_{25}$ contours are also found in our \chandra\ catalog.  The  smaller fraction of sources found outside the D$_{25}$ contours in our catalog is simply due to the fact the many of the {\em XMM} sources are not in the FOV of our \chandra\ images.  Due to the lower spatial resolution of {\em XMM}, several of the {\em XMM} sources, D13-090, D13-093, and D13-130,  align with more than one \chandra\ source; we  list only the closest association in Table \ref{table_ids}.

\subsection{Historical Supernovae}

M83 has been the site of six historical SNe, two of which coincide in position  with X-ray sources in our survey.  The first of these is X279 = SN1957D, which was discussed in detail by \cite{long12}, for which we conclude an active pulsar is most likely powering the relatively hard emission observed.  

The second is SN1968L, which is spatially coincident with a bright X-ray source (X216)   in the nuclear region of the galaxy.    SN1968L was a poorly observed SN, thought to be of Type II \citep{wood74}, which \cite{dopita10} may have recovered as a faint \oiii-dominated emission nebula with \hst.   If SN1968L is the source associated with X216, it would be one of the more luminous young SNRs known, with L$_X$ of about \EXPU{7}{37}{\LUM}.  It is not detected as a radio source, which is not surprising due to the bright diffuse emission from the nuclear region.  It does have an X-ray spectrum dominated by a thermal plasma.  However, as discussed in more detail in Section \ref{sec_nucleus}, it is quite possible, perhaps likely, that X216 is simply a peak in the hot thermal gas permeating the nuclear region.  As a result,  more study will be required to establish whether X216 is really the X-ray counterpart to SN1968L.   

We do not detect X-rays from any of the other historical SNe in M83, including SN1950B, which we do detect with ATCA (A024) and which had previously been detected at radio wavelengths by \cite{maddox06}, or SN1923A, whose identification with the ATCA source  A096 is discussed in Section \ref{sec_radio}.

\subsection{Supernova Remnants \label{sec_snrids}}

\subsubsection{Optical Supernova Remnants}

Optical SNRs in nearby galaxies are usually identified on the basis of high \sii:\HA\ ratios compared to \hii\ regions in interference filter imagery.  The most recent study of M83 was carried out by \cite{blair12}, who  identified 225 SNR candidates in M83 based on this technique.  (This paper specifically excluded the nuclear region, which had previously been covered by \citet{dopita10} using \hst\ data.)   Of these 225 objects, 67 (73) lie within 1\arcsec\ (2\arcsec)  of sources in our new  \chandra\ catalog.  Almost all of these  are soft X-ray sources, based on their hardness ratios.  
Their identification as X-ray sources provides strong support that these objects are indeed SNRs.   

A small number of the positional coincidences might be by chance.  To estimate the number of chance coincidences, we have adopted the following simple Monte Carlo procedure, which accounts for the fact that sources are spread in a non-uniform manner across M83.  After having calculated the number of coincidences at the positions in two lists, we shift the positions of one of the lists in a random direction by 3 $\times$ the positional offset that we allow, and recalculate the number of coincidences.  For the SNR catalog described above we find on average 2.1 (8.5) coincidences for a position mismatch of 1\arcsec\ (2\arcsec). Indeed, other source characteristics indicate that a handful of the X-ray sources that coincide in position with optical SNRs are not true associations.  
For example,  the very bright X-ray source X321,  which is spatially coincident with B12-179, has a hard featureless spectrum  and is most likely a black-hole binary observed in the high/soft state, as is discussed in Appendix \ref{appendix_bright}.   

The optical SNRs identified from their \sii:\HA\ line ratios are usually dominated by radiative shocks in the ISM.  Very young SNRs, such as Cas A and E0102--72, which are still dominated by shocks within the ejecta, show strong \oiii\ emission instead.   \cite{blair12} identified 46 emission nebulae that are possibly of this type, but only 6 (8) of those sources lie within 1\arcsec\ (2\arcsec) of an X-ray source.
Of these sources, one  is SN1957D (X279), and therefore does not represent a new SNR identification.  Several others, X044, X054, X135,  have hard X-ray spectra, which makes a SNR identification less likely.  
An optical spectrum we recently obtained of B12-307, which is coincident with X044, indicates the X-ray source is {\em not} a SNR but  a background AGN instead.   None except SN1957D was detected as a radio source.  However, several other oxygen-dominated emission nebulae are likely SNRs.  The source X243, in the nuclear region of the galaxy, is spatially coincident with B12-321, which \cite{dopita10} reported as the first good candidate for a Cas A-like SNR in M83.  Three other sources, X110, X341, and X360, also have soft X-ray spectra which support an identification as a SNR (even though a supersoft source designation cannot completely be ruled out). 

\cite{dopita10}, in an optical search for SNRs using \hst/WFC3 interference filter images of  the nuclear and inner eastern spiral arm regions of M83, identified 60 emission nebulae as likely SNRs, including the Cas A-like object coincident with X243, mentioned above.  Of the 60 emission nebulae, 20 are located within the bright nuclear starburst region (within 200 pc of the nucleus), where the determination of [S II]:\HA\ ratios was especially difficult.\footnote{The number 20 includes the Cas A-like source S03-70 which is not contained in the tables in \cite{dopita10}, but discussed separately.  We ignore the five additional objects with strong [O~III] emission discussed by Dopita et al.\ as these are likely W-R nebulae instead of SNR candidates.}   
Of the 60 nebulae, 29 lie within 1\arcsec\ of sources in our \chandra\ survey. Of these, 14 are duplicates of sources we have  already identified as SNRs based on the list compiled by \cite{blair12}, and 15  have not been considered here previously.  There are 5.3 chance associations expected in this crowded region of M83.  Even given the small maximal position offset we have adopted (1\arcsec), two sources, D10-N08 and D10-N10, are associated with one X-ray source, X223.  Most of the new identifications (12 of 15) are in the nuclear region.  Because of the crowding near the nucleus, even as viewed with \hst, and the difficulty of measuring [S II]:\HA\ ratios in these confused regions with bright backgrounds, we have chosen to limit the objects we identify as SNRs based on the \hst\ imagery to those we feel on other grounds are most likely SNRs.

\subsubsection{Radio Supernova Remnants}

Of the 109 sources detected with ATCA, 36 (47) are coincident with  X-ray sources within 1\arcsec\ (2\arcsec).  By the criterion outlined above, we would expect 1.7 (4.9) of these coincidences to occur by chance.  There is one X-ray source, X116 that could be associated with two radio sources, A031 and A032, but A032 is most likely the correct association given a position error of only 0$\farcs$3 compared to 1$\farcs$8 for A031. Thus there are 46 X-ray sources that have ATCA counterparts. Of these X-ray sources, 21 are also in the list of sources already identified as SNRs from a comparison to the optical data.  Their detection as radio sources solidifies their identification as SNRs.  

All but one of the  ATCA sources coincident with [S II] bright emission nebulae in \cite{blair12} are also detected as X-ray sources. This suggests, as do the total number of sources in the two surveys, that the X-ray observations have an overall sensitivity advantage over ATCA for detecting SNRs.  The exception is A004 which lies within 2\arcsec\ of the  [S~II] dominated optical emission nebula,  B12-004, but does not have an X-ray counterpart.  This identification is recorded in Table \ref{table_radio}; there is no reason to believe that A004 is not a SNR but further analysis is required to determine whether the upper limit on the X-ray flux at that position is significantly lower than the observed X-ray luminosities of other ATCA SNRs. 

This leaves 25  ATCA sources coincident with  X-ray sources that are {\em not} identified with [S~II] dominated optical SNRs.  Some of these, such as A108 and A109, are AGN or background galaxies; one is SN1957D \citep{long12};  and one, as discussed briefly later, one, A062 (=X237), is a newly-identified micro-quasar \citep{soria14}.  The hardness ratios of the X-ray sources identified with ATCA sources is shown in Figure \ref{fig_hardness_radio}.  Most have soft X-ray spectra, including essentially all of the objects also identified with optical SNRs.  The sources identified with AGN, the micro-quasar, and SN1957D are the outliers.  This suggests that a number of the other sources identified with radio sources are also likely to be SNRs.   We have included seven of these sources in our list of candidate SNRs in M83, based on detection in the ATCA sample, their soft X-ray spectra and no indication from other data that they are not SNRs.  These sources are  X067, X078, X095, X097,  X104,  X169, and  X275.

\subsubsection{Summary of SNR Identifications}

Thus of the 378 X-ray sources within the D$_{25}$ contours of M83, we identify a total of 87 as likely SNRs, based on the characteristics of their X-ray spectrum, an association with either an optical emission nebulae with the characteristics of a SNR and/or an association with a radio source in our ATCA data, and no strong contra-indications such as obvious X-ray variability.   With the possible exception of M33, where  \cite{tuellmann11} identified 45 SNRs using similar techniques to those employed here and where \cite{long10}  identified 88 based on a targeted search for SNRs, this is the largest sample of X-ray SNRs in any external galaxy.  A full discussion of the properties of these SNRs will be presented elsewhere.

\subsection{Foreground and Background sources  \label{sec_otherids}}

Our deep X-ray images also show a number of  X-ray sources that are background AGN and foreground stars.  Here we discuss the identification of specific sources that can be recognized based on their appearance in other wavelength bands.  The identification of AGN, the biggest source of contamination, is discussed in a statistical sense in Section \ref{sec_lum}.   

Foreground stars are generally associated with soft X-ray sources and are visible as point sources in optical images.  The numbers of such objects are expected to be relatively small given that M83 is approximately 31$\degr$ from the Galactic plane.    Background AGN generally have hard X-ray spectra similar to many XRBs and are often visible in optical images as long as a region is not confused by other emission.   

In some cases, we were able to identify objects based on existing catalogs. For example, there are eight X-ray objects that are coincident with bright 2MASS sources well away from the optically bright portions of the galaxy;  seven of these sources are clearly stars in our Magellan images. The other  one is a background galaxy/AGN, listed in the Hyperleda catalog \citep{paturel03}  and the 6dF Galaxy Survey \citep{jones09}.   
A few other sources are also previously known AGN or background galaxies in these catalogs.  In addition, as mentioned earlier, as part of an on-going study of the spectra of SNR candidates, we obtained a Gemini GMOS spectra of X044 and found it to be an AGN.

To obtain a more complete census of such objects, we have inspected the positions of the sources in the multiband Magellan images of M83, and made tentative identifications of a group of other objects.  When we found an optical counterpart within the error circle of an X-ray source, we identified it as a likely AGN if the X-ray-to-optical flux ratio is between $\sim 0.1$ and $\sim 10$,
and if the optical counterpart is an isolated blue object (under the assumption that isolated OB stars or young star clusters are not likely to be found far away from star-forming regions).  In a few other cases, we found objects that appear  extended in the Magellan images, and are thus consistent with a background galaxy, even though they appeared to be uncatalogued. Conversely if the isolated object appeared to be a star, and the X-ray to optical luminosity ratio was appropriate, we suggest a foreground star identification for the object.  When the optical counterpart has unequivocal evidence of proper motion, for example in the USNO-B1.0 Catalog \citep{monet03}, or the NOMAD Catalog \citep{zacharias04}, or the PPMXL Catalog \citep{roeser10}, we identified it as a foreground star. In several cases, we also relied on the ``probability of identification'' (star, galaxy, quasar) listed in the Atlas of Radio/X-ray associations \citep{flesch10}.

Not surprisingly the majority of the objects we classified in this fashion are objects outside the main body of M83.   In the end, we found evidence that six of the X-ray sources in the survey are stars, and that 56 are AGN/active galaxies.  Many other objects are unidentified because they have no obvious optical counterparts. 

In Table \ref{table_id_summary}, we provide a census of the various source identifications in M83.  Most of the objects identified within the D25 contour of M83 are SNRs, while most of those outside are identified with AGN/gaalxies.  

\section{Characteristics of the Bright Sources \label{sec_bright}}

Most of the 458 sources in the X-ray catalog are faint, and therefore for most of the sources, spectral analysis yields only a little more information than can be obtained from hardness ratios.  However, 29 of the sources have more than 2000 counts, enough that detailed spectral fitting can be carried out.  An additional  50 sources have between 500 and 2000 counts, not enough for really detailed fitting, but enough that one can learn more by fitting the spectrum than can be determined from simple hardness ratios.  Hereafter, we will refer to the sources with $>$2000 counts as ``bright'' sources and those with between 500-2000 counts as ``intermediate brightness'' sources.

The spatial distribution of the bright and intermediate sources is interesting.
Nine of the 29 bright sources and 13 of the 50 intermediate brightness sources are in the crowded starburst nuclear region, which we define as the region inside 0.4  kpc (0$farcm$3) from the center of M83.   
The remainder are distributed throughout the disk.  Except for the concentration of objects in the nuclear region, the disk sources are distributed fairly uniformly across the disk of the galaxy,  and are not particularly concentrated along  the spiral arms or regions of most active star formation.  Interestingly, one of the bright and 9 of the intermediate brightness sources are located outside of the D25 contours; some of these are background sources. However, because of the very extended outer disk structure seen in H~I and by GALEX, these outer sources cannot automatically be disqualified from belonging to M83.  As part of the source identification process discussed above, we have used the Magellan imaging data to search for optical counterparts, and indeed some AGN/background galaxies have been identified.  However, the X-ray spectral analysis described below and the absence of an obvious optical AGN/galaxy identification allows us to identify a number of these outer sources as likely belonging to M83.

A thorough spectral analysis of the intermediate brightness and  bright sources is beyond the scope of this overview report.  However, we have carried out a preliminary analysis of the time-averaged spectra, as created by AE, which is intended as a guide for more detailed analysis to follow by ourselves or others. 

For the 29 bright sources, we followed the following procedure.  By inspection, we first determined whether the spectra showed evidence of lines or were relatively smooth.  We then fit the spectra using XSPEC V12.8 \citep{arnaud96}.  In all cases, we fixed Galactic absorption N$_H$ at $ 4.0\times10^{20}~{\rm cm^{-2}}$ and included a second absorption component 
internal to M83 that was allowed to vary.  We used the absorption model
{\tt tbabs} and  {\tt tbvarabs} for the Galactic and M83 absorption, respectively, and 
adopted the abundances of \cite{wilms00}. A summary of the fit results for each source is given in the Appendix \ref{appendix_bright}; here we focus on the general motivation and physical interpretation of such analysis.

For the group of sources with spectra containing emission lines, with have fit the spectra with an optically-thin thermal-plasma model ({\tt APEC}), in order to get a sense of the dominant temperature (usually in the range $\sim 0.3$--$0.7$ keV). In some cases, a single-temperature model was clearly inadequate and we had to add additional components: either a second thermal component or a power law.  The physical interpretation of this class of models is either a young SNR (especially if relatively isolated) or a bright, compact starforming knot (especially in the nuclear region) which may include a number of unresolved SNRs and high-mass X-ray binaries.

A second group of bright sources (in particular, X237, X248, X284 and X286) showed featureless but strongly curved spectra that were well fit with an absorbed disk-blackbody model, with characteristic luminosity $\sim 10^{38}$ erg s$^{-1}$, peak color temperature $T_{\rm in} \la 1$ keV and characteristic ``apparent'' inner-disk radius $r_{\rm in} \left(\cos \theta\right)^{0.5} \ga 30$ km  (where $\theta$ is the viewing angle to the disk plane, and the ``true'' inner-disk radius $R_{\rm in} \approx 1.2 r_{\rm in}$ \citealt{kubota98}). The physical interpretation of these sources is that they are stellar-mass BHs in the canonical high/soft state.

A third group of bright sources shows no evidence of lines and negligible broad-band curvature. They are well, or at least adequately, fit with an absorbed power-law model with characteristic photon index $\Gamma \sim 1.5$--$1.7$. A handful of those may be background AGN ({\it e.g.}, X038 and X138); the rest are probably XRBs. At first sight, the simplest physical interpretation for their spectral shapes would be that they are either NSs or BHs in the canonical low/hard state, defined by a power-law with similar photon index. However, this interpretation is almost certainly wrong for the most of sources in this sample, because they have X-ray luminosities of $\approx 1$--$2 \times 10^{38}$ erg s$^{-1}$. This is too luminous for systems in the low/hard state, which is generally associated with luminosities $\la$ a few $10^{37}$ erg s$^{-1}$ in stellar-mass systems (unless some of them are intermediate-mass BHs, which we deem highly unlikely). Therefore, we suspect that the simple phenomenological power-law fit mimics a more complex physical model. In particular, based on our experience of XRB modeling, we investigated the possibility that the majority of the sources in this class are NS XRBs in a high state, close to their Eddington limit. Unlike BHs, NSs have a hard surface; therefore, their spectra tend to contain two optically-thick thermal components (possibly modified by Comptonization): one from the inner disk and one from the boundary layer between disk and surface, or from the surface itself. 

A simple way to model the spectra of this third group of sources is is to use a disk-blackbody plus a simple blackbody, or their Comptonized versions ({\it e.g.}, {\tt diskir} and {\tt comptt} in XSPEC). \citep[See, for example,][for detailed discussions on NS XRB spectral models.]{lin07, barret01, mitsuda89, white88}.  The other characteristic feature of a NS XRB in the high state is that its inner-disk temperature is higher than that for a BH XRB. Compact objects of smaller mass have higher disk temperatures: at a fixed Eddington ratio (for example at $L \approx L_{\rm Edd}$), $T_{\rm in} \propto M^{-1/4}$, while at a fixed luminosity, $T_{\rm in} \propto M^{-1/2}$. When we tried fitting the same sample of sources with absorbed optically-thick thermal components ({\it e.g.}, disk-blackbody plus a simple blackbody), we obtained equally good fits as with an absorbed, hard power-law. Crucially, the characteristic temperatures of the best-fitting disk-blackbody components are $T_{\rm in} \sim 1.5$--$2$ keV and the characteristic sizes are $r_{\rm in} \left(\cos \theta\right)^{0.5} \sim 10$ km. Both values are consistent with the expected parameters of NS XRBs in their high state. It is intuitive to see why a hot disk-blackbody component may be mistaken for a hard power-law at low or moderate signal-to-noise. The peak of the disk-blackbody emission occurs at $E \sim 5$--$6$ keV; the continuum below the peak energy can be approximated with an absorbed, hard power-law, while the Wien curve above the disk-blackbody peak occurs in an energy range where {\it Chandra}'s sensitivity is very low, and does not constrain the model significantly. To quantify this hypothesis, we used the {\tt fakeit} task in XSPEC to simulate the spectrum of a source with disk-blackbody parameters $T_{\rm in} = 1.8$ keV, $r_{\rm in} \left(\cos \theta\right)^{0.5} = 10$ km, total $N_{\rm H} = 8 \times 10^{20}$ cm$^{-2}$, $0.35$--$8$ keV unabsorbed luminosity $\approx 2 \times 10^{38}$ erg s$^{-1}$ (Eddington luminosity of a NS XRB) and simulated exposure time chosen to produce $2300$ counts. We then re-fitted the spectrum with the original model and with an absorbed power-law model, and found that  both models produce statistically acceptable fits; for the best-fit power-law model, the photon index was $\Gamma = 1.61 \pm 0.11$ and the inferred luminosity $\approx 3 \times 10^{38}$ erg s$^{-1}$. We conclude that apparent ``hard-power-law'' sources are indeed consistent with being NS XRBs in the high state. In some cases, though, the interpretation remains ambiguous (Appendix \ref{appendix_bright}).

For the intermediate brightness sources, our approach has been more abbreviated. We  performed automated spectral fits in 
{\tt XSPEC} with a non-thermal model ({\tt powerlaw}), an optically-thin thermal-plasma model ({\tt APEC}), and an accretion disk 
model ({\tt diskbb}).  As for the brightest sources, each  
model included an absorption component for the Galaxy fixed at  
$N_{\rm H}$ of $ 4.0\times10^{20}~{\rm cm^{-2}}$, another absorption component 
internal to M83 that was allowed to vary. We used the absorption model
{\tt tbabs} and  {\tt tbvarabs} for the Galactic and M83 absorption, respectively, and 
adopted the abundances of \cite{wilms00}.   

We examined the fit results
and classified the spectra by which model resulted in the best fit and physically reasonable
parameters.  For example, we selected the thermal-plasma model for fits that resulted
in power-law indices $\Gamma > 3.0$ and fitted temperatures for the
thermal-plasma model $kT <2.5$~keV, even if the fit statistic for the
power-law model might have been slightly lower. Conversely, we
selected the power-law model for fits that resulted
in photon indices $< 3.0$ and fitted plasma temperatures $>2.5$~keV.  For some spectra with a distinctly curved continuum, the disk-blackbody model
resulted in a significantly better fit than either the power-law or thermal-plasma model.  This typically
occurred for spectra with a large number of counts such that a
deviation from a pure power-law could be detected.

As for the brightest sources, the intermediate-luminosity sources that are best fitted by a pure thermal-plasma model are probably either individual SNRs or compact knots of hot gas in dense star-forming regions. The sources for which the
disk-blackbody model provides the dominant component are most likely XRBs in a
high state. The sources that are best fitted by a pure
power-law (typically, with photon index $\sim 1.5$) are a mixed bag and we cannot give a general physical interpretation based on X-ray modeling alone: some of them may be NS XRBs with optically-thick thermal emission at $T \ga 1.5$ keV as discussed earlier; others (especially sources with only $\sim 500$ counts) may be BH XRBs in the low/hard state, at $L \sim$ a few $10^{37}$ erg s$^{-1}$; a few others might be high-energy pulsars and pulsar-wind nebulae (Crab-like sources), which have a synchrotron power-law X-ray spectrum and can also reach X-ray luminosities $\sim$ a few $10^{37}$ erg s$^{-1}$; and finally, some of the power-law sources are background AGN.

In conclusion, we summarize our results by classifying the sources into six spectral categories depending on their best-fit model: ``P'' non-thermal
(power-law); ``T'' optically-thin thermal-plasma; ``D'' disk-blackbody; ``P/D'' either power-law or disk-blackbody with similar statistical likelihood; ``P/T'' either power-law or (hot) thermal-plasma; ``P+D'' power-law
plus disk-blackbody. These classifications are listed in Table~\ref{table_ids}.
Of the 79 sources fitted, 38 were best fit  by a power-law model (P), 18 by a  thermal-plasma model (T), and 11 by a disk-blackbody
model (D).  Six were inconclusive between a power-law and disk blackbody
(P/D), two were inconclusive between a power-law and a thermal-plasma model (P/T),
and four were best fitted by a combination of power-law and disk-blackbody models (P+D).  This type of spectra classification is crude, because (as explained earlier) the power-law model is often degenerate with disk-blackbody or Comptonization models, and thus completely different classes of physical sources can end up in the same group. Nonetheless, it provides a simple spectral classification that can be used for monitoring luminosity and spectral changes in future observations of M83, and to in some cases for comparisons between X-ray source populations in different galaxies
For reference and to show the diagnostic power of the fits, 
Figure \ref{fig_bright_spectra} shows examples of the 
model fits for two bright and two intermediate brightness sources.

\section{Nuclear Region \label{sec_nucleus}}

The bright nuclear region of M83 is a complex and dynamic environment \citep{{harris01}, {dottori08}, {knapen10}, {wofford11}},    
and exhibits a complex X-ray and optical morphology involving point sources of various types as well as diffuse emission in both bands \citep{{soria03}, {hough08}}.  Many massive, young star clusters imply an age gradient and a distributed burst of star formation, much of which is hidden from direct view by dust, especially to the north. 
\cite{dopita10} have identified some 20 SNRs in the nuclear region using {\em HST}/WFC3 imagery, as well as tentatively identifying the optical counterpart of SN1968L, as mentioned earlier. 
For our purposes, we choose the nuclear region to be represented by the inner 0.4 kpc ($\sim$0$\farcm$3) of the galaxy, which includes the entire bright starburst region, as well as a population of older stars that comprise most of the bulge the of the galaxy \cite[e.g.\ Figure \label{fig_clf_reg} of][]{knapen10}.\footnote{We discuss the cumulative luminosity function of sources in the nuclear region, and the rest of the bulge in Section \ref{CLF_bulge}.}
This region is shown in Figure \ref{fig_nuc}, where three-color versions of the Magellan broadband and \chandra\ data (unsmoothed and smoothed) are provided, with scaling to show the brightest emissions that are mostly saturated in the wider field images provided earlier.  

A reddish optical cluster and X-ray source X233 are coincident with the optical nucleus (ON), which is identified by the red circles in Figure \ref{fig_nuc}.   An optical SNR \cite[Table 3 \#15 from][]{dopita10} also nearly aligns with the ON.  The kinematic center (KC) of the nuclear region lies some 3\farcs7 to the WSW  \citep[see, e.g.][and references therein]{thatte00, knapen10, piqueras12}. Furthermore, a second mass concentration, dubbed the ``hidden nucleus'' or HN by \cite{dottori08}, was identified by \cite{mast06}  some 3\farcs9 west of the ON. The relative positions of the KC and HN, reconstructed from the original references, are shown by the blue circles in the left panels of Figure \ref{fig_nuc}.\footnote{Note that the relative positions of the ON, KC, and HN as shown in Figure 2 of Dottori et al. (2008) do not agree with the positions shown in \cite{thatte00} and \cite{mast06}.  We have carefully reconstructed the relative positions by referring back to the original figures in these references.}  Neither of these two positions has a specific X-ray counterpart, although both are within the bright, clumpy soft X-ray emission in the nuclear region energized by both the SNR population and the strong stellar winds from all of the young massive stars.

We have identified a large number of X-ray point sources within the nuclear region.  However, given both the complexity and the appearance of the region, not all of these may be physically distinct individual objects and we also may be missing sources that would be detected if they were more isolated.
iAs indicated in Figure \ref{fig_nuc}, there is very strong diffuse emission in the nuclear region; some of these sources (those whose spectra show strong emission lines)  may simply be peaks in the bright diffuse thermal emission  the permeates the nuclear region.  Nine of the sources in the nuclear region are in our bright ($\> 2000$ counts) group, and 13 are in the intermediate group discussed above, and are clearly detected  above the bright diffuse X-ray background in the region.
Significant dust lanes are also evident in the Magellan data, and these are no doubt also responsible for shaping the character of the X-ray emission.  Since we don't know where each X-ray source is along the line of sight, some sources may appear harder in X-rays than they really are, depending on the foreground absorption.  However, large regions are nearly free of overlying dust, and so some sources, especially on the near side of the nucleus, may not have excessive column densities.

Figure \ref{fig_nuc} shows an overlay of X-ray source regions, and the potential for confusion of identification of individual objects is obvious.  Despite the source crowding even at \chandra\ resolution, we have inspected the positions of all of the nuclear sources against the optical (Magellan) broadband data, as well as the \cite{dopita10}  SNR list to provide context for the possible identifications.  For the bright sources, brief descriptions are provided as part of the notes in Appendix \ref{appendix_bright}.  Even for these bright sources, we describe how alignments with optical SNRs or other structures may be chance coincidences, given the fits to the X-ray spectra. (For instance, see sections on objects such as X216, X233, and others.)  The spatial alignments for nuclear sources are annotated in Table \ref{table_ids}.

In Figure \ref{fig_nuc} (right panels) we show green and magenta regions that correspond to the \chandra\ nuclear sources.  Green circles are identified with regions of star formation, including a few that align with actual clusters.  Magenta circles indicate those sources that align with SNRs. (Note: some sources align with both star formation and SNRs.)   Sources X186, X207, X212 and X250 all align with \cite{dopita10} SNRs and are relatively soft sources on the fringes of the nuclear region.  These are probably fairly solid IDs with SNRs.  X243=S03-70 is a special case identified by D10: a compact oxygen-enhanced emission nebula that is a likely cousin to  Cas A  in our Galaxy.  X237, the magenta circle to the upper left of the field center, corresponds to a relatively hard  X-ray source (turquoise in the figure).  It aligns with an optical source that had been identified as a SNR by \cite{dopita10}; however,  our multiwavelength assessment (including  radio, \chandra, and \hst\ data) now points to its being a newly-identified microquasar MQ1 \citep{soria14}.  The apparently hard spectrum is primarily due to the fact that the source lies behind considerable absorption.  Its intrinsic spectrum can be modeled in terms of disk-blackbody models with kT of about 0.7 keV.  Several green circles align with bright white (medium-hard) X-ray sources  and are likely XRBs, but without specific optical counterparts, they have been labeled as star-formation regions.  Finally, we highlight two hard (blue) X-ray sources, X242 on the SE fringe of the nucleus and X225 in the north, both of which are projected onto dark dust lanes.  At least part of the spectral hardness of these sources may be due to absorption, assuming these sources are behind the dust.

 Because of the overall complexity of the nuclear region of M83, a future dedicated study using all available multi-wavelength data is warranted.

\section{Other Soft and Very Soft Sources in the Disk\label{sec_soft}}

As discussed in Section \ref{sec_snrids}, we have classified 87 objects as likely SNRs, based on existing catalogs of SNRs in M83 and absence of evidence that would suggest an object is not a SNR.   Of these, 79 SNRs lie within the D$_{25}$ contours of
the galaxy but outside the nuclear region, and of these, 72 have (M--S)/T hardness ratios of -0.5 or less.\footnote{There are seven SNRs,  including X279, identified with SN1957D, that have higher values of (M--S)/T. There are eight SNRs within the nuclear region.}  There are a total of 154 sources in the catalog that have (M--S)/T
hardness ratios of -0.5 or less and that lie within the D$_{25}$ contours of
the galaxy but outside the nuclear region.  Here we discuss the nature of the 82 soft sources in the disk, which are not classified as SNRs.  Soft sources in the vicinity of the nucleus have been discussed in Section \ref{sec_nucleus}.

What are the remaining soft sources likely to be?   There are at least three possibilities:
(a) currently unidentified
SNRs, (b) true supersoft sources, that is binary X-ray sources with very soft spectra, or (c) peaks in
the diffuse gas emission, possibly due to specific star formation
regions that have been characterized as point-like.   We note in passing that some of these same questions have been asked previously by \cite{distefano04} who analyzed the archival data for M83 and a number of other galaxies in a search for supersoft X-ray binaries. Using hardness ratio based criteria, they identified 53 very soft sources (VSSs) in M101, 23 VSSs in M51, 19 VSSs in the elliptical galaxy NGC4697, and (using the archival data) 54 VSSs in M83.\footnote{The energy bands that \cite{distefano04} used were S = 0.1-1.1 keV, M=1.1-2.2 keV, H=2-7 keV. They required a VSS to have (M-S)/(M+S)$<$0.8 and (H-S)$<$0.8.  Although our bands are slightly different, we follow \cite{distefano04} and denote sources that have (M-S)/(M+S) ratios  $<-$0.88 in our bands as very soft sources, to distinguish them from true supersoft sources. } 

\cite{soria03} identified two sources, S03-068=X240 and S03-096=X312 as
supersoft sources, based on all their counts occurring at energies below 1 keV. Both have
(M--S)/T hardness ratios  of $-$1 in the full dataset, consistent with this identification. Interestingly,
both appear to be variable, with more flux in the archival data than in
the more recent observations.  This suggests that these two sources, at least, are 
true supersoft sources.  A total of 17 of the  82 soft sources in the disk not identified with SNRs  show time variability, as compared to only seven apparently variable sources among  the 87 objects that we have suggested are SNRs.  X365 appears to be a ``classical transient''; S03-38=X162 and  S03-124=X393 were detected in the early observations with \chandra, but not the more recent observations, while  X273 was seen in the new observations but not the archival data. Variability suggests that at least 20\% of the soft sources in the disk are binaries.  The median value of the (M--S)/T hardness ratio of the 17 variable sources is -0.99, consistent with many of them being supersoft sources.

Both the SNRs and the other soft sources
show marked concentrations in the spiral arms.  Of the objects with
(M--S)/T$\le-$0.5, those that are not identified with SNRs tend to have
lower hardness ratios than those that are identified with SNRs.  
The median hardness ratio for the soft sources (M--S/T$\le-$0.5) which
are not identified with SNRs is $-$0.88 while the median for SNRs is $-$0.83 (or $-$0.85 if the SNRs with (M--S)/T$>-$0.5 are excluded). The SNRs are somewhat harder on average than the other sources.  Of the 87 SNRs, there are 49 with hardness ratios $> -$0.88 and 28 with values less than this. However, a two-sided Kolgoramov-Smirnov test comparing the hardness ratios of SNRs with (M--S)/T$\le-$0.5 to other sources with M--S/T$\le-$0.5, yields a probability (p-value) of 11\%, implying we cannot distinguish the underlying source populations on the basis of hardness ratios, and providing some support for the hypothesis that many of the other soft sources are unidentified SNRs.

We have created composite, or more specifically summed, spectra for the objects
identified with SNRs and those which are not, dividing each of these groups into those that 
are soft $-$0.5$>$(M--S)/T$>$$-$0.88 and those that are very soft $-$0.88$>$(M--S)/T.  The four
spectra are shown in Figure \ref{fig_soft_spectra}.  It is quite clear
that the composite spectrum of the soft sources (with (M--S)/T between
$-$0.88 and $-$0.5) not identified with SNRs  (``Soft, not SNRs'') is very similar to spectra from those
identified with SNRs (``Soft SNRs''), suggesting that many of the ``Soft, not SNRs'' sources are actually SNRs whose
optical and radio emission is too faint to have been detected. The shape of
spectrum of the very soft sources not identified with SNRs (``Very Soft, Not SNRs'')
differs from the very soft sources identified with SNRs (``Very Soft SNRs''), particularly near 0.5 keV.  Although there could be a substantial contribution from unidentified SNRs to the ``Very Soft, Not SNRs'' spectrum, there must also be a large contribution from  some class of sources
whose spectrum is similar to a blackbody. A crude blackbody fit to the``Very Soft, Not SNRs''  spectrum, fixing foreground absorption at \EXPU{4}{20}{cm^{-2}}, yields kT of about 0.1 keV. 

In the absence of a much more detailed multi-wavelength analysis, we cannot  rule out the possibility that some of the unidentified soft and very soft sources are simply peaks in the diffuse emission, which arises from hot gas, and is also concentrated  in the spiral arms.  By definition, we do not have ancillary information on these objects to show they are SNRs.  A preliminary extraction of the spectrum of the diffuse emission, with the point sources removed,  yields a (M--S)/T hardness ratio of -0.93,  similar to many of the softest point sources, and which might be associated with the feature near 0.5 keV in the ``Very Soft, Not SNRs'' spectrum.  However, the diffuse emission spectrum shows features due to  the lines of \ion{O}{7} and
\ion{O}{8} at 0.56 and 0.65 keV, as well as a feature at 0.82 keV, which is likely due to \ion{Fe}{17}, but these are not  not seen in the ``Very Soft, Not SNRs'' spectrum.  In the absence of more evidence, our belief is that this group of sources is dominated by  a mixture of SNRs and supersoft sources, with the possibility that a few of the softest sources are peaks in the diffuse emission.

\section{Luminosity Function \label{sec_lum}}

The cumulative luminosity function (CLF) of the sources in M83 reflects the evolution history of the types of objects that produce X-ray sources, which, with the exception of the stars that produce SNRs, are almost entirely binaries \citep[see, e.g.,][for a recent review of normal galaxy X-ray populations]{fabbiano06}.  There are currently several approaches to such analyses.  One approach involves modeling the star formation history itself, \citep[e.g.][]{tzanavaris13} which is beyond the scope of the present effort.  Here, we consider two simpler techniques, more appropriate for a first look at the CLF of M83. 

The first, developed by \cite{grimm02,grimm03} and by \cite{gilfanov04}, utilizes differences in ``canonical'' luminosity functions of low-mass X-ray binaries (LMXB) and high-mass X-ray binaries (HMXB)  to estimate the fraction of each of these source types in different galaxies. The  CLF  of HMXB is described by a power law $N(>L) \propto L^{-\alpha}$, with an index $\alpha = 0.64\pm0.15$ and no obvious cutoff at high luminosity. The CLF for LMXB has a flatter slope $\alpha = 0.26\pm0.08$ and a distinct cutoff  at about \POW{37.5}{\LUM} \citep{grimm02}. These canonical luminosity functions have been applied to strongly star-forming galaxies, thought to be dominated by HMXB \citep{grimm03} and older stellar systems, thought to be dominated by LMXB \citep{gilfanov04}. Mixed systems should be characterized by luminosity functions that are a combination of these.

The second focuses simply on changes in the index of a power law fit to the CLF. For example, \cite{kilgard02,kilgard05}, using a small sample of spiral galaxies, found that the CLFs of starburst galaxies, such as M82 and the Antennae, have an index of $\sim0.5$, while ``star-forming'' galaxies, a group comprising M51, M83, and M94, have indices of $\sim0.6$, similar to that of the canonical HMXB CLF determined by \citet{grimm02}.  Spiral galaxies with lower star-formation rates have higher indices, typically $\sim1.2$. Since many luminosity functions do not extend below \POW{37}{\LUM},  evidence for the `break' in the power law due to LMXBs can be confused by completeness issues. In these cases, the index may then reflect the joint contribution from the  index characteristic of HMXB and higher index characteristic of LMXB above the break.

The CLF of M83 above about \EXPU{5}{36}{\LUM}  has been described by \citet{soria03} and by \citet{kilgard05} using the early \chandra\ observations.  Both concluded that the overall CLF is a broken power law. \citet{soria03} find an index of $\sim0.6$ below the break, and steeper index, 1.6,  above the break\footnote{Despite obtaining indices from fitting differential luminosity functions, we quote the indices for the cumulative luminosity function in order to be consistent with current literature, and to allow ready comparison with the plots.}. \citet{kilgard02} found an index of 1.4 above the break and, ignoring the break,  a mean slope of 0.6. \cite{soria03} found that the CLF of the inner disk and nucleus (within 60\arcsec\ of the nucleus) could be fit with a pure power law,with a slope of 0.7 and no break, which they suggested was due to the influence of a large population of HMXBs. They also suggested that the break in the CLF of the disk outside this region might be due to multiple epochs of star formation. 

With our new  \chandra\ observations, we can now extend the luminosity function to luminosities below \POW{36}{\LUM}, can remove the contributions from foreground and background objects, and can explore the CLF as a function of position in more detail.  For this analysis, we consider only the sources within the D$_{25}$ contours, beyond which, as shown in Figure~\ref{fig_ps_rad_prof}, the surface density of sources becomes indistinguishable from that expected from the background. To permit effective comparisons with other studies, we have calculated luminosity functions in three bands, 0.35-8 keV, 0.5-2.0 keV, and 2-8.0 keV.  Unless otherwise indicated, however, our comments refer to the 0.35-8 keV broad band.

We constructed CLFs from the mean fluxes listed in Table~\ref{table_fluxes} for the combined \chandra\ data. We converted fluxes to luminosities using the power law spectrum described in Section \ref{sec_fluxes}.  Due to telescope vignetting and the change of roll angles between observations, there is some variation in the sensitivity of our observations across the face of M83. To account for this variation, we calculated the point source detection limit as a function of position. Using this method, we calculated the fraction of the area over which a source of a given luminosity would have been detected, and used this assessment to correct the luminosity function for incompleteness. We have set our sensitivity limit to be the luminosity for which the area correction factor was less than 10\%.  This varies from region to region, ranging from \EXPU{6.4}{35}{\LUM} for the bulge region to \POW{36}{\LUM} for the outer disk. (See Table~\ref{table_pops}).

The total CLF for the 0.35-8 keV band is shown in Figure
\ref{fig_band_compare}. The luminosity scale for sources in M83 is shown
on the upper axis.  The total CLF includes not only sources within M83,
but also a contribution from background AGN and foreground Galactic
stars, which must be removed to determine the CLF of M83. The total CLF also includes SNRs, which are
usually ignored in attempts to explain X-ray sources populations in normal
galaxies. The SNRs also need to be removed in order to study  the
X-ray binary populations.  

\subsection{Removing Background Sources from the CLF \label{sec_agn}}

Although we have identified a number of AGN and background galaxies as point sources in our M83
sample, most of these are outside of the D25 ellipse.  The hardness ratios of AGN are expected to be similar to
those of the XRBs and we do not have enough ancillary information to identify most of the AGN that are expected 
within the D25 ellipse.  Therefore AGN have to be removed statistically.
The number of background AGN expected in our catalog depends
upon the intrinsic flux distribution of the background AGN as well as
the point source detection limit and galactic absorption as a function
of position in the galaxy.  For our estimate, we used the intrinsic flux function
(log N-log S) obtained by \citet{kim07}, since they provide number counts for
our bands of interest. We estimated the absorption due to M83 from
a map of total hydrogen column density obtained by \cite{lundgren04}.

To establish the background point source limit at each position in M83,
we approximated the process used by {\it AE} to create the source
catalog. Specifically, at each $0\farcs492 \times 0\farcs492$ pixel in our
image mosaic, we first determined the local background rate using the
same method as {\it AE}. We then calculated the minimum number of
counts that a source at that location would need in order to have a
probability of no source $<5\times10^{-6}$ in the sum of the individual
observations. Finally, we converted counts into
fluxes using the exposure maps of each chip for each observation,
to produce point source detection limit maps for our three bands of
interest. From this point source detection limit, we calculated the
number of AGN at each flux in $\Delta\log{F}=0.1$ bins, using the
intrinsic flux functions in \cite{kim07} at each image pixel. 

From this calculation, we estimate that 97 AGN are among the 378 sources
detected with \chandra\ in the 0.35-8.0 keV band within the
D$_{25}$ region. The intrinsic broad band luminosity function of M83,
with the contributions from AGN, known foreground stars, and known SNRs all subtracted, is shown as
the dashed line in Figure \ref{fig_clf}.
There is some cosmic variance in the normalization of the AGN
function at different locations in the sky. To see if this is a
problem in the direction of M83, we calculated  the  predicted number of sources
outside the D$_{25}$ region    using the \citet{kim07} flux
function to get 59, reasonably consistent with the 54 actually identified. In Figure~\ref{fig_ps_rad_prof}, we show the radial
profile of the observed sources in M83 along with the number of sources
predicted from the flux function. If anything, we seem to have
overestimated the AGN contamination slightly.

\subsection{Removing SNRs from the CLF}

In most X-ray observations of galaxies, the numbers of SNRs detected are relatively small, and as a result the effect of SNRs on the CLF is usually ignored. However, SNRs constitute a significant fraction of the sources in M83, and must be removed from the CLF, if we want to examine the CLF of XRBs.

The left panel in  Figure \ref{fig_clf} shows the relative shapes of the CLFs for
all  the sources, for  AGN alone, and for SNRs alone.  The right hand panel shows the effects 
of removing the AGN and foreground stars from the CLF and of removing the SNRs in addition. The slope of the SNR-only CLF is very steep and does not significantly affect the broad band luminosity function above \POW{36.8}{\LUM}, but it it flattens the CLF at lower luminosities.   In the discussion of XRBs in M33 below, we have removed only the 87 sources we have specifically identified as SNRs. We will return to the effect of removing the ``Soft, Not SNR'' and ``Very Soft, Not SNR'' sources from the CLF for XRBs in Section \ref{sec_lumfunc_soft}, simply noting here that this would flatten the CLF even more.

\subsection{The CLF of XRBs in M83}

Having removed the AGN and (the majority of) SNRs, we now consider the CLF of the XRBs
in M83.  The corrected
CLF (the green histogram in the right hand panel of Figure \ref{fig_clf}) resembles that of \citet{soria03} and
\citet{kilgard05} above \POW{37}{\LUM}, as expected, even though \cite{soria03} did not
remove  AGN and neither removed SNRs, both of which become important at
luminosities below \POW{37}{\LUM}. The shape of the CLF for all sources within
the D$_{25}$ radius can be roughly characterized as a power law below
\POW{37.5}{\LUM} with progressively higher indices at higher luminosities.

To quantify our results, we have fit the differential luminosity function with a maximum
likelihood method similar to that outlined in \citet{kilgard05}.\footnote{We, as do many others, perform fits to the differential luminosity function rather than the CLF because the points in a cumulative luminosity function are not statistically independent of one another.  We quote results in terms of the cumulative luminosity function since this is the historical practice.} Since
the AGN must be removed statistically, and the maximum likelihood method
requires integral sources \citep{crawford70}, each fit cited here is the result of 1000
trial fits, each of which is made to the data from which the AGN have
been removed.
We have used the distribution of the fits of individual trials to set the uncertainties.

The fit to the corrected CLF within D$_{25}$
yields an index of $0.41\pm0.02$. Restricting the fit to luminosities
below \POW{37.5}{\LUM} yields an index of $0.50\pm0.02$. This slope is
intermediate  between that expected for a pure HMXB population and a
pure LMXB population, and the difference in slope reflects the break  expected for LMXBs.  
Had we excluded all of the soft and very sources (``Soft, Not SNRs'' and ``Very Soft, Not SNRs'') from the CLF, which would be the correct thing to do if most of these sources are unrecognized SNRs, then the slope for XRBs below \POW{37.5}{\LUM} would have been shallower
still, with an index of $0.15\pm0.03$.  This would imply, within the context majority of canonical descriptions of X-ray CLFs, that the majority of XRBs in M83 are LMXBs, a point we discuss in more detail Section
\ref{sec_lumfunc_soft}.

The mixture of high and low mass X-ray binaries typically varies as a function of
position in galaxies, depending on the local star formation history.
In particular, bulges and disks generally reveal different luminosity
functions, e.g., M31  \citep{kong03} or M81 \citep{tennant01}, with bulges
more dominated by LMXBs, and star-forming disks more dominated by HMXBs.  
We have  constructed CLFs of the XRBs  for various regions of
M83 to investigate this. The results are shown in the upper left, upper right, and
lower left panels  of Figure \ref{fig_clf_reg} for the 0.35-8 keV, 0.5-2
keV, and 2-8 keV bands, respectively.  In constructing the various CLFs, we have only used sources that were detected with high confidence (probability-of-no-source $<$ \EXPN{5}{-6}) in that energy band. Unless otherwise indicated, the following
discussion refers to the 0.35-8 keV band.

\subsubsection{Comparison of the CLFs in Different Energy Bands}

As shown in  Figure \ref{fig_clf_reg}, the CLF derived from the 0.5-2 keV range is almost identical to that in the full spectral range, 0.35-8 keV. This is not surprising since Chandra is most sensitive in the range 0.5-2 keV, which is contained within the broader band. The CLF derived from the 2-8 keV band has less structure than the broad and 0.5-2 keV CLFs and follows a simple power law to nearly its highest luminosities. Similar behavior is observed in NGC300 \citep{binder12}.   However, we do not find a difference in slope between the 0.35-8.0 and 2-8 keV band CLFs; for M83 both have indices of $0.61\pm0.03$. \chandra\ is less sensitive in the 2-8 keV range, and as a result, the CLF does not extend to as faint a limit in this band. 

\subsubsection{The CLF of the Bulge and Nucleus of M83 \label{CLF_bulge}}

Although the nuclear starburst is expected to produce a significant
number of young high-mass X-ray binaries, one expects a substantial
number of low-mass binaries from the old stellar population forming the
bulge. As in Section 7, we take the nuclear starburst to occupy the
inner $0\farcm3$. In the optical, the surface brightness of the bulge
has circular isophotes out to at least r=0$\farcm$6 and has a profile
approximating that of an $r^{1/4}$ power law \citep{jensen81}. Since the
bulge has a broader distribution than the nuclear starburst, one might
expect the mixture of populations, and thus the CLFs to vary with
radius. Given the strength of the starburst, one expects the young
population to dominate at smaller radii.

As shown in Figure \ref{fig_clf_reg}, the 0.35-8 keV CLF of the combined
bulge and nuclear region is relatively flat with increasing indices
above \POW{37.5}{\LUM}. This CLF stands in apparent contrast with the
simple power law found by \citet{soria03} within $1\farcm0$, however,
the break in our CLF is near the limiting luminosity of the
\citet{soria03} CLF. A power law fit to the data below \POW{37}{\LUM}
yields a power law index of $0.19\pm0.04$. The power law indices are not
significantly steeper in the 0.5-2 keV band but are in the 2-8 keV
band. The break is also more abrupt in the 2-8 keV CLF, than in the
other energy bands.  Thus, in terms of the simple prescriptions
described by \cite{grimm02}, it appears that the CLF of the bulge and
nuclear region of M83 resembles that expected for an X-ray source
population dominated by LMXBs. This result suggests that the bulge
dominates over the starburst.

In light of this unexpected conclusion, we constructed the CLFs
separating the inner starburst region from the outer region which is
presumably more bulge dominated. The outer bulge region was set to be
$0\farcm3<r<0\farcm6$, while the starburst was set to be $r<0\farcm3$.
The CLFs are shown in Figure~\ref{fig_bul_elf}. The 0.35-8 keV CLF of
the inner nuclear starburst is quite flat with with  a break at
$\sim$\POW{37.3}{\LUM}, while the 0.35-8 keV CLF for the outer bulge
region is a well fit by a power law with a rather steep index of
$0.84\pm0.02$. A Kolmogorov-Smirnov test shows that the two CLFs are not
drawn from the same distribution to a high degree of confidence. The CLF
of the inner region resembles that of a canonical LMXB CLF, while the
outer region resembles that expected from HMXB.

That the CLF of the outer bulge is not dominated by an old population
suggests that the bulge population is either weak or more centrally
concentrated. That the CLF of the starburst region is dominated by an
old population is more problematic, and requires a closer consideration
of observational factors. If the nuclear region were an intrinsic power
law CLF, then our observed CLF would be missing faint sources. Missing
faint sources is not unreasonable given the bright diffuse emission and
confusion in the center of the nuclear region. Both issues are of more
concern for the soft band than the hard band. To avoid source confusion
issues,  one requires less than 3-10\% of the region to be occupied by
sources depending on the slope of the CLF \cite[see,
e.g.][]{takeuchi04}. Assuming source radii of an arc second, sources
cover 12\% and 7\% in the soft and hard bands respectively. The source
coverage increases to $\sim23$\% in both bands as the radius of the
nuclear region is decreased to $0\farcm1$. Thus we have significant confusion
issues, particularly in the inner nucleus. Thus the true shape of the
CLF in the nuclear starburst is ambiguous.

\subsubsection{The CLF of the Disk}

Outside the nuclear region, one might expect the situation to be simpler, with HMXBs prominent in the spiral arms and LMXBs in the interarm regions.  We define the disk as the region outside of the bulge, but inside the D$_{25}$ radius. As shown in Figure \ref{fig_clf_reg}, the disk CLF does not appear to be a simple power law with a cutoff. Instead, the slope is steeper below \POW{36.5}{\LUM} (96 sources, index $\sim$1.0) than between between \POW{36.5} and \POW{38}{\LUM} (56 sources, index $\sim$0.4), where there is a break and the CLF steepens once again (19 sources, index $\sim$0.7). These features are not due to small-number statistics. The detailed structure structure cannot be explained easily within the context of the canonical LMXB/HMXB luminosity functions, but the
break near \POW{37.5}{\LUM} is the break associated with LMXB.

We have attempted to clarify our understanding of  the CLF for the disk by considering subregions shown in Figure \ref{fig_key}.  As shown in Figure \ref{fig_clf_reg}, the inner disk (the sum of the arm and inter-arm CLFs) and the outer disk have similar CLFs, save for the sharp increase in the number of sources in the lowest bin of the outer disk. The inner and outer disk CLFs are even more similar in the 0.5-2 keV band. Significant differences do appear in the 2-8 keV band; the inner and outer disks have very different indices, with the outer disk having a steeper index of $0.83\pm0.06$ and the inner disk having a shallower index of $0.24\pm0.02$. The inner disk does show the same type of roll-off at the higher luminosities seen in other bands. The difference between the 0.5-2 and 2-8 keV bands suggests a color difference between the inner and outer sources, and that one should be aware of band pass effects when comparing results of different observers using different bandpasses. Any radial effects, such as changes in metallicity and star-formation history, are indistinguishable in the CLFs. 

We also divided the inner disk into arm and inter-arm regions (see Figure \ref{fig_key}). In the 0.35-8 and 0.5-2 keV bands, the arm CLF shows the increased index below \POW{36.5}{\LUM} that is seen for the overall disk CLF, while the inter-arm CLF does not. Thus the \POW{36.5}{\LUM} inflection is due to sources in the arms. Neither the arm nor inter-arm CLF has a slope as steep as expected for an HXRB-dominated population. Thus, a perhaps na\"{i}ve  expectation of finding HMXBs preferentially in the arms is not confirmed.

\subsubsection{The Effect of Soft Sources on the CLF \label{sec_lumfunc_soft}}

In constructing the CLF of X-ray binaries in M83, we have excluded the 87 objects identified as likely SNRs. However, many more of the soft and very soft sources in our sample may also be SNRs (see Section \ref{sec_soft}).  The lower right hand panel of Figure~\ref{fig_clf_reg} shows what happens to the CLF if all sources with hardness ratio (M--S/T)$<-0.8$ are removed from the sample.  Specifically, the disk CLF in the broad band becomes a broken power law with an index of $0.24\pm0.08$ below the break, a break at \POW{37.6}{\LUM}, and an index of $0.61\pm0.05$ above the break.  Since the soft sources are preferentially faint, the main effect of their removal is to flatten the CLFs, making the corrected  CLF of M83  look even more like that expected for an LMXB population.

This effect is particularly strong in the disk, since many of the soft sources are located there.  Moreover, the CLF shapes from the different subregions of the disk look much more similar. There is, for example, virtually no difference in the CLF for sources in the arm and inter-arm regions once the soft sources have been removed. This is consistent with the hypothesis that a majority of the binary X-ray sources in M83 are associated with older source populations.   The corollary is that soft sources are sensitive to environment (and possibly to the existence of a young stellar population), and are thus found preferentially in the arms. It should also be noted that the bulk of these soft sources are not detected in the hard band, so removing these sources has little effect on the hard band CLFs.

\subsection{Summary and Further Discussion of the CLF of M83 \label{clf_discussion}}

Our results on the CLF of M83 can be summarized as follow:

\begin{itemize}

\item
The CLF of the nuclear starburst and bulge of M83, as derived from our point source catalog, resembles, {\it prima facie}, that of a population of LMXB. However, this result appears to be misleading, reflecting, primarily, the effects of source confusion in the innermost 0$\farcm$3 and the bright, diffuse, strongly structured background there.  In fact, the outer bulge (the region between 0$\farcm$3 and 0$\farcm$6 of the center) surrounding the inner nuclear starburst region has the power law luminosity function expected from a HMXB population. 
\item
The CLF of the disk shows a complex shape, with an inflection at \POW{36.5}{\LUM} and a break at \POW{37.5}{\LUM}. The \POW{37.5}{\LUM} break is common to all the subregions of the disk, and appears to be a fundamental characteristic of the M83 XRB population. As this break is at roughly the luminosity expected for the break in the canonical LMXB luminosity function, the interpretation may be straight-forward. The \POW{36.5}{\LUM} inflection is stronger in the arms than in the inter-arm regions, and disappears if one removes the soft and very soft sources. We suspect that a large fraction of the soft sources and part of the very soft source population (not currently identified) are in fact SNRs, and their association with spiral arms supports this hypothesis. The true XRB CLF below \POW{37.5}{\LUM} is thus likely to be more like the CLF with the soft and very soft sources removed than the CLF. The CLF for the XRBs in the disk of M83 is thus well characterized as a broken power law. The index below the break is consistent with that of canonical LMXB luminosity functions, as is the break luminosity. That the index above the break does not roll off as quickly as the canonical LMXB luminosity function indicates the presence of {\it some} HMXB despite the dominance of LMXB. We would have significantly overestimated the HMXB contribution had we not removed  (the majority of) the SNRs.
\end{itemize}

Since M83 clearly has a mixture of HMXBs and LMXBs, we may test the extent to which  scaling of the canonical HMXB and LMXB luminosity functions \citep{gilfanov04,grimm03} provides a reasonable match to our observed CLFs. \citet{gilfanov04} provides the canonical LMXB luminosity function with a normalization per galactic stellar mass (as measured from the K band luminosity) while \citet{grimm03} provides the canonical HMXB luminosity function with a normalization per unit star-formation rate. The Milky Way version of these functions are only slightly different and extend to  \EXPU{\sim3}{35}{\LUM}. Our estimate of the mass in stars follows directly from from the K-band magnitude, which we take to be  4.62 from the 2MASS Large Galaxy Catalog \citep{jarrett03}, the distance to M83,  the absolute K magnitude of the Sun (3.39), and a mass-to light-ratio estimate of 0.59 for M83 based on the B -- V color of the galaxy \citep{bell01}. We expect M83 to have a stellar mass of \EXPU{4}{10}{\MSOL}. We take the star-formation rate of 3.5{\MSOL}yr$^{-1}$ from the value of 3-4 {\MSOL}yr$^{-1}$ estimated by \cite{boissier05}. In Figure~\ref{fig_gg}, we compare the broad band CLF for the entire galaxy within the D$_{25}$ after the removal of the AGN and SNRs (both with and without the removal of the soft and very soft sources) with the Grimm \& Gilfanov canonical luminosity functions, scaled to the mass and star-formation rate of M83. The scaled combination of canonical luminosity functions fails to match the one we have derived from the observations rather badly.   

For the CLF including the soft and very soft sources that we propose are dominated by SNRs, a better agreement can be found if the star-formation is only $\sim2$ {\MSOL} yr$^{-1}$ and the stellar mass increased by  a factor of 1.5-2.0.  However, it is not clear that either of these is allowed; the alternative is that the scaling factors provided by \cite{gilfanov04} are simply averages and vary significantly from galaxy to galaxy, depending on the star formation history and perhaps the metallicity.  Better agreement would be obtained if one also allows the LMXB function to shift to higher luminosities by factors of two to four; variation in the cut-offs of individual galaxies have been observed previously \citep{gilfanov04}.  In any event, the X-ray CLF suggests that the current starburst has provided a very thin icing of HMXB over a substantial population of LMXB that has built up over a long period of steady star-formation over the last Gyr. 

A recent study of the luminosity functions of star-forming galaxies \citep{mineo12} examined only galaxies with a star-formation rate/stellar mass ratio $>10^{-10} yr^{-1}$. M83 falls just below this value, suggesting that it is not HMXB dominated. Using the Gilfanov scaling from stellar mass to LMXB rate, \citet{mineo12} found an unexpectedly large number of LMXB in their ``HMXB-dominated'' galaxies. M83 suggests that LMXB scaling may not need to be reduced so much for the galaxies at the low end of the specific star-formation rate, and that the HMXB scaling may need to be reduced as well. Overall, M83 falls below their $L_X$-SFR relation, but within the observed scatter for that relation. 

As we noted at the beginning of this Section, our current examination of the CLF in M83 was intended primarily to provide an initial exploration of the properties of the CLF.  That the canonical luminosity functions do not describe the M83 CLF {\it well} was probably to be expected; more sophisticated models based on calculating the evolution of stellar populations \citep[e.g. StarTrack,][]{belczynski08} show that even simple star formation histories can produce complex CLFs \citep{luo12,fragos08,tzanavaris13}.  We anticipate applying these more sophisticated models to these data in the near future. However, even before applying them, it is clear that the faint populations in M83 may pose some significant challenges.  Further studies, including those we are carrying out with HST \citep{blair14}, are needed to identify as many of the soft and very soft sources as possible. Since a small variation in the number of high luminosity sources can make a significant change in the shape of the CLF, and those source should be dominated by HMXB, optical efforts to identify HMXB hold are likely to shed some light on this mystery.

\section{Discussion}

We have surveyed point sources in M83 to a depth of about \POW{36}{\LUM} in the 0.35-8 keV band, finding 378 sources within the D$_{25}$ contours of the galaxy.  As reported by \cite{soria03}, the sources are concentrated in the nuclear region of the galaxy, and in the spiral arms. The association with the spiral arms is mostly due to the sources with soft spectral indices; the harder sources, which tend to be brighter, are more evenly distributed across the face of the galaxy.  

In an attempt to understand what fraction of the X-ray emission of the galaxy that has been resolved into point sources by our observations, we have carried out a simple two-component thermal fit to the spectrum obtained from all of the source counts within the D$_{25}$ contours of the galaxy.  Based on this we estimate the intrinsic 0.35-8 keV X-ray luminosity of M83 to be \EXPU{1.85\pm0.20}{40}{\LUM}.  Of this, \EXPU{5.8}{39}{\LUM} or 32\% arises from the nuclear/bulge region of the galaxy, and  \EXPU{1.27}{40}{\LUM} or 68\%  arises from the disk. About half of the luminosity of the disk, or \EXPU{6.3}{39}{\LUM} arises from sources brighter than \POW{38}{\LUM}.    The total intrinsic luminosity of the point sources, calculated by summing up all of the point source fluxes within the D25 contours, and fitting the resulting spectrum, is \EXPU{1.30\pm0.15}{40}{\LUM}, or 70\% of the total.  

The remaining 30\% of the X-ray emission is most likely dominated by a combination of emission from hot diffuse gas and unresolved point sources. If we exclude regions containing point sources within the D$_{25}$ contours, and construct a spectrum of M83, the spectrum shows a very strong soft X-ray component with obvious emission from a hot plasma.  If we carry out a two-component thermal fit, we find that about \EXPU{6}{39}{\LUM} arises from a soft component, clearly associated with emission from hot diffuse gas, and \EXPU{1-2}{39}{\LUM} arises from a hard component, which we suspect is unresolved point sources.\footnote{Analyzed in this manner, the sum of the point source luminosity \EXPU{1.30}{40}{\LUM} and the two components to the unresolved emission, \EXPU{6}{39}{\LUM}  and  \EXPU{1-2}{39}{\LUM} is \EXPU{2}{40}{\LUM}, which exceeds the total luminosity obtained by simply fitting all the counts within the D25 contours,  \EXPU{1.85\pm0.20}{40}{\LUM} slightly.  Small differences of this type are to be expected given the relatively crude way in which these spectra have been extracted from the data and then fit.}   A large portion of the unresolved emission, both in the soft component and the hard component, arises from the nuclear region where only 31\% of the  luminosity is resolved into point sources.

The number of SNRs we appear to have found in M83 is remarkable: 87 if we count only sources with supporting information that they are SNRs, and $\sim 130$ if we assume that half of the unidentified sources with (M--S)/T hardness ratios less than $-$0.5 are SNRs.  There are 378 sources within the D$_{25}$ contour of M83.  Of these, we estimate that  97 are AGN and we identify 2 as foreground stars.  Thus of the approximately 279 sources that we believe are actually in  M83 within the D$_{25}$ contours, we estimate 31\% are SNRs, or 47\% if we include half the unidentified soft sources as likely SNRs.  

Is the large number of SNRs unusual?   M33, which is much nearer \cite[817$\pm$58 kpc, ][]{freedman01}, much less massive \cite[\EXPU{3-6}{9}{\MSOL},][]{corbelli03},  a lower SF rate \cite[0.45$\pm$0.10 $\MSOL yr{^-1}$,][]{verley09} and similar foreground absorption \cite[\EXPU{6}{20}{cm^{-2}},][]{dickey90}, is one obvious point of comparison, since it was the subject of a very deep set of \chandra\ observations totaling 1.4 Msec of observing time \citep{plucinsky08}.  Two separate analyses of this data set have been carried out.  In a targeted search for SNRs,  \cite{long10} found excess ($>$2$\sigma$) X-ray emission at the positions of  82  of the 135 optically-identified SNRs in M33.  They estimated that there are no undiscovered SNRs in the region covered by the \chandra\ observations brighter than \EXPU{4}{35}{\LUM} in the 0.35-2 keV band.  As a result, M33 is the only spiral galaxy with comparable numbers of identified X-ray SNRs to M83.   Separately, in an analysis of  point sources in M33 with detection criteria very similar to those adopted here, \cite{tuellmann11} found 45 SNRs out of a total of 662 sources, to a luminosity limit of \EXPU{1.2}{35}{\LUM} in the 0.5-2 keV band.      In the case of M33, AGN constitute a much larger fraction of the sources in the survey than in M83; roughly 80\% of the sources are background AGN, \cite[see Figure 10 of][]{tuellmann11}, leaving about 130 sources from M33. Therefore in M33,  as in M83, roughly one third of the sources are SNRs.    Of the 56 sources reported in that study that have an an X-ray luminosity greater than \POW{36}{\LUM}, five were SNRs.\footnote{As calculated by us from using the photon fluxes contained in \cite{tuellmann11} and assuming a power law with photon index -1.9.}   Scaling by the SFR, 0.45$\pm$0.10 $\MSOL yr{^-1}$ vs. 3-4 \MSOL, one would have expected to find 30-60 SNRs in M83 brighter than \POW{36}{\LUM}, approximately what is found.

At a distance of 2 Mpc \citep{dalcanton09}, the Sculptor group spiral NGC300 is similar to M33 in terms of foreground absorption \cite[\EXPU{6}{20}{cm^{-2}},][]{dickey90}and stellar mass \cite[\EXPU{2.4}{9}{\MSOL},][]{puche90}.  
\citet{binder12} found 95 X-ray sources down to a luminosity limit of \POW{36}{\LUM} in a 60 ksec \chandra\ observation.  Of these, they suggest that 11 sources are SNRs based on hardness, of which seven turned out to be within 3.5\arcsec\ of known optical SNR candidates.  They estimate that about 47\% of the sources are AGN, in which case about 22\% of the sources located in NGC300 are SNRs, so it too has a relatively high SNR fraction.

For M31,  the  fraction of SNRs appears to be lower.  \cite{stiele11}, using XMM, found 1897 sources along the line of sight to M31 down to a luminosity limit of \POW{35}{\LUM} in the 0.2-4.5 keV band.  
They detected 25 SNRs and 31 SNR candidates, roughly 3\% of the total. The SNRs, along with the bright XRBs, are concentrated in the arms, and thus near regions of active star formation.  However, the majority of their sources are, not surprisingly, unidentified.  They did find 31 SSSs, 14  transients arising from optical novae, and a number of X-ray binaries, including  36 certain and 16 likely LMXBs in globular clusters.   They classify about 260 as foreground objects, and about 90 as background objects (with various degrees of certainty), based on the sources' characteristics and comparisons to other surveys and optical data.  They did not attempt a statistical analysis of the number of background sources in the sample, deferring this analysis to a future paper; they do note in passing that they expect about 2/3 of the 1247 hard spectrum objects that are unclassified objects to be background objects.  If this is correct, there are a total of 715 sources in their sample located in M31, reflecting the fact that the mass of M31 is much larger than either M33 or NGC300.  The SNRs and SNR candidates thus comprise about 8\% of the sources, much less than the corresponding fraction in M33 or M83.      

The fact that M31 has a lower fraction of SNRs is due to the specific SFR, which is lower in M31 because it is an earlier type galaxy.  The SF rate in M31 is estimated to be 0.4 \MSOL~ yr$^{-1}$ \citep{barmby06}; the stellar mass is \EXPU{10-15}{10}{\MSOL} \citep{tamm12}.  This compares to $\sim$3-4 \MSOL~yr$^{-1}$ \citep{boissier05} and \EXPU{4}{10}{\MSOL} in M83 (see, Section \ref{clf_discussion}) and to 0.45$\pm$0.1 \MSOL~yr$^{-1}$ \citep{verley09} and \EXPU{3-6}{9}{\MSOL}\citep{corbelli03}  in M33. 

The fact that the absolute numbers of SNRs are high in M83, despite its distance of 4.61 Mpc, is almost certainly due to the high star formation rate near $\sim$3-4 \MSOL yr$^{-1}$ \citep{boissier05}, the depth of our \chandra\ observations, and low galactic absorption along the line of sight.  While M83 and M33 have roughly the same number of X-ray detected SNRs, the majority of the SNRs that were detected in M33 would not have been detected at the distance of M83.   M33 has not had a single historical SNe, and M31 has only had one; M83 has had six.   There may also be some differences in the SNR populations in M33 that are not associated simply with the star formation rate.  There is, for example, some indication \citep[see Figure 15 of][] {blair12} that M83 SNRs are systematically brighter in \HA\ than their M33 counterparts, presumably because they are on average expanding into a denser medium.  If this is the case, then one might expect them to be systematically brighter in X-rays as well.   However, a detailed discussion of the characteristics of the SNRs in M33 and a comparison to the properties of SNRs in other galaxies is beyond the scope of this report.

As described in Section \ref{sec_soft}, there are 82 unidentified sources with (M--S)/T hardness ratios of -0.5 or less,  and half of these have ratios $<$-0.88.  Based on their variability, some of these sources, of order 20\%, are  supersoft sources. Some of these are likely to be WDs undergoing nuclear burning as a result of mass transfer, and others are likely BH transients.  Like the SNRs, the supersoft sources  are seen due to the fact that N$_H$ is so low along the line of sight to M83.  Further work is clearly required.  Deeper narrow-band imaging of these sources with {\it HST} and our follow-up JVLA study of M83 may identify many of the SNRs and, by elimination, the remaining supersoft objects in the sample.

\section{Summary \label{sec_summary}}

We have presented an overview of the sources identified from a series of new \chandra\  observations of M83 that total 729 ks of observing time distributed over a year, along with a new radio survey of M83 with ATCA.  Combined with archival observations from a decade earlier, we have found 378 X-ray sources within the D$_{25}$ contours of the galaxy to a limiting luminosity of about \EXPU{8}{35}{\LUM}, of which 45 are coincident with ATCA sources.  About 1/4 of the X-ray sources are seen to be variable in our initial analysis. Despite the sensitivity of the survey, a large majority of the sources are associated with M83, as opposed to background or foreground objects.

The luminosity of the X-ray point sources totals \EXPU{1.30\pm0.15}{40}{\LUM}. The brightest source in the galaxy is a ULX, discussed in detail by \cite{soria12},  located in an interarm region of the galaxy.  The total luminosity of the galaxy is about \EXPU{1.85\pm0.2}{40}{\LUM}, and so about 70\% of the X-ray luminosity has been resolved into  point sources. Our main results are as follows:

\begin{itemize}
\item 
Most of the sources in the galaxy are binary X-ray sources.   The binary X-ray sources in the disk are not strongly correlated with the spiral arms.  The luminosity function of the binary X-ray sources in the disk  resembles that expected from LMXBs, despite the fact that M83 has a high star formation rate of between 3 and 4 \MSOL\ yr$^{-1}$ \citep{boissier05}.  The CLF outer bulge/nuclear region appears to be dominated by high mass X-ray binaries. The CLF of the inner bulge/nuclear starburst has the shape expected for a low mass binary population, contrary to expectations, but this unexpected result may be attributable to the effects of source crowding and strong diffuse emission in the region.

\item 
There is a substantial number of SNRs in the sample, a higher percentage than has been found  in  studies of most (but not all) other galaxies.  A total of about 67 (73) sources lie within 1\arcsec\ (2\arcsec)  of SNRs identified by \cite{blair12} from optical interference filter imagery.  The spectra of these sources are soft, and in X-ray hardness-ratio diagrams they occupy a region where SNRs are expected.   Counting sources from other optical surveys \citep{dopita10} and objects with soft spectra coincident with ATCA radio sources, we find that 87 X-ray sources are most likely SNRs, fully 24\% of the sample within the D$_{25}$ contours of the galaxy, and 31\% after AGN have been statistically removed.  There are 82 other sources in the same region of the hardness-ratio diagram as SNRs; many of these are likely SNRs as well.  The large number of SNRs detected in the survey is most likely due to a combination of factors, including the high sensitivity of the survey to soft sources, the fact that the density in the ISM in M83 is high, making SNRs relatively bright, and the fact that M83 lies along a line of sight with low foreground column.  This is also consistent with the high SF and SN rates in M83.  Unlike the binary X-ray sources, the SNRs (and indeed the soft sources as a whole) are concentrated in the spiral arms of the galaxy.
\end{itemize}

This data set represents a resource for understanding the X-ray properties of galaxies with active star formation, which we plan to exploit in conjunction with our on-going efforts to acquire additional data with \hst\ and the JVLA.  Future reports will address areas such as the global properties of the supernova remnant population in M83, the characteristics of the diffuse emission, and X-ray emission from the nuclear region.

\acknowledgements

Support for this work was provided by the National Aeronautics and Space Administration 
through \chandra\ grant number GO1-12115, issued by the \chandra\ X-ray Observatory Center, 
which is operated by the Smithsonian Astrophysical Observatory 
for and on behalf of NASA under contract NAS8-03060.   WPB and KK acknowledge \chandra\ 
grant number GO1-12115C to Johns Hopkins University.
PFW also  acknowledges financial support from the National Science Foundation 
through grant AST-0908566, 
and the hospitality of the Research School of Astronomy and Astrophysics, 
Australian National University, during a portion of the work presented here.

\appendix
\section{Specific Comments on the 29 Brightest X-ray Sources\label{appendix_bright}}

As discussed in the Section \ref{sec_bright}, there are 29 X-ray sources with more that 2000 counts that allow a more detailed treatment of there X-ray spectral fitting.  Below we present these sources in increasing RA order, and provide a brief description of their position in the galaxy, any possible optical counterparts, and comments about the spectral results and what they say about the possible identification of the source type.  The source luminosities quoted below are derived from our individual XSPEC spectral fits, and differ from those in Table \ref{table_fluxes} which assume a simple power-law model with fixed photon index and column density:

\begin{itemize}

\item  Source X029 (=S03-005) is a persistent X-ray source, with little evidence of time variability.  It is located along the line of sight to a small star-forming clump or OB association.  It has a hard, featureless X-ray spectrum that can be fit (phenomenologically) with a power law with photon index $\Gamma = 1.58 \pm 0.08$ and $0.35$--$8$ keV luminosity \EXPU{2-3}{38}{\LUM}. We explained in Section 6 that the hard power-law spectrum cannot be physically interpreted as a canonical low/hard state, and is more likely the result of two optically-thick thermal (or thermal-Comptonization) components often found in the spectra of NS XRB at high luminosities. In this case, an equally good fit is obtained with a disk-blackbody with $T_{\rm in} = 1.81^{+0.14}_{-0.10}$ keV and $r_{\rm in} \left(\cos \theta\right)^{0.5} \approx 10$ km. Thus, we interpret this source as a NS XRB near its Eddington limit.

\item Source X038 is a persistent source that is coincident with an AGN (z=0.078) in the 6dF galaxy survey \cite{jones09}.   The source varies by a factor of about four on timescales of months. 
Consistent with this identification, the spectrum is well fit with a power law spectrum with a photon index $\Gamma = 2.4 \pm 0.1$.  The luminosity of the source is about \EXPU{2.1}{42}{\LUM} at this redshift.

\item Source X138 (=S03-027) is a hard X-ray source, whose spectrum is well modeled in terms of a simple power law with $\Gamma = 1.24 \pm 0.10$ (unusually hard for an XRB).  It is located well away from the star forming regions of M83 near the southern edge of the bright optical disk, and there is no evidence of emission from a background AGN in the optical images.  The source varied by $\pm$30\% in late 2010 and 2011, but was considerably brighter in one of the archival observations.  We also tried a disk-blackbody plus simple blackbody model, and a Comptonized disk model ({\tt diskir}) but they give slightly worse fits than a simple power-law. If the source is in M83, it has an average $0.35$--$8$ keV luminosity of \EXPU{1.1}{38}{\LUM}, too high for a BH system in the low hard state. Thus, it is possible that this source is a NS XRB in the high state, but its true nature remains unclear; the difficulty in obtaining a good fit to the average spectrum may be due to spectral variability between observations.

\item Source X145 (=S03-031) is a persistent X-ray source with a $0.35$--$8$ keV luminosity of about \EXPU{1.5}{38}{\LUM},  located north of the nucleus within one of the outer spiral arms. The source varies by $\pm$20\% between observations.  There are no stellar clusters at the position of the source. As we discussed for several other sources, it is well fit by a hard power-law model ($\Gamma = 1.53 \pm 0.07$) which we regard as purely phenomenological.
A disk-blackbody plus simple blackbody model provides a more physical interpretation, but has too many free parameters to constrain the fit. A disk-blackbody model provides an acceptable fit (statistically equivalent to the simple power-law model) with $T_{\rm in} = 1.80^{+0.13}_{-0.11}$ keV and $r_{\rm in} \left(\cos \theta\right)^{0.5} \approx 8$ km, as expected from a near-Eddington NS XRB.

\item Source X152 (=S03-033) is a persistent source, varying by $\pm$20\% between observations; it is located in the southern spiral arm, but there is no obvious optical counterpart. As for several other cases already discussed, a power-law model gives a hard slope $\Gamma = $1.45$\pm$0.07, but a more physical interpretation is to be found in a disk-blackbody plus blackbody model. The disk parameters are $T_{\rm in} = 1.97^{+0.27}_{-0.15}$ keV and $r_{\rm in} \left(\cos \theta\right)^{0.5} \approx 7$ km, and the unabsorbed $0.35$--$8$ keV luminosity is \EXPU{1.5}{38}{\LUM}. Once again, we conclude this system is probably a NS XRB, with surface and disk emission, radiating near the Eddington limit.

\item Source X168 was not detected in either 2000 or 2001, but appears as a bright X-ray source in observations in late 2010; it then faded gradually by at least a factor of 10 when observed in 2011 September, before brightening again in 2011 December.  Its average $0.35$--$8$ keV luminosity is \EXPU{8}{37}{\LUM}.The source is located in the inner portion of one of the spiral arms just south of the nucleus, but has no obvious specific optical counterpart in the Magellan images.  The time-averaged spectrum is best modeled in terms of a standard disk-blackbody with $T_{\rm in} = 1.01 \pm 0.05$ keV and $r_{\rm in} \left(\cos \theta\right)^{0.5} \approx 20$ km. This suggests that it is more likely a BH transient reaching the high/soft state in some of the observations, but a NS XRB cannot be completely ruled out.

\item Source X185 (=S03-44), one of the brightest in M83, is a persistent source located in one of the spiral arms due north of the nucleus, without an obvious optical counterpart.  The flux from the source rose by about 10\% gradually throughout 2011.  The spectrum appears slightly curved, and is inconsistent with either a simple disk-blackbody or a simple power-law model. It is well fitted either by a broken power-law or by a disk-blackbody plus power-law model. The disk parameters ($T_{\rm in} = 2.2 \pm 0.1$ keV, $r_{\rm in} \left(\cos \theta\right)^{0.5} \approx 10$ km) suggest a NS and are inconsistent with those expected for a BH XRB. On the other hand, the $0.35$--$8$ keV luminosity is \EXPU{5-7}{38}{\LUM}, depending upon the model used, which is extremely high for a NS.  We suggest that the system is more likely to be a NS XRB radiating at 2 or 3 times its Eddington luminosity, but further studies are merited.

\item Source X193 (=S03-047) is a bright, modestly variable source just SW from the brightest portion of the nuclear region.  The spectrum is moderately curved and cannot be fit by a simple power-law. Several more complicated models (broken power-law, or disk-blackbody plus power-law, or a variety of comptonized disk models) produce satisfactory fits, and yield consistent $0.3$--$8$ keV luminosities of about \EXPU{1.3}{38}{\LUM}. In the disk-blackbody plus power-law model, the best-fit parameters are $\Gamma = 1.9^{+0.8}_{-0.5}$, $T_{\rm in} = 1.7 \pm 0.3$ keV and $r_{\rm in} \left(\cos \theta\right)^{0.5} \approx 7$ km, consistent with a NS XRB in the high state.

\item Source X198 (=S03-049)  is one of a grouping of sources NW of the optical/IR nucleus and generally projected onto a region of high optical extinction.  By our conservative definition of variability, the source appears to have been constant throughout the year. It can be fitted with a simple power-law model with $\Gamma = 1.54 \pm 0.09$ and total $N_{\rm H} =$ \EXPU{1.5\pm0.3}{21}{cm^{-2}}; however, there is a hint of curvature in the spectrum at about 3 keV. A disk-blackbody model provides a better fit, with only line-of-sight absorption, $T_{\rm in} = 1.75^{+0.08}_{-0.16}$ keV and $r_{\rm in} \left(\cos \theta\right)^{0.5} \approx 7$ km. In this model, the unabsorbed $0.3$--$8$ keV luminosity of \EXPU{9}{37}{\LUM}. Once again, the luminosity and spectral properties point to a bright NS XRB.

\item Source X216 (=S03-056) is a bright nuclear source with a thermal spectrum, showing prominent emission lines from Mg XI at 1.3 keV, Si XIII at 1.8 keV and S XV at 2.4 keV. The source flux appears constant with time, as one would expect from spatially extended hot plasma. A two-temperature thermal plasma model with $T_1 \approx 0.26$ keV and $T_1 \approx 0.72$ keV provides a good fit but only if the lower-temperature component sees a much lower absorption (line-of-sight only) than the one seen by the higher-temperature plasma ($N_{\rm H} \approx 5 \times 10^{21}$ cm$^{-2}$). The unabsorbed X-ray luminosity is about \EXPU{1.3}{38}{\LUM}. The source is consistent with the position of SN1968L, as reconstructed by \cite{dopita10}, and if it is the remnant of SN1968L would be one of the brightest X-ray SNRs known. However, the connection of the SN to the X-ray source is far from conclusive. It is located in a region of extremely active star formation, directly adjacent to one of the brightest young star clusters.  Two other SNRs listed in \cite{dopita10} lie close by (nuclear SNRs \#5 to the NW and \#6 to the SSW). In addition, there are several other nearby sources with very similar thermal-plasma spectra, and the general bright diffuse emission from hot gas in the nuclear region could significantly contaminate the spectrum of X216.

\item Source X220 (=S03-057) is within the bright optical nucleus directly W of the optical/IR nucleus, with a thermal spectrum, showing Si XIII at 1.8 kev and S XV at 2.4 keV, very similar to X216.   The spectrum requires both a multi-temperature thermal-plasma ($T \sim 0.4$--$0.7$ keV) and a steep power-law ($\Gamma = 3.1 \pm 0.3$) component, with intrinsic absorption of order \EXPU{1}{21}{cm^{-2}}.  The $0.35$--$8$ keV luminosity is about \EXPU{1}{38}{\LUM}.  No known SNR candidates align with this source. There is some evidence that the source was brighter in the archival observations, which suggests that we are seeing a combination of (one or more) unresolved XRBs in a compact knot of thermal-plasma emission.  

\item Source X223  is a nuclear source in close proximity to X216 and X220, and has a very similar thermal spectrum, although it is somewhat less luminous, \EXPU{5}{37}{\LUM}. It is probably the same as S03-059.  Dopita et al. (2010)'s nuclear SNR \#8 is essentially coincident, and \#10 is just to the ESE, but given the similar discussion of some of the other sources in the nucleus, the alignment with the SNR could be by chance.  This source could again be a bright peak in  thermal emission in the region rather than a single physical object. The source is not variable.

\item Source X227 (=S03-60) is another hard, luminous source, that when fitted with a power law yields a photon index $\Gamma = 1.46 \pm 0.05$ but is too luminous for the canonical BH low/hard state. Once again, we favour a physical model based on simple blackbody plus disk-blackbody; the disk parameters are $T_{\rm in} = 2.0^{+0.3}_{-0.1}$ keV and $r_{\rm in} \left(\cos \theta\right)^{0.5} \approx 7$ km, and the intrinsic $0.35$--$8$ keV luminosity is \EXPU{2}{38}{\LUM}. In two of the observations (one from 2011 and one from 2001), X227 was considerably fainter than in all the other epochs. The source is located in a southern spiral arm of M83, but is not associated with any particular star cluster or optical source.  Two optical SNRs, B12-133 and B12-136 from the Magellan list \citep{blair12} are within a few arcsec to the south, but X227 is most likely a luminous NS XRB.

\item Source X228 (=S03-062) is a bright, variable, and relatively hard source on the southern edge of the bright optical nuclear region.  It is adequately fit in terms of a simple power-law with $\Gamma = 1.67 \pm 0.10$, but also (more physically) by a disk-blackbody with $T_{\rm in} = 1.47 \pm 0.10$ keV, $r_{\rm in} \left(\cos \theta\right)^{0.5} \approx 8$ km, and intrinsic luminosity \EXPU{7.4}{37}{\LUM}. Statistically, the best fit is provided by a broken-power-law model, with a break at $E = 4.1 \pm 1.2$ keV, or by a bremsstrahlung model with $T_e = 8 \pm 2$ keV. Dopita et al. (2010)'s nuclear SNR \#13 is very close to X228 and could contribute to its X-ray emission (for example via a pulsar wind nebula). This confusion leaves the true nature of the object uncertain, although a NS identification seems plausible.

\item Source X229 technically misses out on our ``bright source'' classification, because it has only $\approx 1900$ counts. However, this is due to its relatively high intrinsic absorption (total $N_{\rm H} \approx 2.4 \times 10^{21}$ cm$^{-2}$). It has a soft thermal-plasma spectrum, with a temperature range $\sim 0.1$--$0.6$ keV and an intrinsic luminosity of about \EXPU{1.1}{38}{\LUM}. SNR \#12 in \cite{dopita10} is quite close to X229, in the crowded nuclear region.

\item Source X231 is a non-variable, soft source in the nuclear region, located just WSW of the optical/IR nucleus. It has a highly absorbed thermal-plasma spectrum, with obvious lines from Mg XI and Si XIII at 1.3 and 1.8 keV respectively. We fit the spectrum with a two-temperature thermal-plasma model ($T_1 = 0.26 \pm 0.03$ keV, $T_2 = 0.75 \pm 0.06$ keV), with intrinsic absorption of \EXPU{3.5\pm0.6}{21}{cm^{-2}}. The unabsorbed luminosity is \EXPU{\sim 7.1}{37}{\LUM}. The spectral properties of X231 are consistent with an SNR, and it might be identified with nuclear SNR \#11 in \cite{dopita10}; if this identification is correct, this would be one of the most luminous X-ray SNRs known. We cannot identify a compact radio source at this position, but this may simply be because an individual radio SNR would be swamped by the strong, extended radio emission (both synchrotron and free-free) from the starburst nuclear region.

\item Source X233 (=S03-063) corresponds to the optical/IR galactic nucleus(red circle in the panels of Figure 9) and is also a radio source. The nucleus is the site of a massive 10$^{8}$ year-old star cluster, whose dynamical mass is estimated to be about \EXPU{1.3}{7}{\MSOL} \citep{thatte00}. X233 varied by 15\% in 2010 and 2011, but was somewhat brighter in the 2000 and 2001 observations. Its X-ray spectrum is well fitted with a simple absorbed power-law, with N(H)$_{Gal}$ fixed at \EXPU{4}{20}{cm^{-2}}, intrinsic $N_{\rm H}$ of \EXPU{1.9\pm0.3}{21}{cm^{-2}}, photon index $\Gamma = 1.45\pm0.05$ and average $0.35$--$8$ keV luminosity of \EXPU{3.2\pm0.2}{38}{\LUM}. These parameters are at least consistent with a supermassive BH at a luminosity $\sim 10^{-7} L_{\rm Edd}$, although we cannot rule out a luminous stellar-mass XRB at the same location as the nuclear star cluster.

\item Source X234 (=S03-064) is located directly S of the optical/IR nucleus, at the SE end of the star forming bar in the nucleus.  It aligns with a bright blue patch of optical emission, indicating intense star formation in the region.  However, its X-ray properties show that it is not an SNR. The source varied in flux by a factor of about 5 during 2010--2011. Its average spectrum has a photon index $\Gamma = 2.65 \pm 0.07$ when fit with a simple power-law; this is typical of a BH in the steep-power-law state \citep{remillard06}. However, the source can also be fit with a power-law plus disk-blackbody model, consistent with other states of luminous XRBs. In the latter model, $\Gamma = 2.9 \pm 0.2$, $T_{\rm in} = 1.2 \pm 0.3$ keV, $r_{\rm in} \left(\cos \theta\right)^{0.5} \approx 9$ km, and the intrinsic $0.35$--$8$ keV luminosity is \EXPU{3.3}{38}{\LUM}. We cannot rule out a BH XRB, but we consider a NS XRB more likely, based on the small inner-disk radius.

\item Source X236 (=S03-065) is located in a dense stellar field due south of the nucleus, near several young star clusters and an SNR. It was brightest in December 2010, and faded as 2011 progressed. As for many other sources discussed in this Appendix, a simple power-law model gives a hard photon index $\Gamma = 1.49 \pm 0.10$, but a disk-blackbody plus blackbody model (or more complex two component models with Comptonized thermal emission) provides a more physical interpretation, consistent with a NS XRB. The disk parameters are $T_{\rm in} = 1.7^{+0.3}_{-0.2}$ keV and $r_{\rm in} \left(\cos \theta\right)^{0.5} \approx 6$ km; the intrinsic $0.35$--$8$ keV luminosity is \EXPU{7}{37}{\LUM}.

\item Source X248 (=S03-72) is a bright source that varies by about a factor of two over the \chandra observations.  It is located in a dense stellar field in the southern spiral arm, near a dust lane, but with no obvious optical counterpart.  The spectrum is fairly soft, and is well-modeled by a disk-blackbody, with $T_{\rm in} =0.85 \pm 0.02$ keV and $r_{\rm in} \left(\cos \theta\right)^{0.5} \approx 52$ km.  With an X-ray luminosity of \EXPU{3.5}{38}{\LUM}, this is almost certainly a BH system in the canonical high/soft state.  

\item Source X251 (=S03-73) is a highly variable source located in the inner bar of M83, northeast of the nuclear region.  The source flux sometimes changes by more than an order of magnitude on time-scales of less than an hour.  A faint blue clump of optical emission on the edge of a dust lane appears to align with the X-ray source, and a cluster of young stars is directly adjacent to the NW.  The spectrum is slightly curved, and is not well fitted by either a simple power-law or a simple disk-blackbody model. It is, instead, well fitted by a broken power-law (breaking at $E = 3.3^{+0.5}_{-0.4}$ keV) or by a blackbody plus disk-blackbody model; in the latter case, the disk parameters are $T_{\rm in} = 1.40^{+0.07}_{-0.10}$ keV and $r_{\rm in} \left(\cos \theta\right)^{0.5} \approx 13$ km; the intrinsic $0.35$--$8$ keV luminosity is \EXPU{1.8}{38}{\LUM}, all consistent with a near-Eddington NS XRB. Occultation by a disk rim or rapid outflows may accounting for the intra-observational variability. This is another source that merits more detailed studies.

\item Source X252 (=S03-075) is located at the outer edge of the stellar disk almost due north of the nucleus, and has a spectrum that is well fitted by a power-law with photon index $\Gamma = 1.73 \pm 0.10$, high intrinsic absorption $N_{\rm H} =$ \EXPU{3.8\pm0.5}{21}{cm^{-2}} and intrinsic $0.35$--$8$ keV luminosity \EXPU{1.6}{38}{\LUM}. Two-component optically-thick thermal or thermal Comptonization models provide similarly good fits, with a variety of parameters often not well constrained (several local minima in the Cash statistics) and inconsistent from model to model. A Comptonized disk model ({\tt diskir}) works for a ranget of disk temperatures $T_{\rm in} \sim 0.2$--$0.8$ keV and $r_{\rm in} \left(\cos \theta\right)^{0.5} \sim 25$--$225$ km. The main reason why it is hard to constrain the spectral model and parameters is that most of the emission below 1 keV is wiped out by the high absorbing column density.  The identification is therefore very uncertain.

\item Source X258 (=S03-078) is located in a southern interarm region, but with a very young stellar population $\sim$12\arcsec\ to the S and SE.  Modestly variable ($\pm$30\%), its spectrum is not well fitted either in terms of a simple power-law or a disk-blackbody.  Disk-blackbody plus simple blackbody fits approximate the shape of the spectrum better, and yield an X-ray luminosity of \EXPU{1.7}{38}{\LUM}, but the implied radius of the NS, $\approx 18$ km,  and location of the inner edge of the disk, $r_{\rm in} \left(\cos \theta\right)^{0.5} \approx 170$ km, are both larger than expected.  If it is a NS XRB, it may have an extended photosphere and a truncated inner disk. We conclude that we have not yet found a satisfying spectral model for this source. 

\item Source X281 (=S03-085) is a persistent, somewhat variable, source with a hard X-ray spectrum (power-law phton index $\Gamma = 1.38 \pm 0.07$) but an intrinsic X-ray luminosity (\EXPU{1.2}{38}{\LUM}), too high for the canonical BH low/hard state. It is located in an inner spiral arm NE of the nucleus in a general region of very active star formation but with no specific optical counterpart.  A cluster of optical SNRs, including the remnant of SN1957D, are adjacent within 10--20\arcsec\ to the S and SE. Double-component thermal/thermal Comptonization models ({\it e.g.}, disk-blackbody plus blackbody) approximate the spectrum fairly well,  although the implied NS radius of $\approx 20$ km is a bit larger than expected.  Nevertheless, the object is almost certainly a NS system.

\item Source X284 (=S03-086) is a persistent source with a relatively soft X-ray spectrum located in the outer region of of the spiral arm SE of the nucleus.  It was considerably (50\%) brighter in 2000 and 2001 and  2012 December, than in the other observations. The spectrum of the source is well-fitted in terms of a pure disk-blackbody spectrum with a $0.35$--$8$ keV luminosity of \EXPU{3}{38}{\LUM}. The best-fit parameters are $T_{\rm in} = 0.80 \pm 0.02$ keV and $r_{\rm in} \left(\cos \theta\right)^{0.5} \approx 56$ km, consistent with a BH XRB in the high/soft state. Interestingly, the object lies on the edge of a young stellar association, surrounded by a large ($\approx 150$ pc in radius) ionized bubble or ring-like feature, as seen in our Magellan \HA\ and [O III] images.  Two optical SNRs are also directly adjacent (within a few arcsec), B12-154 and B12-158 from the Magellan SNR list \citep{blair12}.

\item Source X286 (=S03-088) is another source with a soft X-ray spectrum that is well-modeled with an absorbed disk-blackbody, with an average luminosity \EXPU{4.6}{38}{\LUM}. It varied by $\pm$15\% in 2010 and 2011, but appears to have been brighter in 2001. The best-fit parameters are $T_{\rm in} = 0.96 \pm 0.02$ keV and $r_{\rm in} \left(\cos \theta\right)^{0.5} \approx 47$ km: almost certainly another BH XRB in the high/soft state. It is located in bright interarm region $\sim$1\arcmin NE of the nucleus, and is less than 1\arcmin\ north of the ULX source X299.  

\item Source X299 is the ULX that appeared in our new observations of M83, which have discussed previously in detail \citep{soria12}.  It is located in an interarm region 1\arcmin\ ESE of the nucleus, and its spectrum is consistent with that expected from binary system containing a black hole with mass between 10 and 40 \MSOL\  and a low-mass companion.  We identified a V$\simeq$24 blue stellar counterpart to this source that was not present prior to the turn-on of the source, indicating X-ray reprocessing is probably the likely source of the blue light.  A recent XMM-Newton observation from 2013 August (about 3 years after the likely start of the outburst) indicates that the source is still active, with an intrinsic X-ray luminosity $\approx 2 \times 10^{39}$ erg s$^{-1}$, a factor of 2 lower than the average luminosity in the 2010--2011 \chandra\ observations.

\item Source X321 (=S03-104) is located in the inner spiral arm, NE of the nucleus, directly adjacent to a region of massive star formation and \HA\ emission.  The X-ray position is less than $0\farcs5$ from the SNR B12-179, but the X-ray properties of X321 are not compatible with an SNR interpretation.  The source flux varies by a factor of four over the {\it Chandra} observations.  The average spectrum is featureless and is well fitted by a disk-blackbody model, with  $T_{\rm in} = 1.40 \pm 0.05$ keV, $r_{\rm in} \left(\cos \theta\right)^{0.5} \approx 27$ km and intrinsic $0.35$--$8$ keV luminosity of \EXPU{6.5}{38}{\LUM}. Such parameters are consistent with a stellar-mass BH near its Eddington limit; Galactic BH transients with similarly high disk temperatures and relative low apparent radii are sometimes said to be in the ``apparently standard regime'' \citep{kubota04}. The source is quite absorbed, with an intrinsic $N_{\rm H} =$ \EXPU{8.7\pm0.4}{21}{cm^{-2}}. 

\item Source X386 (=S03-121) is located on the fringes of the stellar disk, NE of the nucleus and beyond the bright spiral arms; no specific optical counterpart is present. It is a hard X-ray source, with a power-law photon index $\Gamma = 1.42 \pm 0.06$. As in many other cases, we can more physically model the spectrum with a disk-blackbody plus blackbody component ($T_{\rm in} = 1.93^{+0.17}_{-0.08}$ keV, $r_{\rm in} \left(\cos \theta\right)^{0.5} \approx 8$ km), but broken power-law or Comptonized disk models are also acceptable. Its average $0.35$--$8$ keV luminosity of \EXPU{2.2}{38}{\LUM} is still consistent with a near-Eddington NS XRB. The source flux decreased through the first half of 2011, brightening by a factor 2 in the second semester. 

\item Source X403 is a very bright, variable X-ray source at the edge of the stellar disk. Its spectrum is well modeled by a disk-blackbody with $T_{\rm in} = 1.10 \pm 0.17$ keV, $r_{\rm in} \left(\cos \theta\right)^{0.5} \approx 32$ km,  plus a steep power-law with $\Gamma = 2.5^{+0.4}_{-0.3}$. Its intrinsic luminosity is \EXPU{1.2}{39}{\LUM}. These values are typical of a stellar-mass BH at the top of its high/soft state or apparently standard regime, near the Eddington limit. This source was originally classified as a ULX with {\it ROSAT} \citep{immler99}, when it was associated with a very bright optical source, suggested to be a compact HII region or a globular cluster. In the latter scenario, this would have been the first ULX associated with a globular cluster. However, by comparing our \chandra\ and Magellan images, we find that X403 is about 3\arcsec\ from the optical source. Moreover, the optical source appears to be a background galaxy rather than a globular cluster, with a faint \HA\ extension toward the X-ray source position. We conclude that X403 is probably a BH XRB in M83, with a low-mass donor star. This object was not discussed by \cite{soria03} because they only analyzed data from the S3 chip, which did not cover this region.
\end{itemize}
\clearpage

\pagestyle{empty} 

\begin{deluxetable}{rcr}
\tablecaption{ACIS-S Observations of M83 }
\tablehead{\colhead{Obs.~ID} & 
 \colhead{Date} & 
 \colhead{Exposure} 
\\
\colhead{~} & 
 \colhead{~} & 
 \colhead{(s)} 
}
\tabletypesize{\scriptsize}
\tablewidth{0pt}\startdata
793$^a$ &  2000-04-29 &  50981 \\ 
2064$^a$ &  2001-09-04 &  9842 \\ 
12995 &  2010-12-23 &  59291 \\ 
13202 &  2010-12-25 &  98780 \\ 
12993 &  2011-03-15 &  49398 \\ 
13241 &  2011-03-18 &  78963 \\ 
12994 &  2011-03-23 &  150058 \\ 
12996 &  2011-03-29 &  53044 \\ 
13248 &  2011-04-03 &  54329 \\ 
14332 &  2011-08-29 &  52381 \\ 
12992 &  2011-09-05 &  66286 \\ 
14342 &  2011-12-28 &  67103 \\ 
\tablenotetext{a}{ Archival observation of M83}
\enddata 
\label{table_obs}
\end{deluxetable}


\begin{deluxetable}{cccccl}
\rotate\tablecaption{ATCA Radio Sources }
\tablehead{\colhead{Object} & 
 \colhead{RA} & 
 \colhead{Dec.} & 
 \colhead{S$_{5~GHz}$} & 
 \colhead{S$_{9~GHz}$} & 
 \colhead{Notes$^a$} 
\\
\colhead{~} & 
 \colhead{(J2000)} & 
 \colhead{(J2000)} & 
 \colhead{(mJy)} & 
 \colhead{(mJy)} & 
 \colhead{~} 
}
\tabletypesize{\scriptsize}
\tablewidth{0pt}\startdata
A001 &  13:36:40.38 &  -29:51:21.7 &  0.12$\pm$0.03 &  ~ &  ~ \\ 
A002 &  13:36:40.91 &  -29:51:17.9 &  0.07$\pm$0.03 &  ~ &  X019;~B12-003; \\ 
A003 &  13:36:40.99 &  -29:51:57.5 &  0.03$\pm$0.02 &  ~ &  X020:; \\ 
A004 &  13:36:41.63 &  -29:52:16.8 &  0.06$\pm$0.02 &  ~ &  B12-004:; \\ 
A005 &  13:36:42.22 &  -29:52:31.5 &  0.05$\pm$0.03 &  ~ &  ~ \\ 
A006 &  13:36:43.40 &  -29:52:23.8 &  0.16$\pm$0.03 &  ~ &  ~ \\ 
A007 &  13:36:44.16 &  -29:50:35.0 &  0.09$\pm$0.03 &  ~ &  ~ \\ 
A008 &  13:36:45.24 &  -29:52:45.3 &  0.10$\pm$0.03 &  ~ &  ~ \\
\enddata 
\label{table_radio}
\end{deluxetable}


\begin{deluxetable}{lrrcrccc}
\rotate\tablecaption{M83 X-ray Point Sources }
\tablehead{\colhead{Object} & 
 \colhead{RA} & 
 \colhead{Dec.} & 
 \colhead{Pos.~Err.$^a$} & 
 \colhead{Exp.} & 
 \colhead{Rate(0.35-8.0~keV)} & 
 \colhead{(M-S)/T$^b$} & 
 \colhead{(H-M)/T$^c$} 
\\
\colhead{~} & 
 \colhead{(J2000)} & 
 \colhead{(J2000)} & 
 \colhead{(arcsec)} & 
 \colhead{(ksec)} & 
 \colhead{(10$^{-3}$s$^{-1}$)} & 
 \colhead{~} & 
 \colhead{~} 
}
\tabletypesize{\scriptsize}
\tablewidth{0pt}\startdata
X001 &  13:36:13.66 &  -29:59:38.0 &  0.46 &  514.2 &  ~0.963$\pm$0.064 &  -0.50$\pm$~0.08 &  -0.12$\pm$~0.05 \\ 
X002 &  13:36:15.84 &  -29:58:00.1 &  0.30 &  395.6 &  ~3.045$\pm$0.107 &  -0.35$\pm$~0.04 &  -0.10$\pm$~0.03 \\ 
X003 &  13:36:26.77 &  -29:55:36.1 &  0.29 &  514.2 &  ~0.788$\pm$0.050 &  -0.33$\pm$~0.08 &  -0.09$\pm$~0.05 \\ 
X004 &  13:36:28.61 &  -29:57:20.2 &  0.49 &  514.2 &  ~0.252$\pm$0.036 &  -0.15$\pm$~0.16 &  -0.13$\pm$~0.10 \\ 
X005 &  13:36:28.97 &  -29:55:39.3 &  0.18 &  514.2 &  ~1.805$\pm$0.066 &  -0.20$\pm$~0.04 &  -0.09$\pm$~0.03 \\ 
X006 &  13:36:31.72 &  -29:57:09.2 &  0.34 &  514.2 &  ~0.542$\pm$0.044 &  -0.32$\pm$~0.09 &  -0.05$\pm$~0.06 \\ 
X007 &  13:36:33.65 &  -30:00:18.6 &  0.35 &  739.4 &  ~0.528$\pm$0.040 &  ~0.20$\pm$~0.10 &  -0.10$\pm$~0.06 \\ 
X008 &  13:36:35.24 &  -29:52:03.2 &  0.40 &  436.8 &  ~0.092$\pm$0.029 &  ~0.09$\pm$~0.30 &  -0.17$\pm$~0.26 \\
\enddata 
\label{table_xray_cat}
\end{deluxetable}


\begin{deluxetable}{lrrrrrrr}
\rotate\tablecaption{M83 X-ray Point Source Fluxes \& Luminosities }
\tablehead{\colhead{Object} & 
 \colhead{F(0.35-8~keV)$^a$} & 
 \colhead{F(0.35-1.1~keV)$^a$} & 
 \colhead{F(1.1-2.6~keV)$^a$} & 
 \colhead{F(2.6-8~keV)$^a$} & 
 \colhead{F(0.5-2~keV)$^a$} & 
 \colhead{F(2-8~keV)$^a$} & 
 \colhead{L$_X$(0.35-8~keV)$^b$} 
}
\tabletypesize{\scriptsize}
\tablewidth{0pt}\startdata
X001 &  11.100~$\pm$~0.715 &  8.910~$\pm$~0.838 &  2.620~$\pm$~0.202 &  1.100~$\pm$~0.317 &  6.530~$\pm$~0.412 &  1.900~$\pm$~0.350 &  ~63.34~$\pm$~~~4.05~ \\ 
X002 &  20.700~$\pm$~0.713 &  12.400~$\pm$~0.763 &  5.130~$\pm$~0.222 &  3.010~$\pm$~0.300 &  11.700~$\pm$~0.436 &  4.270~$\pm$~0.333 &  117.81~$\pm$~~~4.05~ \\ 
X003 &  5.100~$\pm$~0.307 &  2.840~$\pm$~0.338 &  1.240~$\pm$~0.093 &  0.796~$\pm$~0.128 &  2.820~$\pm$~0.186 &  1.090~$\pm$~0.141 &  ~28.88~$\pm$~~~1.75~ \\ 
X004 &  1.830~$\pm$~0.250 &  0.729~$\pm$~0.227 &  0.505~$\pm$~0.073 &  0.304~$\pm$~0.123 &  0.849~$\pm$~0.127 &  0.515~$\pm$~0.137 &  ~10.39~$\pm$~~~1.42~ \\ 
X005 &  11.100~$\pm$~0.395 &  4.760~$\pm$~0.392 &  2.850~$\pm$~0.130 &  1.980~$\pm$~0.156 &  5.850~$\pm$~0.248 &  2.720~$\pm$~0.177 &  ~62.83~$\pm$~~~2.24~ \\ 
X006 &  3.310~$\pm$~0.257 &  1.790~$\pm$~0.263 &  0.765~$\pm$~0.076 &  0.614~$\pm$~0.118 &  1.710~$\pm$~0.147 &  0.855~$\pm$~0.131 &  ~18.80~$\pm$~~~1.46~ \\ 
X007 &  4.580~$\pm$~0.338 &  0.699~$\pm$~0.271 &  1.290~$\pm$~0.102 &  0.985~$\pm$~0.160 &  2.160~$\pm$~0.179 &  1.370~$\pm$~0.177 &  ~26.09~$\pm$~~~1.92~ \\ 
X008 &  0.507~$\pm$~0.144 &  0.126~$\pm$~0.105 &  0.162~$\pm$~0.045 &  0.098~$\pm$~0.087 &  0.258~$\pm$~0.073 &  0.137~$\pm$~0.091 &  ~~2.86~$\pm$~~~0.82~ \\
\enddata 
\label{table_fluxes}
\end{deluxetable}


\begin{deluxetable}{llllccll}
\rotate\tablecaption{M83 X-ray Point Source IDs and Comments$^{a}$ }
\tablehead{\colhead{Object} & 
 \colhead{Other~X-ray$^{b}$} & 
 \colhead{Radio~ID$^{c}$} & 
 \colhead{SNR/SNe$^{d}$} & 
 \colhead{P/T$^{e}$} & 
 \colhead{Var.$^{f}$} & 
 \colhead{Type$^{g}$} & 
 \colhead{Notes$^{h}$} 
}
\tabletypesize{\scriptsize}
\tablewidth{0pt}\startdata
X001 &  ~ &  ~ &  ~ &  ~ &  ~ &  AGN &  ~ \\ 
X002 &  ~ &  ~ &  ~ &  P &  ~ &  AGN &  Bright~power~law~source~outside~D25~contour \\ 
X003 &  D13-030: &  ~ &  ~ &  ~ &  ~ &  AGN? &  ~ \\ 
X005 &  D13-033 &  ~ &  ~ &  P &  ~ &  AGN &  Bright~power~law~source~outside~D25~countour;~blueish~optical \\ 
~ &  ~ &  ~ &  ~ &  ~ &  ~ &  ~ &  point~source \\ 
X006 &  D13-040: &  ~ &  ~ &  ~ &  ~ &  AGN? &  ~ \\ 
X007 &  D13-041: &  ~ &  ~ &  ~ &  ~ &  ~ &  ~ \\ 
X009 &  D13-046: &  ~ &  ~ &  ~ &  AN &  ~ &  ~ \\
\enddata 
\label{table_ids}
\end{deluxetable}


\begin{deluxetable}{rrrrr}
\tablecaption{Census of Identifications$^a$ }
\tablehead{\colhead{Object~List} & 
 \colhead{Comparison~List} & 
 \colhead{Number~in~Comparison~List} & 
 \colhead{Number~IDs} & 
 \colhead{Notes} 
}
\tabletypesize{\scriptsize}
\tablewidth{0pt}\startdata
ATCA &  VLA$^b$ &  55 &  47 &  ~ \\ 
ATCA &  X-ray &  458 &  46 &  ~ \\ 
ATCA &  [SII]-SNRs$^c$ &  225 &  21 &  ~ \\ 
X-ray &  \cite{soria03} &  127 &  124 &  ~ \\ 
X-ray &  XMM$^d$ &  46 &  40 &  Within~D25 \\ 
X-ray &  [SII]-SNRs$^c$ &  225 &  73 &  ~ \\ 
SNE &  ATCA &  5 &  3 &  ~ \\ 
SNRs &  ATCA &  109 &  28 &  ~ \\ 
SNe &  X-ray &  5 &  1 &  ~ \\ 
SNRs &  X-ray &  458 &  87 &  ~ \\ 
Stars &  X-ray &  458 &  6 &  ~ \\ 
AGN/Background &  X-ray &  458 &  56 &  ~ \\ 
\tablenotetext{a}{ See text for details, as some sources are chance coincidences}
\tablenotetext{b}{ \cite{maddox06}}
\tablenotetext{c}{ \cite{blair12}}
\tablenotetext{d}{ \cite{ducci13}}
\enddata 
\label{table_id_summary}
\end{deluxetable}


\begin{deluxetable}{lcrrrrrr}
\tablecolumns{8}
\tabletypesize{\footnotesize}
\tablecaption{Source Counts in the Various Regions}
\tablewidth{0pt}
\tablehead{
\colhead{Region} &
\colhead{Definition} &
\colhead{Limit$^{a}$} &
\colhead{Sources$^{b}$} &
\colhead{AGN} &
\colhead{SNR} &
\colhead{Star} &
\colhead{XRB} }
\startdata
\cutinhead{0.35-8.0 keV}
Bulge                  & $r<0\farcm6$                         & $6.4$ &  53 &  0.8 &  7 & 0 & 45.2 \\
~~Nuclear Region$^{c}$              & $r<0\farcm3$                         &  6.4 &  38 &  0.2 &  6 & 0 &  31.8 \\
~~Outer Bulge          & $0\farcm3<r<0\farcm6$                &  6.4 &  15 &  0.7 &  1 & 0 &  13.3 \\
All Disk               & $r>0\farcm6 ~and~R_{proj}<6\farcm5$ & $8.0$ & 265 & 96.3 & 69 & 1 & 98.7 \\
Inner Disk             & $r>0\farcm6 ~and~R_{proj}<3\farcm5$ & $6.4$ & 185 & 34.0 & 60 & 0 & 91.0 \\
~~Arm                  &                                      & $6.4$ &  84 &  9.8 & 37 & 0 & 37.2 \\
~~Inter-Arm            &                                      & $6.4$ &  101 & 24.3 & 23 & 0 & 53.7 \\
Outer Disk             & $3\farcm5<R_{proj}<6\farcm5$         & $10.0$ &  89 & 63.0 & 11 & 1 & 14.0 \\
\enddata
\label{table_pops}
\end{deluxetable}


\begin{figure}
\plotone{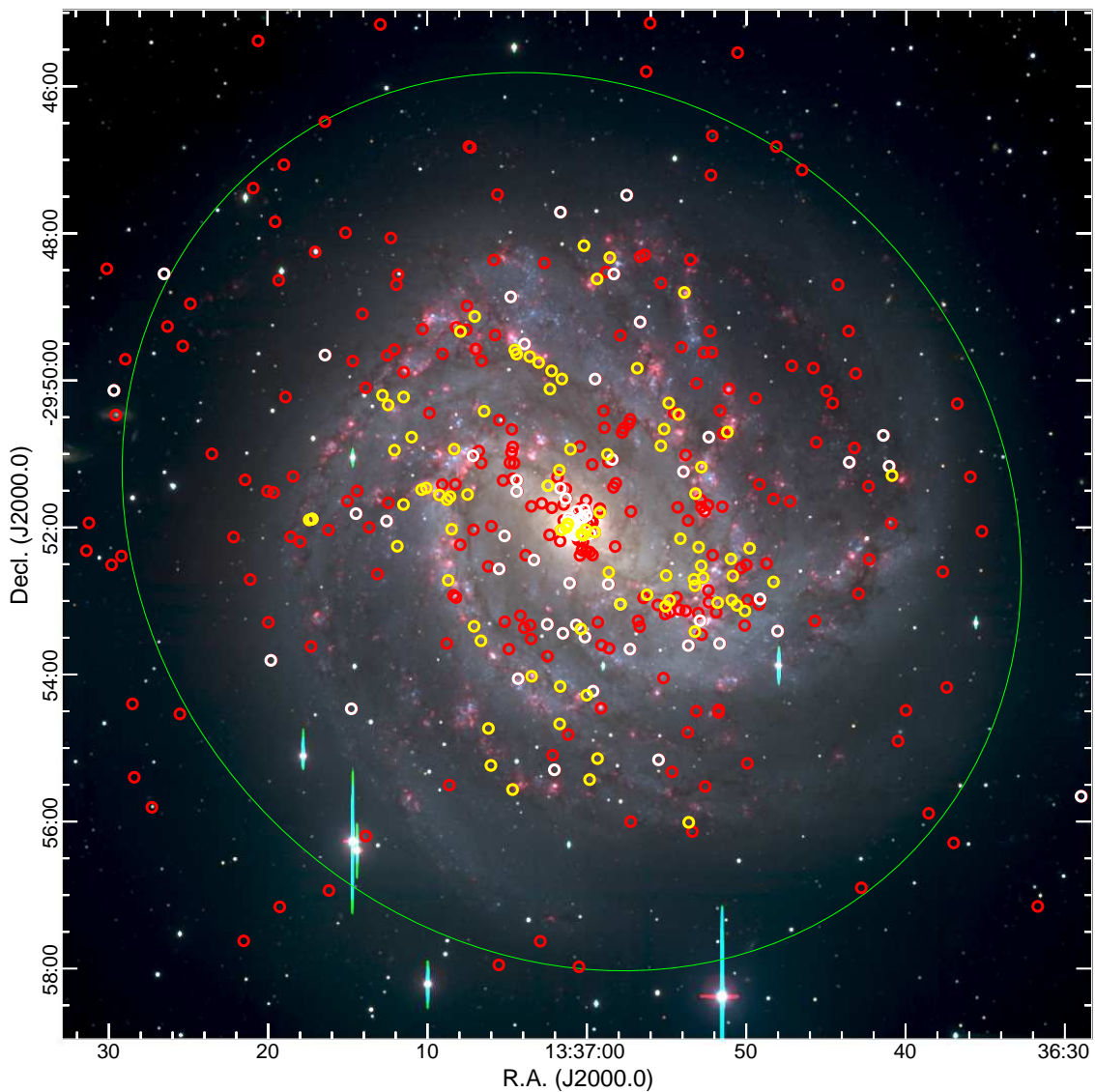}
\figcaption{X-ray sources detected in our \chandra\ survey of M83 are plotted on the same 3-color optical image as shown in Figure \ref{fig_radio}. The field is identical for Figures \ref{fig_xray}, \ref{fig_radio} and \ref{fig_point}. The D$_{25}$ ellipse is again shown in green. Sources identified as SNRs are identified in yellow, the brighter sources (those with $>500$ counts in our survey) are in white, and fainter sources in red. Most of the sources outside of this boundary are foreground or background objects, mainly AGN.
\label{fig_point}}
\end{figure}

\begin{figure}
\plotone{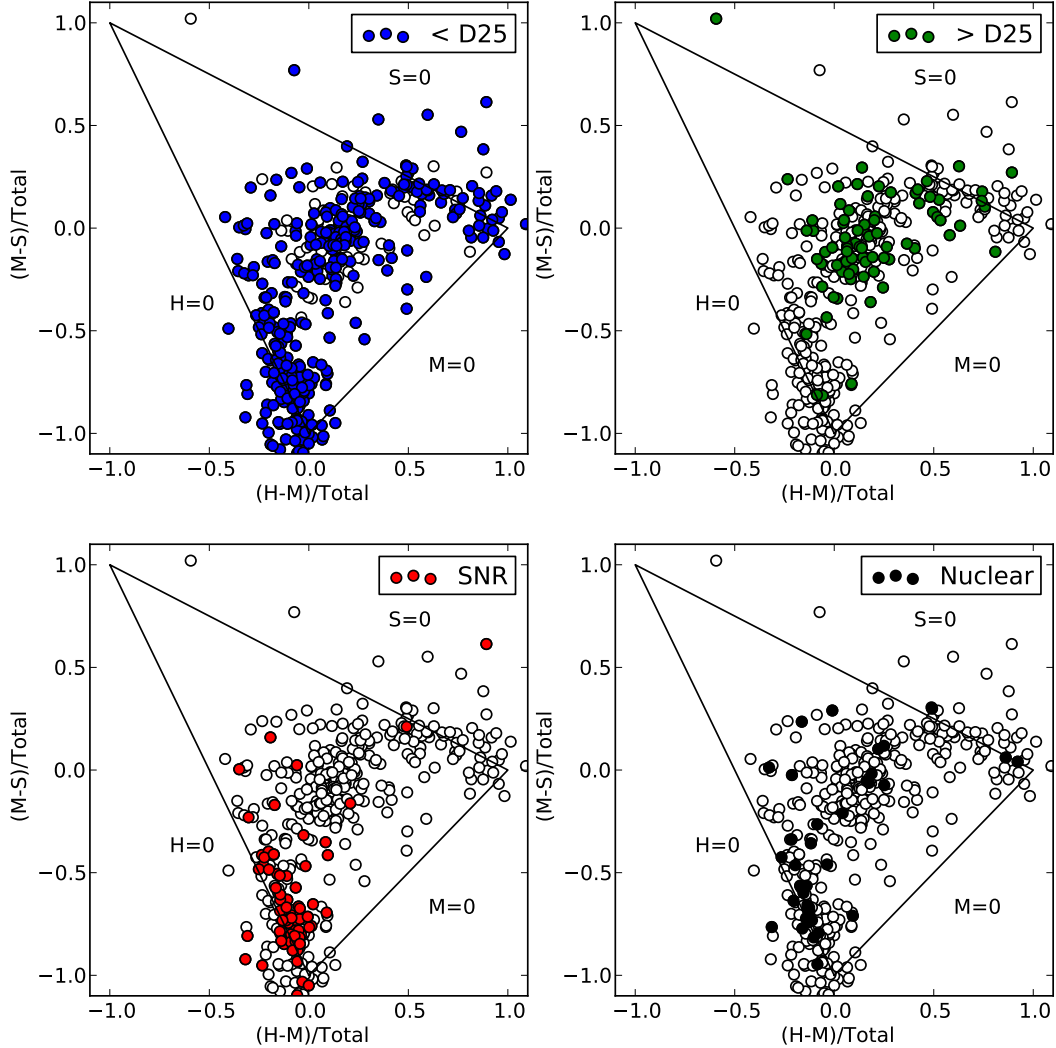}
\figcaption{A hardness-ratio plot of the sources detected in M83, with different subsets highlighted in the different panels. Hardness ratios of individual sources are listed in Table \ref{table_xray_cat}.  The soft (S), medium (M) and hard (H) bands correspond to energies of 0.35-1.1 keV, 1.1-2.6 keV and 2.6-8 keV, respectively.  As a result of counting statistics, some sources have ratios that fall outside of the triangular region where all sources should lie.  Sources shown in blue (upper left panel) denote ones within the D$_{25}$ ellipse, while those in green (upper right) denote ones outside D$_{25}$.  
Sources shown in red (lower left) coincide in position  with SNRs and SNR candidates reported by \citet{blair12}.  The black filled circles (lower right) denote sources in the nuclear region, within 0.5 kpc of M83's center.  Some of these nuclear-region sources  are relatively hard, typical of  X-ray binaries, while others are quite soft, consistent with what one might expect from SNRs.    \label{fig_hardness}
}
\end{figure}

\begin{figure}
\plotone{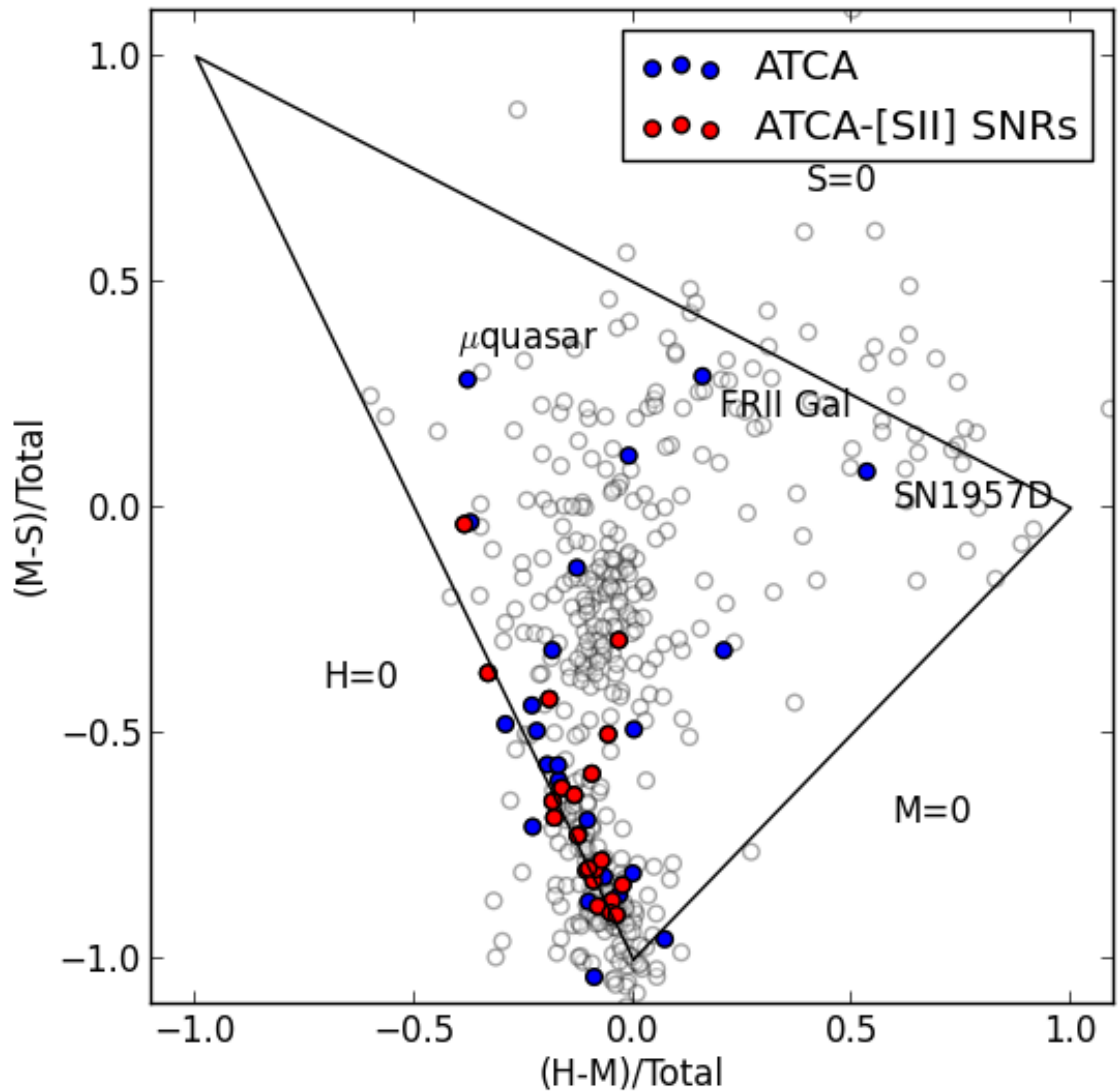}
\figcaption{
A hardness ratio plot of X-ray sources identified with ATCA sources.  Sources which are identified with optically identified SNRs are shown in red, and the remainder of the sources are plotted in blue. The entire sample is also shown for context.  The hardness ratios of several specific sources are labeled.
\label{fig_hardness_radio}
}
\end{figure}

\begin{figure}
\epsscale{0.9}
\plotone{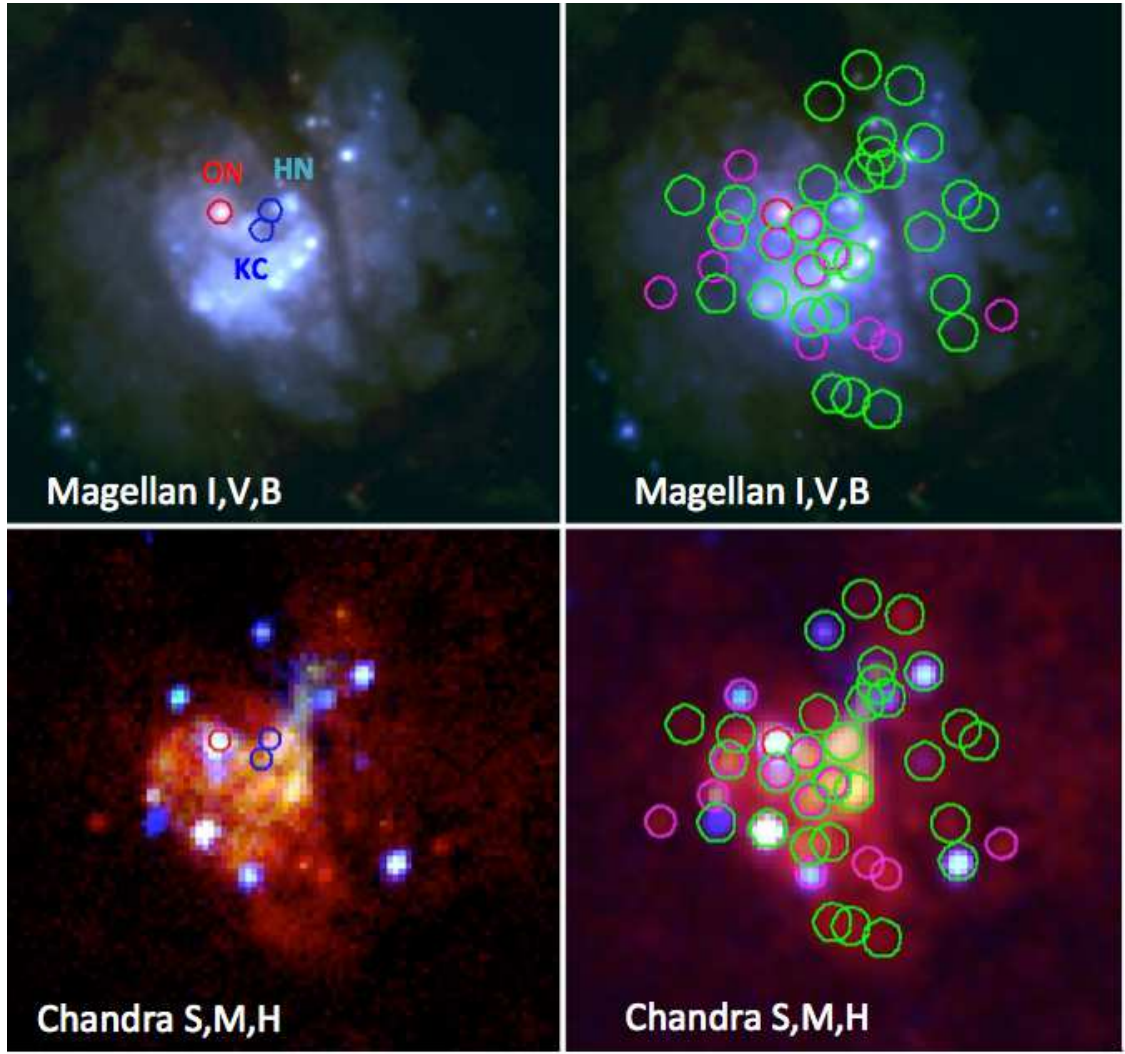}
\caption{A 35\arcsec\ region near the nucleus of M83 in optical and X-ray.  Top two panels show the Magellan broadband data, where red = I band, green = V band, and blue = B band.  The bottom two panels show the \chandra\ data in three-color format, with red = 0.35-1.2 keV (soft), green = 1.2-2.6 keV (medium), and blue = 2.6-8.0 keV (hard).  The left panel is at the full  \chandra\  resolution, while the right panel has had a small amount of smoothing applied. The three small regions shown in the left two panels represent the positions of the optical nucleus (red), the kinematic center determined by \cite[][ lower blue circle]{thatte00}, and the so-called ``hidden nucleus'' inferred by \cite{mast06}.  The two right panels show source positions as found by {\em AE}.  Green circles typically have no specific optical source but rather correspond to high star formation regions, or in a few cases actual young clusters.  Magenta circles align closely with SNRs identified by \citet{dopita10} in {\em HST}/WFC3 data, although in some cases the alignment may be fortuitous.   A red circle in the right two panels indicates source X233, which corresponds to the optical nucleus.  Some of the harder (blue and white) sources in both green and magenta circles may actually be X-ray binaries, but no optical identifications with point sources have been made.} 
\label{fig_nuc}

\end{figure}

\begin{figure}
\plotone{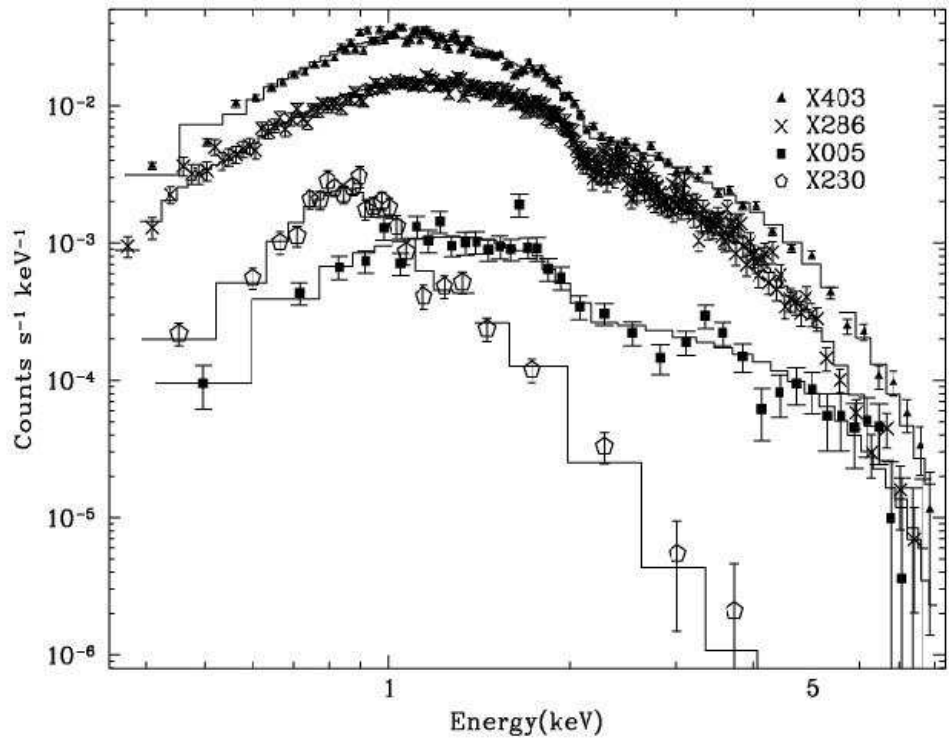}
\figcaption{
Examples of the different types of spectra observed in the bright and moderately bright sources.  X403 and X286 are examples of bright sources.  X403 had more than 12,000 counts and is modeled in terms of a disk-blackbody + power law spectrum, while X286 is just a disk-blackbody.  X005 and X230 are examples of moderately bright sources, both having of order 1000 counts total.  These two were selected to show the difference between a thermal and non-thermal spectrum with this number of source counts. \label{fig_bright_spectra}
}
\end{figure}

\begin{figure}
\plotone{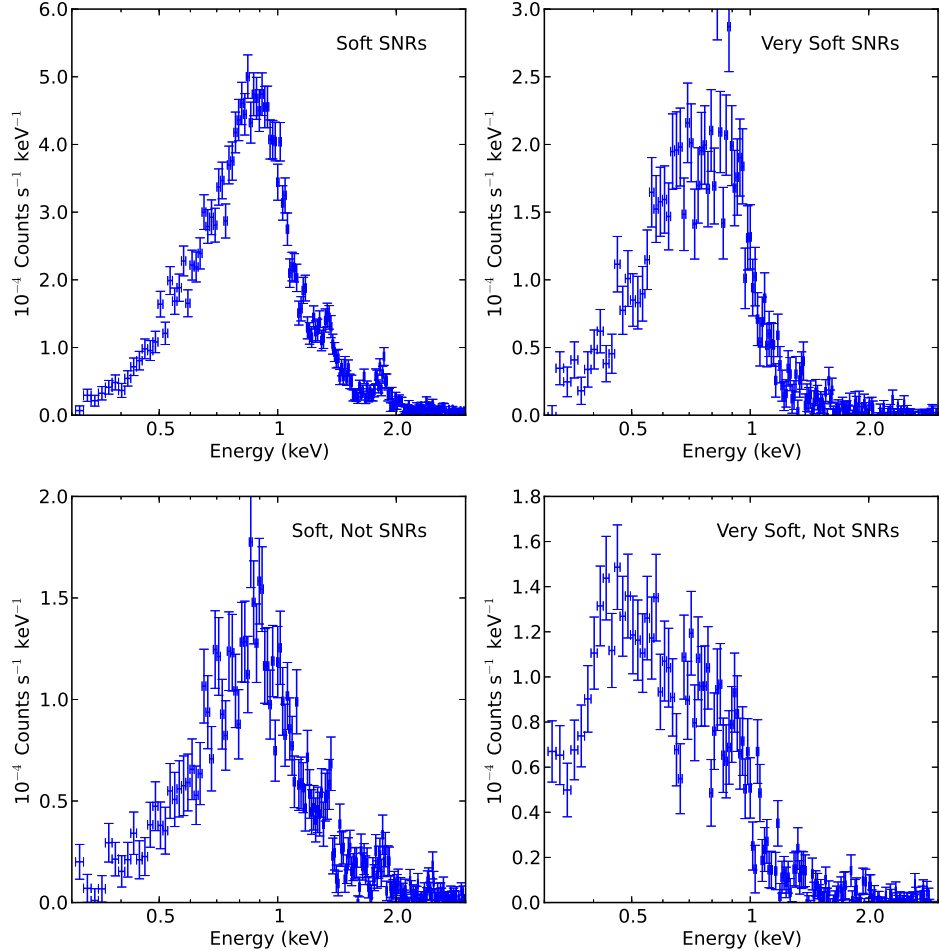}
\figcaption{
A comparison of the summed spectra of various groups of soft sources. The sources have been divided into four groups.  The upper left panel shows the summed spectra for objects identified as SNRs which have hardness ratios (M--S/T) between -0.88 and -0.5.   The lower left panels shows summed spectra for objects which are not identified as SNRs but have hardness ratios between -0.88 and -0.5.  The upper right and lower right panels show objects which have hardness ratios less than -0.88, split into objects that are identified as SNRs and objects that are not identified as SNRs.  The composite spectra of the ``Soft, Not SNRs'' objects and the ``Soft SNRs'' objects are very similar, showing the same spectral features, suggesting that some of the `Soft, Not SNRs'' objects are actually SNRs as well.  The composite spectrum of ``Very Soft, Not SNRs''  objects shows and an excess at 0.5 keV, not seen in the ``Very Soft SNRs'' group, which may indicate the some of these objects are true supersoft sources.   \label{fig_soft_spectra}
}
\end{figure}

\begin{figure}
\plotone{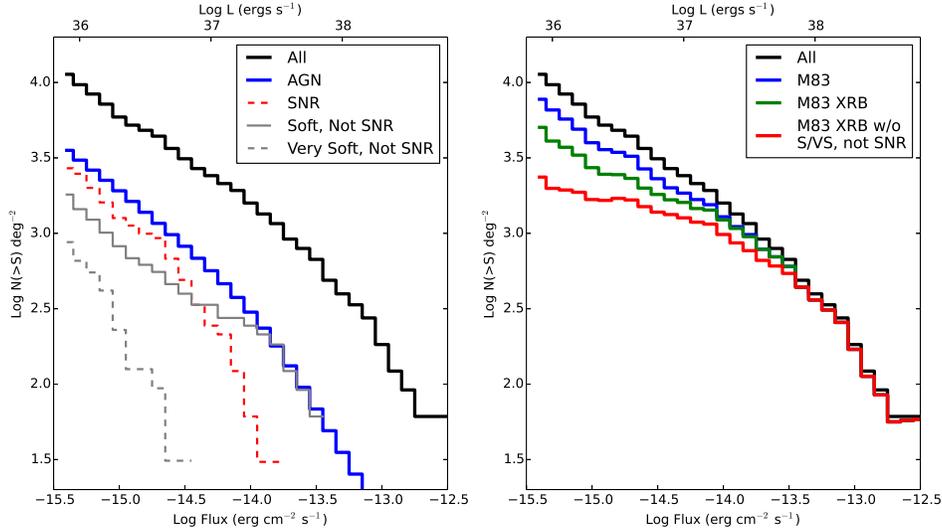}
\caption{
{\it Left: } The broad-band (0.35 - 8 keV) CLF of sources within the D$_{25}$ region of M83 before subtraction of foreground and background sources, and various subpopulations of the CLF,  namely AGN,  X-ray sources identified as  SNRs, other soft sources (``Soft, not SNRs''), and other very soft sources (``Very Soft, Not SNRs'').  The  contribution of AGN to the CLF  was estimated statistically, as discussed in Section \ref{sec_agn}.  Foreground stars are not an important contributor to the CLF.  The nature of the ``Soft, not SNRs'' and ``Very Soft, Not SNRs'' sources is discussed in Section \ref{sec_soft}.\label{fig_clf} {\it Right: } The broad-band CLF before subtraction, the M83 CLF obtained by  removing AGN and foreground stars, and the XRB CLF obtained by also removing SNRs from the sample, and finally the CLF, with background and foreground sources, SNRs, and other soft and very soft sources subtracted.  We suspect that  some, perhaps the majority, of the ``Soft, not SNRs'' and ``Very Soft, Not SNRs'' sources are SNRs. If so, this has a significant affect on the CLF for XRBs in M83.\label{fig_band_compare}
}
\end{figure}

\begin{figure}
\plotone{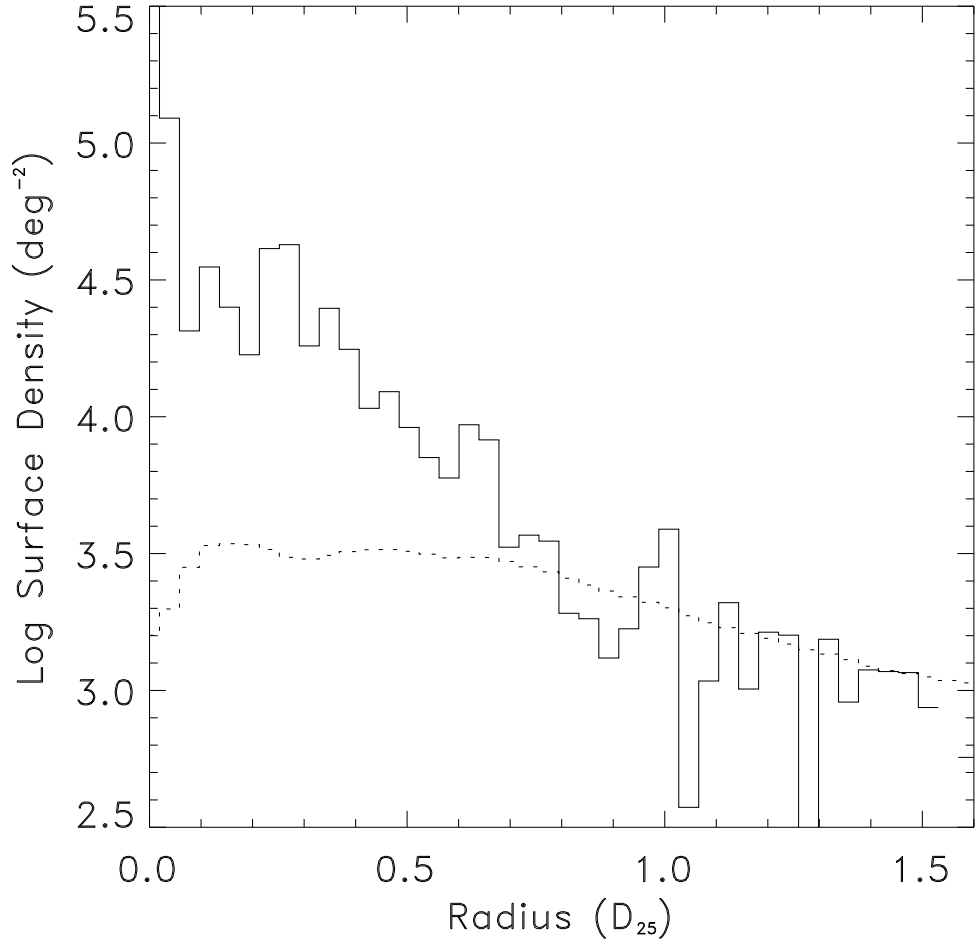}
\caption{
The surface density of all point sources with $\log{F}\ge -6.5\PHOTFLUX$ in the 0.35-8 keV band. The dotted line is the radial profile of the AGN calculated as described in the Section \ref{sec_agn}. The decline in the predicted density of AGN towards the center of the galaxy is due to the absorption by M83, while the decline towards the outer parts of the galaxy is due to the decrease in sensitivity of the observations. 
\label{fig_ps_rad_prof}}
\end{figure}

\clearpage

\begin{figure}
\plotone{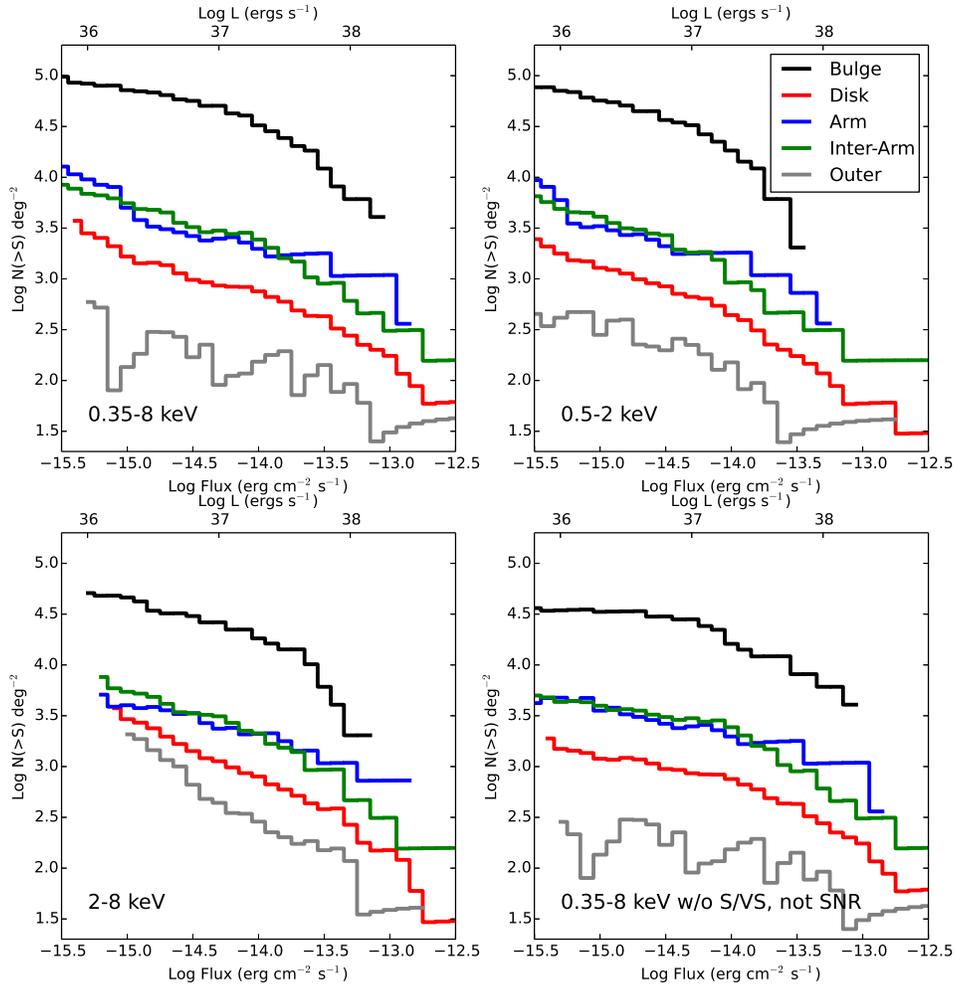}
\caption{XRB CLFs in the 0.35-8 keV (upper left), 0.5-2.0 keV (upper right) and 2.0-8 keV (lower left) bands for a variety of different regions in M83. The number of sources differ in the various bands because only sources with probability-of-no-source $<$ \EXPN{5}{-6}  in that band are included in the CLF for that band. In
all panels,  the black lines shows the CLF  for the bulge/nuclear region of the galaxy, while the red line
shows the CLF of the disk.   The   CLF for the arm and inter-arm regions and outer disk are  shown separately.  Source density decreases
with radius, which explains why the CLF for arm and inter-arm regions inner disk disk are always greater than that for the outer disk.  The various regions are shown graphically
in Figure \ref{fig_key}.
\label{fig_clf_reg}
The lower right hand panel shows the 0.35-8 keV CLF  with all soft sources removed, not just SNRs.  Removing the soft sources (``Soft, not SNRs'' and ``Very Soft, not SNRs'') flattens the CLF at lower luminosities, and makes all of the CLFs look more similar aside from an overall normalization.  The densities of sources in the arm and inter-arm regions are very similar, indicating that there is not a large concentration of XRBs in the spiral arms.
}
\end{figure}

\begin{figure}[]
\plotone{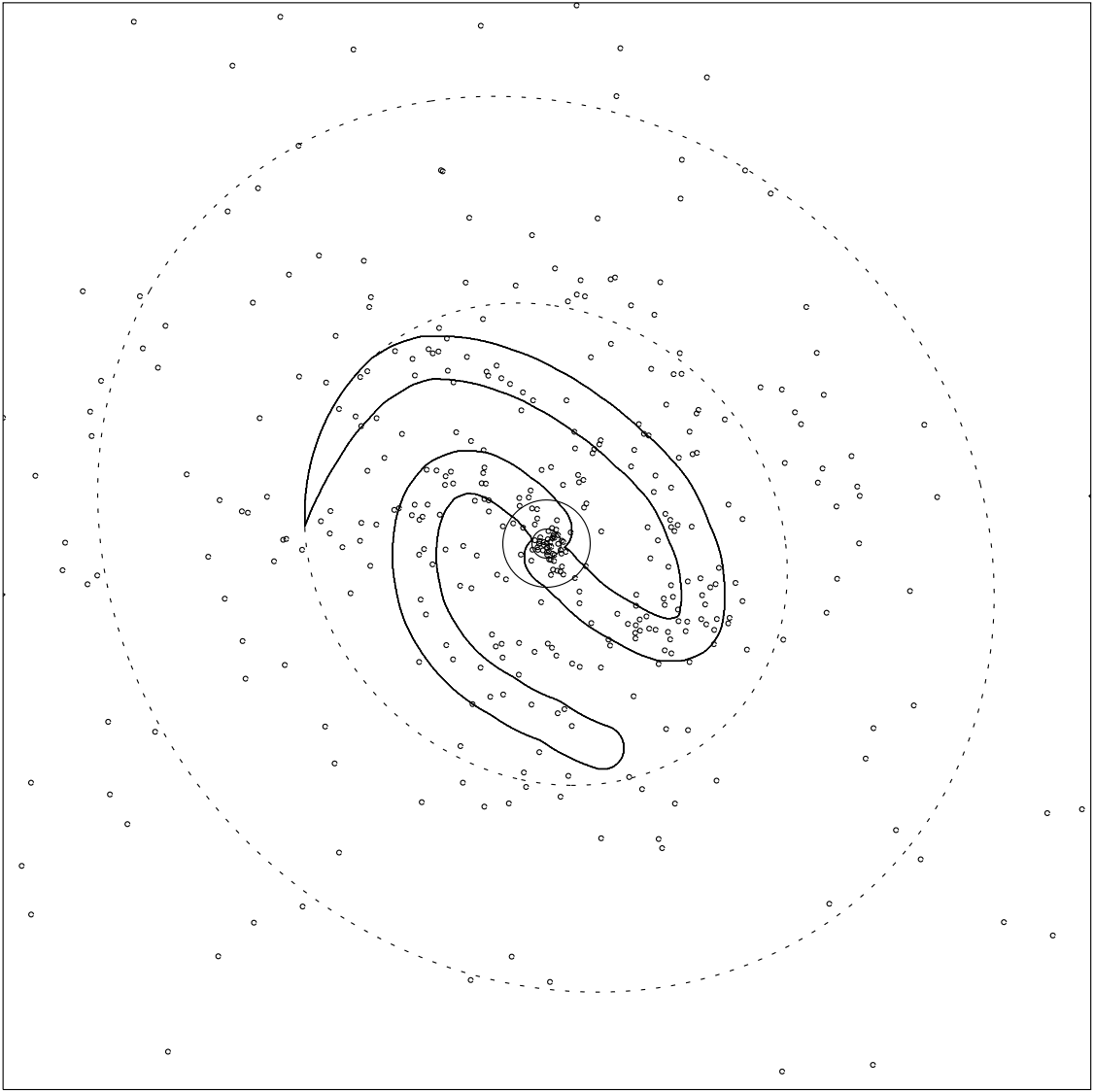}
\caption{
The regions over which CLFs were extracted.   The positions of the sources detected in the 0.35-8 keV band are also shown.  The inner ellipse at $R_{proj}=3\farcm5$ is the boundary between the inner and outer disk; the outer ellipse at  $R_{proj}=6\farcm5$ corresponds to the D$_{25}$ contour.   The inner circles have radii of $0\farcm3$ and $0\farcm6$ and define the nucleus and outer bulge regions.  
Although the arms extend beyond $R_{proj}=3\farcm5$, we have only used the region between  $0\farcm6$ and
 $R_{proj}=3\farcm5$ for the purpose of extracting the arm and inter-arm CLFs. \label{fig_key}}
\end{figure}

\begin{figure}[]
\plotone{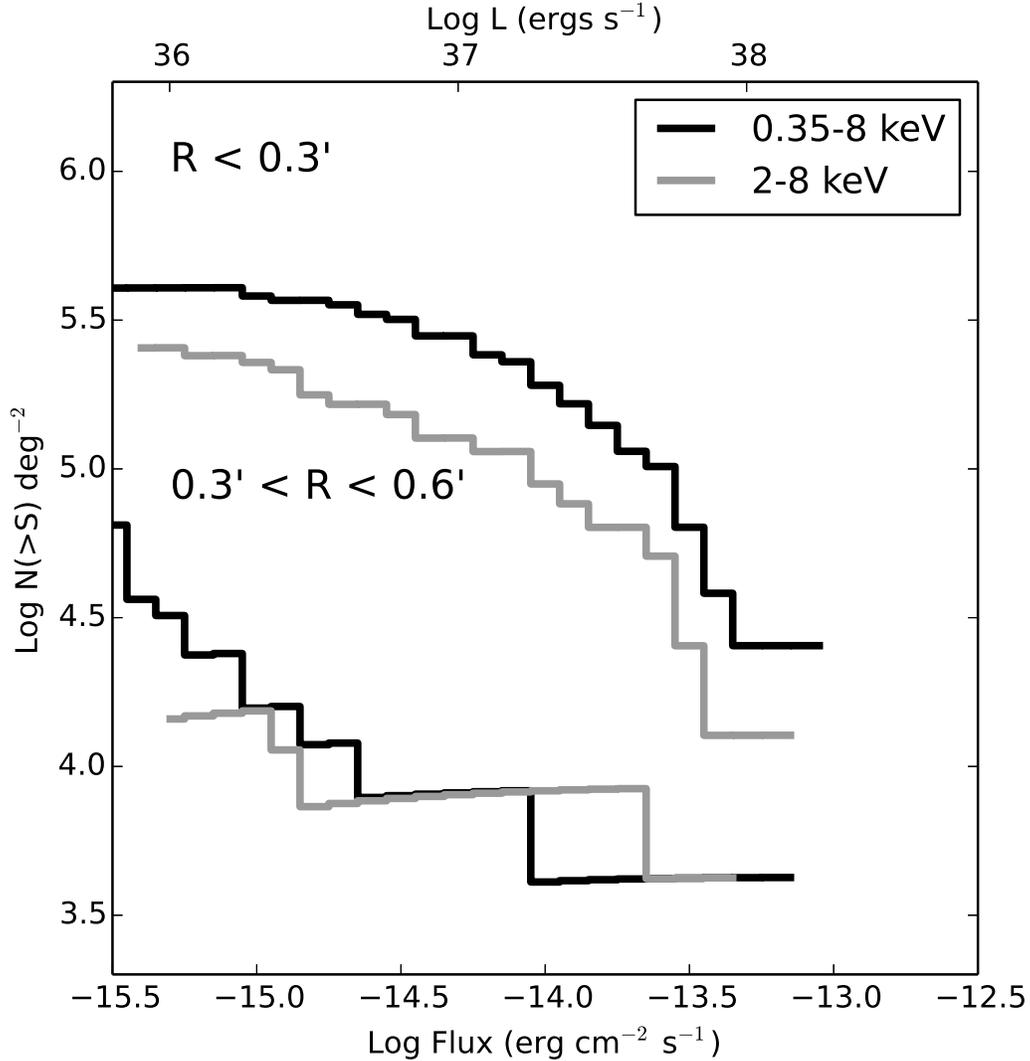}
\caption{The CLFs for the inner bulge or nuclear region ($R<0\farcm3$) and the outer bulge region ($0\farcm3<R<0\farcm6$). The heavier lines indicate the 0.35-8.0 keV band;  lighter lines,  the 2.0-8.0 keV band. The 2.0-8.0 keV band CLFs have been adjusted to the 0.35-8.0 keV band luminosities.  At 0.35-8 keV, the CLF for the outer bulge extends to lower luminosities  than that for the inner nucleus, due to the strong diffuse emission in the nucleus and to
 source crowding. \label{fig_bul_elf}
}
\end{figure}

\begin{figure}[]
\plotone{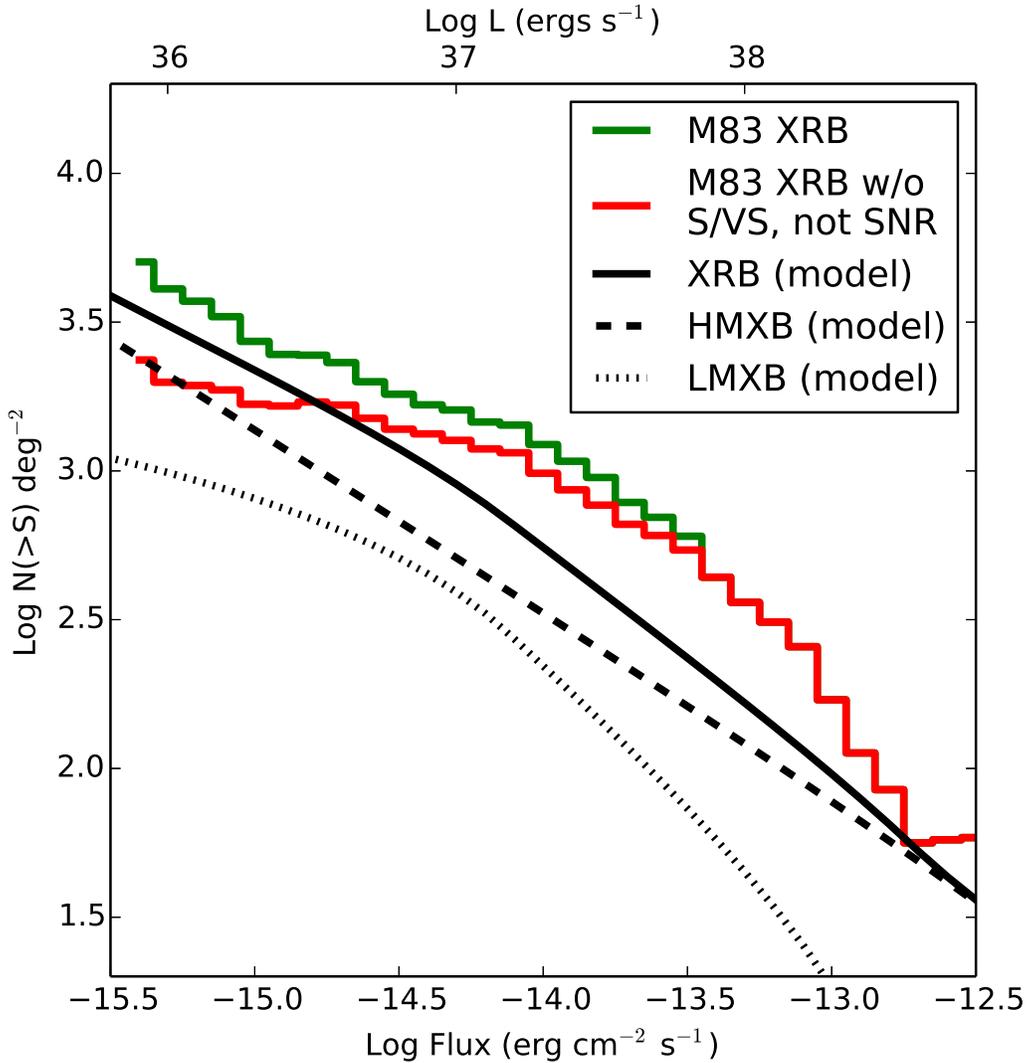}
\caption{
The 0.35-8.0 keV band CLF of the galaxy within the D$_{25}$ contour after the removal of the
AGN, foreground stars and SNRs (in green), and, in addition, the ``Soft, Not SNR'' and ``Very Soft, Not SNR'' sources (in red). The remaining lines are  the predicted CLF based on the model of  \cite{gilfanov04} for XRBs (the solid black line) and the contributions from HMXBs (the dashed black line) and LMXBs (the dotted black line).  The predicted CLF has been scaled from  the star-formation rate and stellar mass
of M83.  The predictions are not a very good match to the CLF observed in M83.  
\label{fig_gg}
}
\end{figure}


\begin{thebibliography}{}
  
\bibitem[Arnaud(1996)]{arnaud96} Arnaud, K.~A. 1996, in Astronomical Society of the Pacific Conference Series,
  Vol. 101, Astronomical Data Analysis Software and Systems V, ed.
  {G.~H.~Jacoby \& J.~Barnes}, 17

\bibitem[Barmby et al.(2006)]{barmby06} Barmby, P., et al.\ 2006, \apjl, 650, L45 

\bibitem[Barret(2001)]{barret01} Barret, D.\ 2001, Advances in Space Research, 28, 307 

\bibitem[Belczynski et al.(2008)]{belczynski08} Belczynski, K., Kalogera, V., Rasio, F.~A., Taam, R.~E., Zezas, A., Bulik, 
T., Maccarone, T.~J., \& Ivanova, N.\ 2008, \apjs, 174, 223 

\bibitem[Bell \& de Jong(2001)]{bell01} Bell, E.~F., \& de Jong, R.~S.\ 2001, \apj, 550, 212 

\bibitem[Bertelli et 
al.(1994)]{bertelli94} Bertelli, G., Bressan, A., Chiosi, C., Fagotto, F., \& Nasi, E.\ 1994, \aaps, 106, 275 


\bibitem[Bertelli et al.(2009)]{bertelli09} Bertelli, G., Nasi, E., Girardi, L., \& Marigo, P.\ 2009, \aap, 508, 355 

\bibitem[Binder et al.(2012)]{binder12} Binder, B., et al.\ 2012, \apj, 758, 15 


\bibitem[Blair \& Long(2004)]{blair04} Blair, W.~P., \& Long, K.~S.\ 2004, \apjs, 155, 101 

\bibitem[Blair et al.(2013)]{blair14} Blair, W.~P., et al.\ 2014, \apj, submitted

\bibitem[Blair et al.(2012)]{blair12} Blair, W.~P., Winkler, P.~F., \& Long, K.~S.\ 2012, \apjs, 203, 8 

\bibitem[Boissier et al.(2005)]{boissier05} Boissier, S., et al.\ 2005, \apjl, 619, L83 

\bibitem[Broos et al.(2010)]{broos10} Broos, P.~S., Townsley, L.~K., Feigelson, E.~D., Getman, K.~V., Bauer, F.~E.,  \& Garmire, G.~P.\ 2010, \apj, 714, 1582 



\bibitem[Burstein \& Heiles(1982)]{burstein82} Burstein, D., \& Heiles, C.\ 1982, \aj, 87, 1165 


\bibitem[Chandra et al.(2009)]{chandra09} Chandra, P., Dwarkadas, V.~V., Ray, A., Immler, S., 
\& Pooley, D.\ 2009, \apj, 699, 388 

\bibitem[Chevalier(1982)]{chevalier82} Chevalier, R.~A.\ 1982, \apj, 259, 302 

\bibitem[Corbelli(2003)]{corbelli03} Corbelli, E.\ 2003, \mnras, 342, 199 



\bibitem[Cowan \& Branch(1985)]{cowan85} Cowan, J.~J., \& Branch, D.\ 1985, \apj, 293, 400 

\bibitem[Cowan et al.(1994)]{cowan94} Cowan, J.~J., Roberts, D.~A., \& Branch, D.\ 1994, \apj, 434, 128 

\bibitem[Crawford et al.(1970)]{crawford70} Crawford, D.~F., Jauncey, D.~L., \& Murdoch, H.~S.\ 1970, \apj, 162, 405 

\bibitem[Dalcanton et al.(2009)]{dalcanton09} Dalcanton, J.~J., et al.\ 2009, \apjs, 183, 67 

\bibitem[Dickey \& Lockman(1990)]{dickey90} Dickey, J.~M., \& Lockman, F.~J.\ 1990, \araa, 28, 215 


\bibitem[Di Stefano \& Kong(2004)]{distefano04} Di Stefano, R., \& Kong, A.~K.~H.\ 2004, \apj, 609, 710 

\bibitem[Dopita et al.(2010)]{dopita10} Dopita, M.~A., et al.\ 2010, \apj, 710, 964 

\bibitem[Dottori et al. (2008)]{dottori08} Dottori, H., Diaz, R., \& Mast, D.\ 2008, \aj, 136, 2468

\bibitem[Ducci et al.(2013)]{ducci13} Ducci, L., Sasaki, M., Haberl, F., \& Pietsch, W.\ 2013, \aap, 553, A7 


\bibitem[Dwarkadas \& Gruszko(2012)]{dwarkadas12} Dwarkadas, V.~V., \& Gruszko, J.\ 2012, \mnras, 419, 1515 


\bibitem[Eck et al.(1998)]{eck98} Eck, C.~R., Roberts, D.~A., Cowan, J.~J., \& Branch, D.\ 1998, \apj, 508, 664 




\bibitem[Fabbiano(2006)]{fabbiano06} Fabbiano, G.\ 2006, \araa, 44, 323 

\bibitem[Flesch(2010)]{flesch10} Flesch, E.\ 2010, \pasa, 27, 283 

\bibitem[Fragos et al.(2008)]{fragos08} Fragos, T., et al.\ 2008, \apj, 683, 346 

\bibitem[Freedman et al.(2001)]{freedman01} Freedman, W.~L., et al.\ 2001, \apj, 553, 47 



\bibitem[Fridriksson et al.(2008)]{fridriksson08} Fridriksson, J.~K., Homan, J., Lewin, W.~H.~G., Kong, A.~K.~H., 
\& Pooley, D.\ 2008, \apjs, 177, 465 


\bibitem[Fruchter  \& Sosey et al.(2009)]{multidrizzle}Fruchter, A. \& Sosey, M. et~al. 2009, ``The MultiDrizzle Handbook'', version 3.0, (Baltimore, STScI)

\bibitem[\protect\citeauthoryear{{Fruscione} et~al.}{{Fruscione}  et~al.}{2006}]{fruscione06}
{Fruscione}, A., et~al. 2006, in Society of Photo-Optical Instrumentation
Engineers (SPIE) Conference Series, Vol. 6270, Society of Photo-Optical
Instrumentation Engineers (SPIE) Conference Series

\bibitem[Gilfanov(2004)]{gilfanov04} Gilfanov, M.\ 2004, \mnras, 349, 146 

  
\bibitem[Gradari et al.(2011)]{gradari11} Gradari, S., et al.\ 2011, \mnras, 412, 2689 

\bibitem[Green(2009)]{green09} Green, D.~A.\ 2009, Bulletin of the Astronomical Society of India, 37, 45 

\bibitem[Grimm et al.(2002)]{grimm02} Grimm, H.-J., Gilfanov, M., \& Sunyaev, R.\ 2002, \aap, 391, 923 

\bibitem[Grimm et al.(2003)]{grimm03} Grimm, H.-J., Gilfanov, M., \& Sunyaev, R.\ 2003, \mnras, 339, 793


  
\bibitem[Hamuy et al.(1992)]{hamuy92} Hamuy, M., Walker, A.~R., Suntzeff, N.~B., Gigoux, P., Heathcote, S.~R., 
\& Phillips, M.~M.\ 1992, \pasp, 104, 533 

\bibitem[Harris et al.(2001)]{harris01} Harris, J., Calzetti, 
D., Gallagher, J.~S., III, Conselice, C.~J., \& Smith, D.~A.\ 2001, \aj, 122, 3046


\bibitem[Hirayama et al.(2002)]{hirayama02} Hirayama, M., Nagase, F., Endo, T., Kawai, N., 
\& Itoh, M.\ 2002, \mnras, 333, 603 

\bibitem[Houghton \& Thatte(2008)]{hough08} Houghton, R.~C.~W., \& Thatte, N.\ 2008, \mnras, 385, 1110 


  
\bibitem[Hughes et al.(2007)]{hughes07} Hughes, J.~P., Chugai, N., Chevalier, R., Lundqvist, P., 
\& Schlegel, E.\ 2007, \apj, 670, 1260 
  
\bibitem[Immler et al.(2005)]{immler05} Immler, S., et al.\ 2005, \apj, 632, 283 

\bibitem[Immler et al.(1999)]{immler99} Immler, S., Vogler, A., Ehle, M., \& Pietsch, W.\ 1999, \aap, 352, 415 

\bibitem[Jarrett et al.(2003)]{jarrett03} Jarrett, T.~H., Chester, T., Cutri, R., Schneider, S.~E., 
\& Huchra, J.~P.\ 2003, \aj, 125, 525 

\bibitem[Jensen et al.(1981)]{jensen81} Jensen, E.~B., Talbot, R.~J., Jr., \& Dufour, R.~J.\ 1981, \apj, 243, 716 

\bibitem[Jones et al.(2009)]{jones09} Jones, D.~H., et al.\ 2009, \mnras, 399, 683 




\bibitem[Kalberla et 
al.(2005)]{kalberla05} Kalberla, P.~M.~W., Burton, W.~B., Hartmann, D., Arnal, E.~M., Bajaja, E., Morras, R., {\ P{\"o}ppel}, W.~G.~L.\ 2005, \aap, 440, 775 

  
\bibitem[Kilgard et al.(2002)]{kilgard02} Kilgard, R.~E., Kaaret, P., Krauss, M.~I., Prestwich, A.~H., Raley, M.~T., 
\& Zezas, A.\ 2002, \apj, 573, 138 


  
\bibitem[Kilgard et al.(2005)]{kilgard05} Kilgard, R.~E., et al.\ 2005, \apjs, 159, 214 

\bibitem[Kim et al.(2007)]{kim07} Kim, M., Wilkes, B.~J., Kim, D.-W., Green, P.~J., Barkhouse, W.~A., Lee, M.~G., 
Silverman, J.~D., \& Tananbaum, H.~D.\ 2007, \apj, 659, 29 


\bibitem[Kim et al.(2012)]{kim12} Kim, H., Whitmore, B.~C., Chandar, R., et al.\ 2012, \apj, 753, 26 

\bibitem[Knapen et al.(2010)]{knapen10} Knapen, J.~H., Sharp, R.~G., Ryder, S.~D., Falc{\'o}n-Barroso, J., Fathi, K., 
\& Guti{\'e}rrez, L.\ 2010, \mnras, 408, 797 



\bibitem[Kong et al.(2003)]{kong03} Kong, A.~K.~H., DiStefano, R., Garcia, M.~R., \& Greiner, J.\ 2003, \apj, 585, 298 



\bibitem[Kubota \& Makishima(2004)]{kubota04} Kubota, A., \& Makishima, K.\ 2004, \apj, 601, 428 

\bibitem[Kubota et al.(1998)]{kubota98} Kubota, A., Tanaka, Y., Makishima, K., Ueda, Y., Dotani, T., Inoue, H., 
\& Yamaoka, K.\ 1998, \pasj, 50, 667 


\bibitem[Lin et al.(2007)]{lin07} Lin, D., Remillard, R.~A., \& Homan, J.\ 2007, \apj, 667, 1073 

\bibitem[Long et al.(2010)]{long10} Long, K.~S., et al.\ 2010, \apjs, 187, 495 
  
\bibitem[Long et al.(2012)]{long12} Long, K.~S., et al.\ 2012, \apj, 756, 18 

\bibitem[Lundgren et 
al.(2004)]{lundgren04} Lundgren, A.~A., Wiklind, T., Olofsson, H., \& Rydbeck, G.\ 2004, \aap, 413, 505 

\bibitem[Luo et al.(2012)]{luo12} Luo, B., et al.\ 2012, \apj, 749, 130 


\bibitem[Ma et al.(2009)]{ma09} Ma, C., et al.\ 2009, IERS Technical Note, 35, 1 




\bibitem[Maddox et al.(2006)]{maddox06} Maddox, L.~A., Cowan, J.~J., Kilgard, R.~E., Lacey, C.~K., Prestwich, A.~H., 
Stockdale, C.~J., \& Wolfing, E.\ 2006, \aj, 132, 310 

\bibitem[Manchester et al.(1993)]{manchester93} Manchester, R.~N., Staveley-Smith, L., 
\& Kesteven, M.~J.\ 1993, \apj, 411, 756 

\bibitem[Mast et al. (2006)]{mast06} Mast, D., Diaz, R., \& Aguero, M.\ 2006, \aj, 131, 1394 

\bibitem[Mateos et al.(2008)]{mateos08} Mateos, S., et al.\ 2008, \aap, 492, 51 

\bibitem[McMullin et al.(2007)]{mcmullin07} McMullin, J.~P., Waters, B., Schiebel, D., Young, W., 
\& Golap, K.\ 2007, Astronomical Data Analysis Software and Systems XVI, 376, 127 

\bibitem[Milisavljevic et al.(2012)]{milisavljevic12} Milisavljevic, D., Fesen, R., Chevalier, R., Kirshner, R., Challis, P., 
\& Turatto, M.\ 2012, arXiv:1203.0006 

\bibitem[Mineo et al.(2012)]{mineo12} Mineo, S., Gilfanov, M., \& Sunyaev, R.\ 2012, \mnras, 419, 2095 


\bibitem[Mitsuda et al.(1989)]{mitsuda89} Mitsuda, K., Inoue, H., Nakamura, N., \& Tanaka, Y.\ 1989, \pasj, 41, 97 

\bibitem[Monet et al.(2003)]{monet03} Monet, D.~G., et al.\ 2003, \aj, 125, 984 


\bibitem[Morse et al.(2006)]{morse06} Morse, J.~A., Smith, N., Blair, W.~P., Kirshner, R.~P., Winkler, P.~F., 
\& Hughes, J.~P.\ 2006, \apj, 644, 188 


\bibitem[Park et al.(2011)]{park11} Park, S., Zhekov, S.~A., Burrows, D.~N., Racusin, J.~L., Dewey, D., 
\& McCray, R.\ 2011, \apjl, 733, L35 

\bibitem[Paturel et 
al.(2003)]{paturel03} Paturel, G., Petit, C., Prugniel, P., Theureau, G., Rousseau, J., Brouty, M., Dubois, P., \& Cambr{\'e}sy, L.\ 2003, \aap, 412, 45 



\bibitem[Patnaude et al.(2011)]{patnaude11} Patnaude, D.~J., Loeb, A., \& Jones, C.\ 2011, \na, 16, 187 

\bibitem[Pennington et al.(1982)]{pennington82} Pennington, R.~L., Talbot, R.~J., Jr., \& Dufour, R.~J.\ 1982, \aj, 87, 1538 



\bibitem[Piqueras L{\'o}pez et al.(2012)]{piqueras12} Piqueras L{\'o}pez, J., Davies, R., Colina, L., 
\& Orban de Xivry, G.\ 2012, \apj, 752, 47 

\bibitem[Plucinsky et al.(2008)]{plucinsky08} Plucinsky, P.~P., et al.\ 2008, \apjs, 174, 366 


\bibitem[Puche et al.(1990)]{puche90} Puche, D., Carignan, C., \& Bosma, A.\ 1990, \aj, 100, 1468 


\bibitem[Primini et al.(1993)]{primini93} Primini, F.~A., Forman, W., \& Jones, C.\ 1993, \apj, 410, 615 


\bibitem[Remillard \& McClintock(2006)]{remillard06} Remillard, R.~A., \& McClintock, J.~E.\ 2006, \araa, 44, 49 

\bibitem[Reynolds et al.(2009)]{reynolds09} Reynolds, S.~P., Borkowski, K.~J., Green, D.~A., Hwang, U., Harrus, I., 
\& Petre, R.\ 2009, \apjl, 695, L149 

\bibitem[Roeser et al.(2010)]{roeser10} Roeser, S., Demleitner, M., \& Schilbach, E.\ 2010, \aj, 139, 2440 




\bibitem[Saha et al.(2006)]{saha06} Saha, A., Thim, F., Tammann, G.~A., Reindl, B., 
\& Sandage, A.\ 2006, \apjs, 165, 108 

\bibitem[Sault et al.(1995)]{sault95} Sault, R.~J., Teuben, P.~J., 
\& Wright, M.~C.~H.\ 1995, Astronomical Data Analysis Software and Systems IV, 77, 433 


\bibitem[Schlegel et al.(1998)]{schlegel98} Schlegel, D.~J., Finkbeiner, D.~P., \& Davis, M.\ 1998, \apj, 500, 525 



\bibitem[Soria et al.(2012)]{soria12} Soria, R., Kuntz, K.~D., Winkler, P.~F., Blair, W.~P., Long, K.~S., Plucinsky, 
P.~P., \& Whitmore, B.~C.\ 2012, \apj, 750, 152 

\bibitem[Soria et al.(2014)]{soria14} Soria, R., Long, K.~S., Blair, W.~P., et al., Science, in press

\bibitem[Soria \& Perna(2008)]{soria08} Soria, R., \& Perna, R.\ 2008, \apj, 683, 767 

\bibitem[Soria \& Wu(2002)]{soria02} Soria, R., \& Wu, K.\ 2002, \aap, 384, 99 


\bibitem[Soria \& Wu(2003)]{soria03} Soria, R., \& Wu, K.\ 2003, \aap, 410, 53 

\bibitem[Stiele et 
al.(2011)]{stiele11} Stiele, H., Pietsch, W., Haberl, F., Hatzidimitriou, D., Barnard, R., Williams, B.~F., Kong, A.~K.~H., \& Kolb, U.\ 2011, \aap, 534, A55 


\bibitem[Sutherland \& Dopita(1995)]{southerland95} Sutherland, R.~S., \& Dopita, M.~A.\ 1995, \apj, 439, 381 

\bibitem[Stockdale et al.(2006)]{stockdale06} Stockdale, C.~J., Maddox, L.~A., Cowan, J.~J., Prestwich, A., Kilgard, R., 
\& Immler, S.\ 2006, \aj, 131, 889 

\bibitem[Stockdale et al.(2007)]{stockdale07} Stockdale, C.~J., Williams, C.~L., Weiler, K.~W., Panagia, N., Sramek, R.~A., 
Van Dyk, S.~D., \& Kelley, M.~T.\ 2007, \apj, 671, 689 

\bibitem[Takeuchi \& Ishii(2004)]{takeuchi04} Takeuchi, T.~T., \& Ishii, T.~T.\ 2004, \apj, 604, 40 



\bibitem[Talbot et al.(1979)]{talbot79} Talbot, R.~J., Jr., Jensen, E.~B., \& Dufour, R.~J.\ 1979, \apj, 229, 91 

\bibitem[Tamm et al.(2012)]{tamm12} Tamm, A., Tempel, E., Tenjes, P., Tihhonova, O., \& Tuvikene, T.\ 2012, \aap, 546, A4 




\bibitem[Tennant et al.(2001)]{tennant01} Tennant, A.~F., Wu, K., Ghosh, K.~K., Kolodziejczak, J.~J., 
\& Swartz, D.~A.\ 2001, \apjl, 549, L43 


\bibitem[Thatte et al. (2000)]{thatte00} Thatte, N., Tecza, M., \& Genzel, R.\ 2000, \aap, 364, 47 



\bibitem[T{\"u}llmann et al.(2011)]{tuellmann11} T{\"u}llmann, R., et al.\ 2011, \apjs, 193, 31 



\bibitem[Trinchieri et al.(1985)]{trinchieri85} Trinchieri, G., Fabbiano, G., \& Palumbo, G.~G.~C.\ 1985, \apj, 290, 96 



\bibitem[Turatto et al.(1989)]{turatto89} Turatto, M., Cappellaro, E., \& Danziger, I.~J.\ 1989, The Messenger, 56, 36 

\bibitem[Tzanavaris et al.(2013)]{tzanavaris13} Tzanavaris, P., Fragos, T., Tremmel, M., Jenkins, L., Zezas , A., Lehmer, B.~D., Hornschemeier, A., Kalogera, V., Ptak, A., \& Basu-Zych, A.~R.\ 2013, \apj, 774, 136

\bibitem[Verley et al.(2009)]{verley09} Verley, S., Corbelli, E., Giovanardi, C., \& Hunt, L.~K.\ 2009, \aap, 493, 453 



\bibitem[Weiler et al.(2002)]{weiler02} Weiler, K.~W., Panagia, N., Montes, M.~J., \& Sramek, R.~A.\ 2002, \araa, 40, 387 

\bibitem[Weiler et al.(2007)]{weiler07} Weiler, K.~W., Williams, C.~L., Panagia, N., Stockdale, C.~J., Kelley, M.~T., 
Sramek, R.~A., Van Dyk, S.~D., \& Marcaide, J.~M.\ 2007, \apj, 671, 1959 


\bibitem[Weisskopf et al.(2010)]{weisskopf10} Weisskopf, M.~C., Guainazzi, M., Jahoda, K., Shaposhnikov, N., O'Dell, S.~L., 
Zavlin, V.~E., Wilson-Hodge, C., \& Elsner, R.~F.\ 2010, \apj, 713, 912 


\bibitem[White et al.(1988)]{white88} White, N.~E., Stella, L., \& Parmar, A.~N.\ 1988, \apj, 324, 363 







\bibitem[Williams et al.(2008)]{williams08} Williams, B.~J., et al.\ 2008, \apj, 687, 1054 

\bibitem[Wilms et al.(2000)]{wilms00} Wilms, J., Allen, A., 
\& McCray, R.\ 2000, \apj, 542, 914 

\bibitem[Wofford et al.(2011)]{wofford11} Wofford, A., Leitherer, C., \& Chandar, R.\ 2011, \apj, 727, 100 

\bibitem[Wood \& Andrews(1974)]{wood74}Wood, R. \& Andrews, P. J. 1974, \mnras, 167, 13



\bibitem[Wu(2001)]{wu01} Wu, K.\ 2001, \pasa, 18, 443 


\bibitem[Zacharias et al.(2004)]{zacharias04} Zacharias, N., Monet, D.~G., Levine, S.~E., Urban, S.~E., Gaume, R., 
\& Wycoff, G.~L.\ 2004, Bulletin of the American Astronomical Society, 36, 1418 



\end{thebibliography}
\end{document}